\documentclass[aps,showpacs,amsmath,dvips,showkeys,preprint]{revtex4}

\usepackage{graphicx}
\usepackage{color}
 
\DeclareMathOperator*{\diag}{\tt diag}

\def \be  {\begin{equation}}
\def \ee  {\end{equation}}
\def \ee  {\end{equation}}
\def \bea {\begin{eqnarray}}
\def \eea {\end{eqnarray}}

\newcommand{\nn}{\nonumber}

\begin{document}

\preprint{ECTP-2014-06\hspace*{0.5cm}and\hspace*{0.5cm}WLCAPP-2014-06}
\title{SU(3) Polyakov Linear $\sigma$-Model in Magnetic Field: Thermodynamics, Higher-Order Moments, Chiral Phase Structure and Meson Masses}

\author{Abdel~Nasser~Tawfik\footnote{http://atawfik.net/}}
\affiliation{Egyptian Center for Theoretical Physics (ECTP), Modern University for Technology and Information (MTI), 11571 Cairo, Egypt}
\affiliation{World Laboratory for Cosmology And Particle Physics (WLCAPP), Cairo, Egypt}

\author{Niseem~Magdy} 
\affiliation{World Laboratory for Cosmology And Particle Physics (WLCAPP), 11571 Cairo, Egypt}
\affiliation{Brookhaven National Laboratory (BNL) - Department of Physics, P.O. Box 5000, Upton, NY 11973-5000, USA}

\begin{abstract}

Effects of an external magnetic field on various properties of the quantum chromodynamics (QCD) matter under extreme conditions of temperature and density (chemical potential) have been analysed. To this end, we use SU(3) Polyakov linear $\sigma$-model and assume that the external magnetic field ($e B$) adds some restrictions to the quarks energy due to the existence of free charges in the plasma phase. In doing this, we apply the Landau theory of quantization, which assumes that the cyclotron orbits of charged particles in magnetic field should be quantized. This requires an additional temperature to drive the system through the chiral phase-transition. Accordingly, the dependence of the critical temperature of chiral and confinement phase-transitions on the magnetic field is characterized. Based on this, we have studied the thermal evolution of thermodynamic quantities (energy density and trace anomaly) and the first four higher-order moment of particle multiplicity. Having all these calculations, we have studied the effects of the magnetic field on the chiral phase-transition. We found that both critical temperature $T_c$ and critical chemical potential increase with increasing the magnetic field, $e B$. Last but not least, the magnetic effects of the thermal evolution of four scalar and four pseudoscalar meson states are studied. We concluded that the meson masses decrease as the temperature increases till $T_c$. Then, the vacuum effect becomes dominant and rapidly increases with the temperature $T$. At low $T$, the scalar meson masses normalized to the lowest Matsubara frequency rapidly decrease as $T$ increases. Then, starting from $T_c$, we find that the thermal dependence almost vanishes. Furthermore, the meson masses increase with increasing magnetic field. This gives characteristic phase diagram of $T$ vs. external magnetic field $eB$. At high $T$, we find that the masses of almost all meson states become temperature independent. It is worthwhile to highlight that the various meson states likely have different critical temperatures.

\end{abstract}


\maketitle
\tableofcontents
\makeatletter
\let\toc@pre\relax
\let\toc@post\relax                 
\makeatother 

\section{Introduction}

It is believed that at high temperatures and densities there should be phase transition(s) between confined  nuclear matter and the quark-gluon plasma (QGP), where quarks and gluons are no longer confined inside hadron bags \cite{Rischke:2003mt}. Various theoretical studies have been devoted to tackle the possible change in properties of the strongly interacting matter, when the phase transition(s) between hadronic and partonic phases takes place under the effect of an external magnetic field  \cite{Marco2010,Marco2011,Marco2011-2,Fraga2013,Jens13,Skokov2011}. It is conjectured that the strongly interacting system (hadronic or partonic) can response to the external magnetic field with magnetization, $M$, and  magnetic susceptibility, $\chi_M$ \cite{Claudio2013}. Both quantities characterize the magnetic properties of the system of interest. Thus, the effects of the external magnetic field on the chiral condensates should be reflected in the chiral phase-transition \cite{lattice-delia}.  Also, the effects on the deconfinement order-parameter (Polyakov-loop) which includes the confinement-deconfinement phase-transition can be studied  \cite{lattice-delia}.

In an external magnetic field, the hadronic and partonic states are investigated in different models, such as the hadron resonance gas (HRG) model \cite{G_Endrödi}, and other effective models \cite{Klevansky1989,Gusynin1995,Babansky1998,Klimenko1999,Semenoff1999,Goyal2000,Hiller2008,Ayala2006,Boomsma2010}. The Nambu - Jeno-Lasinio (NJL) model \cite{Klevansky1992,Menezes2009,Menezes_2009}, the chiral perturbation theory \cite{Shushpanov1997,Agasian2000,Cohen2007}, the quark model  \cite{Kabat2002} and certain limits of QCD  \cite{Miransky2002} are also implemented. Furthermore, there are some studies devoted to the magnetic effects on the dynamical quark masses \cite{Klimenko2008}. The chiral magnetic-effect was studied in context of the Polyakov NJL (PNJL) model  \cite{Fukushima2010}. Recently, it was reported about lattice QCD calculations in an external magnetic field \cite{lattice-maxim,lattice-delia,Braguta:2011hq,Bali:2011qj,Bali:2012zg}. The Polyakov linear $\sigma$-model (PLSM) was implemented to estimate the effects of the magnetic field on the system \cite{Mizher:2010zb,Skokov2011,Marco2012}.

In the present work, we add some restrictions to the quarks energy due to the existence of free charges in the plasma phase. To this end, we apply the  Landau theory (Landau quantization) \cite{Landau}, which quantizes of the cyclotron orbits of the charged particles in the magnetic fields. We notice that this proposed configuration requires an additional temperature to drive the system through the chiral phase-transition. Accordingly, we find that the value of the chiral condensates increase with increasing the external magnetic field  \cite{Fraga2013}. A few remarks are now in order. In many different calculations for the thermal behavior of the chiral condensates and the deconfinement order-parameter (Polyakov-loop) using PNJL or NJL \cite{Marco2010,Marco2011,Marco2011-2}, the external magnetic field was not constant. Also, the dependence of the critical temperatures of chiral and confinement phase-transitions on the magnetic field was analysed \cite{Gabriel2012}. Almost the same study was conducted in PLSM \cite{Fraga2013,Jens13,Skokov2011}. All these studies lead to almost the same pattern, the critical temperature of chiral phase-transition increases with increasing the external magnetic field. But, the critical temperature of the confinement phase-transition behaves, oppositely. The latter behavior agrees - to some extend - with the lattice QCD calculations \cite{lattice-delia}.  In the present work, we study the effects of external magnetic field on the phase transition and deduce the phase-diagram curve using SU(3) PLSM \cite{Tawfik:2014uka}.

In light of this, we recall that the PLSM is widely implemented in different frameworks and different purposes. The LSM was introduced by Gell-Mann and Levy in 1960 \cite{Gell Mann:1960} long time before QCD was known to be the theory of strong interaction. Many studies have been performed with LSM like $\mathcal{O}(4)$ LSM \cite{Gell Mann:1960}, $\mathcal{O}(4)$  LSM at finite temperature \cite{Lenaghan:1999si, Petropoulos:1998gt} and $U(N_f)_r \times U(N_f)_l$ LSM for $N_f=2$, $3$ or even $4$ quark flavors \cite{l, Hu:1974qb, Schechter:1975ju, Geddes:1979nd}. In order to obtain reliable results, Polyakov-loop corrections have been added to LSM, in which information about the confining glue sector of the theory was included in form of Polyakov-loop potential. This potential is to be extracted from the pure Yang-Mills lattice simulations \cite{Polyakov:1978vu, Susskind:1979up, Svetitsky:1982gs,Svetitsky:1985ye}. So far, many studies were devoted to investigating the phase diagram and the thermodynamics of PLSM at different Polyakov-loop forms with two \cite{Wambach:2009ee,Kahara:2008} and three quark flavors \cite{Schaefer:2008ax,Mao:2010,Tawfik:2014uka}. Also, the magnetic field effect on the QCD phase-transition and other system properties are investigated using PLSM \cite{Mizher:2010zb,Skokov2011,Marco2012}.

The present paper is organized as follows.  In section \ref{sec:approaches}, we introduce details about SU(3) PLSM under the effects of an external magnetic field. Section \ref{sec:Results} gives some features of the PLSM in an external magnetic field, such as the quark condensates, Polyakov loop, some thermal quantities, the phase-transition(s) and scalar and pseudoscalar meson masses under the magnetic field effect. In section \ref{sec:conclusion}, the final conclusions and outlook shall be presented.
 
\section{Approach}
 \label{sec:approaches}

The Lagrangian of LSM with $N_f =2+1$ quark flavors and $N_c =3$ color degrees of freedom, where the quarks couple to the Polyakov-loop dynamics, was introduced in Ref. \cite{Schaefer:2008ax,Mao:2010,Tawfik:2014uka},
\begin{eqnarray}
\mathcal{L}=\mathcal{L}_{chiral}-\mathbf{\mathcal{U}}(\phi, \phi^*, T), \label{plsm}
\end{eqnarray}
where the chiral part of the Lagrangian $\mathcal{L}_{chiral}=\mathcal{L}_q+\mathcal{L}_m$ has $SU(3)_{L}\times SU(3)_{R}$ symmetry  \cite{Lenaghan,Schaefer:2008hk}. The Lagrangian with $N_f =2+1$ consists of two parts.  The first part represents fermions, Eq. (\ref{lfermion}) with a flavor-blind Yukawa coupling $g$ of the quarks. The coupling between the effective gluon field and quarks, and between the magnetic field, $B$, and the quarks is implemented through the covariant derivative  \cite{Skokov2011}
\begin{eqnarray}
\mathcal{L}_q &=& \sum_f \overline{\psi}_f(i\gamma^{\mu}
D_{\mu}-gT_a(\sigma_a+i \gamma_5 \pi_a)) \psi_f, \label{lfermion} 
\end{eqnarray}
where the summation $\sum_f$ runs over the three flavors, $f=1, 2, 3$ for $u$-, $d$- and $s$-quark, respectively, $  T_a $ is the Gell-Man matrices. The flavor-blind Yukawa coupling, $g$, should couple the quarks to the mesons \cite{blind}. The coupling of the quarks to the Euclidean gauge field, $A_{\mu}$, was discussed in Ref \cite{Polyakov:1978vu,Susskind:1979up}. For the Abelian gauge field, the influence of the external magnetic field, $A_{\mu}^{M}$, \cite{Mizher:2010zb} is given by the covariant derivative  \cite{Skokov2011},
\begin{eqnarray}
D_{\mu}&=& \partial_{\mu} - i\, A_{\mu} - i\, Q\, A_{\mu}^{EM}, \label{covdiv} 
\end{eqnarray}
where $A_{\mu}=g\, A_{\mu}^a \lambda^a/2$ and $A_{\mu}^{EM}=(0,Bx,0,0)$ and $Q$ is a matrix defined by the quark electric charges $Q=\diag(q_u, q_d, q_s) $  for up, down and strange quarks, respectively. The interaction of charged pion $\pi^{\pm}=(\pi_1\pm i \pi_2)/\sqrt{2}$ with the magnetic field is included by $D_{\mu}=\partial_{\mu} - i\, e\, A_{\mu}^{M}$ with $e$ is the electric charge  \cite{Skokov2011}.

The second part of chiral Lagrangian stands for the the mesonic contribution, Eq. (\ref{lmeson}), 
\begin{eqnarray}
\mathcal{L}_m &=& \mathrm{Tr}(\partial_{\mu}\Phi^{\dag}\partial^{\mu}\Phi-m^2 \Phi^{\dag} \Phi)-\lambda_1 [\mathrm{Tr}(\Phi^{\dag} \Phi)]^2 \nonumber\\
&-& \lambda_2 \mathrm{Tr}(\Phi^{\dag} \Phi)^2+c[\mathrm{Det}(\Phi)+\mathrm{Det}(\Phi^{\dag})] +\mathrm{Tr}[H(\Phi+\Phi^{\dag})].  \label{lmeson}
\end{eqnarray}
In Eq. (\ref{lmeson}), $\Phi$ is a complex $3 \times 3$ matrix, which depends on the $\sigma_a$ and $\pi_a$ \cite{Schaefer:2008hk}, where $\gamma^{\mu}$ are the chiral spinors, $\sigma_a$ are the scalar mesons and $\pi_a$ are the pseudoscalar mesons.

The second term in Eq. (\ref{plsm}), $\mathbf{\mathcal{U}}(\phi, \phi^*, T)$, represents the Polyakov-loop effective potential \cite{Polyakov:1978vu}, which is expressed by using the dynamics of the thermal expectation value of a color traced Wilson loop in the temporal direction  $\Phi (\vec{x})= \langle \mathcal{P}(\vec{x})\rangle/N_c$. 
Then, the Polyakov-loop potential and its conjugate read $\phi = (\mathrm{Tr}_c \,\mathcal{P})/N_c, \label{phais1}$ and
$\phi^* = (\mathrm{Tr}_c\,  \mathcal{P}^{\dag})/N_c, \label{phais2}$, respectively. $\mathcal{P}$, which stands for the Polyakov loop,   can be represented by a matrix in the color space  \cite{Polyakov:1978vu} 
\begin{eqnarray}
 \mathcal{P}(\vec{x})=\mathcal{P}\mathrm{exp}\left[i\int_0^{\beta}d \tau A_4(\vec{x}, \tau)\right],\label{loop}
\end{eqnarray}
where $\beta=1/T$ is the inverse temperature and $A_4 = iA^0$ is the Polyakov gauge \cite{Polyakov:1978vu,Susskind:1979up}. The Polyakov loop matrix can be given  as a diagonal representation \cite{Fukushima:2003fw}. 

In the PLSM Lagrangian, Eq. (\ref{plsm}), the coupling between the Polyakov loop and the quarks is given by the covariant derivative of $D_{\mu}=\partial_{\mu}-i A_{\mu}$  \cite{Mao:2010}.  It is apparent that the PLSM Lagrangian  is invariant under the chiral flavor-group. This is similar to the original QCD Lagrangian \cite{Ratti:2005jh,Roessner:2007,Fukushima:2008wg}. In order to reproduce the thermodynamic behavior of the Polyakov loop for pure gauge, we use a temperature-dependent potential $U(\phi, \phi^{*},T)$. This should agree with the lattice QCD simulations and have $Z(3)$ center symmetry as that of the pure gauge QCD Lagrangian \cite{Ratti:2005jh,Schaefer:2007d}. In case of vanishing chemical potential, then $\phi=\phi^{*}$ and the Polyakov loop is considered as an order parameter for the deconfinement phase-transition  \cite{Ratti:2005jh,Schaefer:2007d}. In the present work, we use $U(\phi, \phi^{*},T)$, Landau-Ginzburg type potential, as a polynomial expansion in $\phi$ and $\phi^{*}$ \cite{Ratti:2005jh,Roessner:2007,Schaefer:2007d,Fukushima:2008wg}
\begin{eqnarray}
\frac{\mathbf{\mathcal{U}}(\phi, \phi^*, T)}{T^4}=-\frac{b_2(T)}{2} \phi\; \phi^* -\frac{b_3
}{6}(\phi^3+\phi^{*3})+\frac{b_4}{4}(\phi\; \phi^*)^2, \label{Uloop}
\end{eqnarray}
where, $\phi$ and $\phi^*$ are introduced previously  and $b_2(T)=a_0+a_1\left(T_0/T\right)+a_2\left(T_0/T\right)^2+a_3\left(T_0/T\right)^3$, where constants are $a_0=6. 75 $, $a_1=-1. 95$, $a_2=2. 625 $, $a_3=-7. 44 $,  $b_3 = 0.75$ and $b_4=7.5$. 

In Eq. (\ref{Uloop}), the Vandermonde Jacobian contribution, $\kappa\, \ln[J(\phi,\phi^*)]$, was ignored due the small value of $\kappa$. In principle, the Vandermonde term comes from the change of variables from vector potential to $\phi$ in the path integral and should guarantee a reasonable behavior of the mean field approximation \cite{simon}, i.e. it was suggested to solve the problem that the normalized Polyakov loop becomes greater than $1$ at very high temperatures.
\bea
J[\phi,\phi^*] &=& \frac{27}{24 \pi^2} \left[1-6 \phi\,\phi^* + 4 (\phi^3 + \phi^{*3}) - 3 (\phi\,\phi^*)^2\right], \nn
\eea
where  $J(\phi,\phi^*)$ is the Vandermonde determinant, which is not explicitly space-time dependent. The dimensionless parameter $\kappa$ would be dependent on the temperature and the chemical potential. Therefore, $\kappa$ should be estimated, phenomenologically. 

In order to reproduce the pure gauge QCD thermodynamics and the behavior of the Polyakov loop as a function of  temperature, we use the parameters listed out above in this section (\ref{sec:approaches}) \cite{Ratti:2005jh}. In calculating the grand potential, we use the mean field approximation \cite{Tawfik:2014uka},
\begin{eqnarray}
\Omega(T, \mu) &=& U(\sigma_x, \sigma_y) + \mathbf{\mathcal{U}}(\phi, \phi^*, T) + \Omega_{\bar{\psi}
\psi} (T;\phi,\phi^{*},B). \label{potential}
\end{eqnarray}

The purely mesonic potential is given as,
\begin{eqnarray}
U(\sigma_x, \sigma_y) &=& \frac{m^2}{2} (\sigma^2_x+\sigma^2_y)-h_x
\sigma_x-h_y \sigma_y-\frac{c}{2\sqrt{2}} \sigma^2_x \sigma_y \nonumber \\
&+& \frac{\lambda_1}{2} \sigma^2_x \sigma^2_y +\frac{1}{8} (2 \lambda_1
+\lambda_2)\sigma^4_x + \frac{1}{4} (\lambda_1+\lambda_2)\sigma^4_y,  \label{Upotio}
\end{eqnarray}
where $m^2$, $h_x$, $h_y$, $\lambda_1$, $\lambda_2$ and $c$  are the model fixed  parameters \cite{Schaefer:2008hk}. 
The quarks and antiquark contribution to the medium potential was introduced in Ref \cite{Tawfik:2014hwa} and based on Landau quantization and magnetic catalysis concepts, App. \ref{appnd:1}, we get
\begin{eqnarray}
\Omega_{\bar{\psi} \psi}(T, \mu_f, eB) &=& -2 \sum_{f}  \dfrac{|q_{f}| B T}{2 \pi} \sum_{\nu=0}^{\infty} \int \dfrac{d p}{2 \pi} \left(2-1 \delta_{0\nu}\right) \nn \\
& &  \hspace*{10mm} \left\{ \ln \left[ 1+3\left(\phi+\phi^* e^{-\frac{(E_f - \mu_f)}{T}}\right)\, e^{-\frac{(E_f - \mu_f)}{T}}+e^{-3 \frac{(E_f - \mu_f)}{T}}\right] \right. \nonumber \\ 
&& \hspace*{8.5mm} \left.  +\ln \left[ 1+3\left(\phi^*+\phi e^{-\frac{(E_f + \mu_f)}{T}}\right)\, e^{-\frac{(E_f + \mu_f)}{T}}+e^{-3\frac{(E_f + \mu_f)}{T}}\right] \right\},   \label{new-qqpotio}
\end{eqnarray}
It is worthwhile to highlight that the chemical potential used everywhere in the manuscript is the quark one, $\mu_f$ with $f$ being the quark flavor. The different variables are elaborated in the App. \ref{appnd:1}.  The potential at vanishing $e B$ reads 
\begin{eqnarray} 
\Omega_{\bar{q}q}(T, \mu_f)&=& -2 \,T \sum_{f=l, s} \int_0^{\infty} \frac{d^3\vec{p}}{(2 \pi)^3} \nn \\
& &  \hspace*{5mm} \left\{ \ln \left[ 1+3(\phi+\phi^* e^{-(E_f-\mu_f)/T})\, e^{-(E_f -\mu_f)/T}+e^{-3 (E_f-\mu)/T}\right] \right. \nonumber \\ 
&& \hspace*{3.2mm} \left.  +\ln \left[ 1+3(\phi^*+\phi e^{-(E_f+\mu_f)/T})\, e^{-(E_f+\mu_f)/T}+e^{-3 (E_f+\mu)/T}\right] \right\}.  \label{thermalOMG}
\end{eqnarray}
This is the system free of Landau quantization.

The Landau theory quantizes of the cyclotron orbits of charged particles in magnetic field. For small magnetic fields, the number of occupied Landau levels (LL) is large and the quantization effects are washed out, while for large magnetic fields, the Landau levels are less occupied and the chiral symmetry restoration occurs for smaller values of the chemical potential.

According to Eqs. (\ref{new-qqpotio}) and (\ref{thermalOMG}), Eq. (\ref{potential}) get an additional term,
\begin{eqnarray}
\Omega(T, \mu_f, eB) &=& U(\sigma_x, \sigma_y) + \mathbf{\mathcal{U}}(\phi, \phi^*, T) + \Omega_{\bar{\psi}
\psi} (T,\mu_f;\phi,\phi^{*},eB) + \delta_{0,eB} \Omega_{\bar{\psi}\psi} (T,\mu_f;\phi,\phi^{*}), \hspace*{10mm} \label{potentialNEW}
\end{eqnarray}
where $\Omega_{\bar{\psi}\psi} (T,\mu_f;\phi,\phi^{*})$ represents the potential term at vanishing magnetic field, $\delta_{0,eB}$ switches between the two systems; one at vanishing and one at finite magnetic field.

We notice that the sum in Eqs. (\ref{Uloop}), (\ref{new-qqpotio}) and (\ref{Upotio}) give the thermodynamic potential density as in Eq. (\ref{potential}). By using the minimization condition, App. \ref{appnd:2}, we can evaluate the parameters. Having the thermodynamic potential, Eq. (\ref{potential}), we can determine all thermal quantities including the higher-order moments of particle multiplicity, and then mapping out the chiral phase-diagram \cite{Tawfik:2014uka}. The meson masses are defined by the second derivative with respect to the corresponding fields of the grand potential, Eq. (\ref{potential}), evaluated at its minimum.

\section{Results}
\label{sec:Results}

The results of the chiral condensates $\sigma_x$ and $\sigma_y$, section \ref{subsec:condensates}, the thermodynamic quantities, section \ref{subsec:thermo}, the non-normalized and normalized higher-order moment of particle multiplicity, section \ref{sec:H-M} and section \ref{sec:NH-M}, respectively, the chiral phase-transition, section \ref{subsec:phase-T} and finally the meson masses, section \ref{sec:masses}, are introduced as follows.

\subsection{Phase transition: quark condensates and order parameters}
\label{subsec:condensates}

The thermal evolution of the chiral condensates, $\sigma_x$ and $\sigma_y$, and the Polyakov order parameters, $\phi$ and $\phi^*$ is calculated from Eq. (\ref{potential}) at finite chemical potential and finite magnetic field using the minimization conditions given in Eq. (\ref{cond1}). The dependence on the four parameters, temperature $T$, chemical potential $\mu$, magnetic field $B$ and minimization parameter with respect to it the minimization condition shall be analysed.

\begin{figure}[htb] \label{sigxy}
\centering{
\includegraphics[width=5.cm,angle=-90]{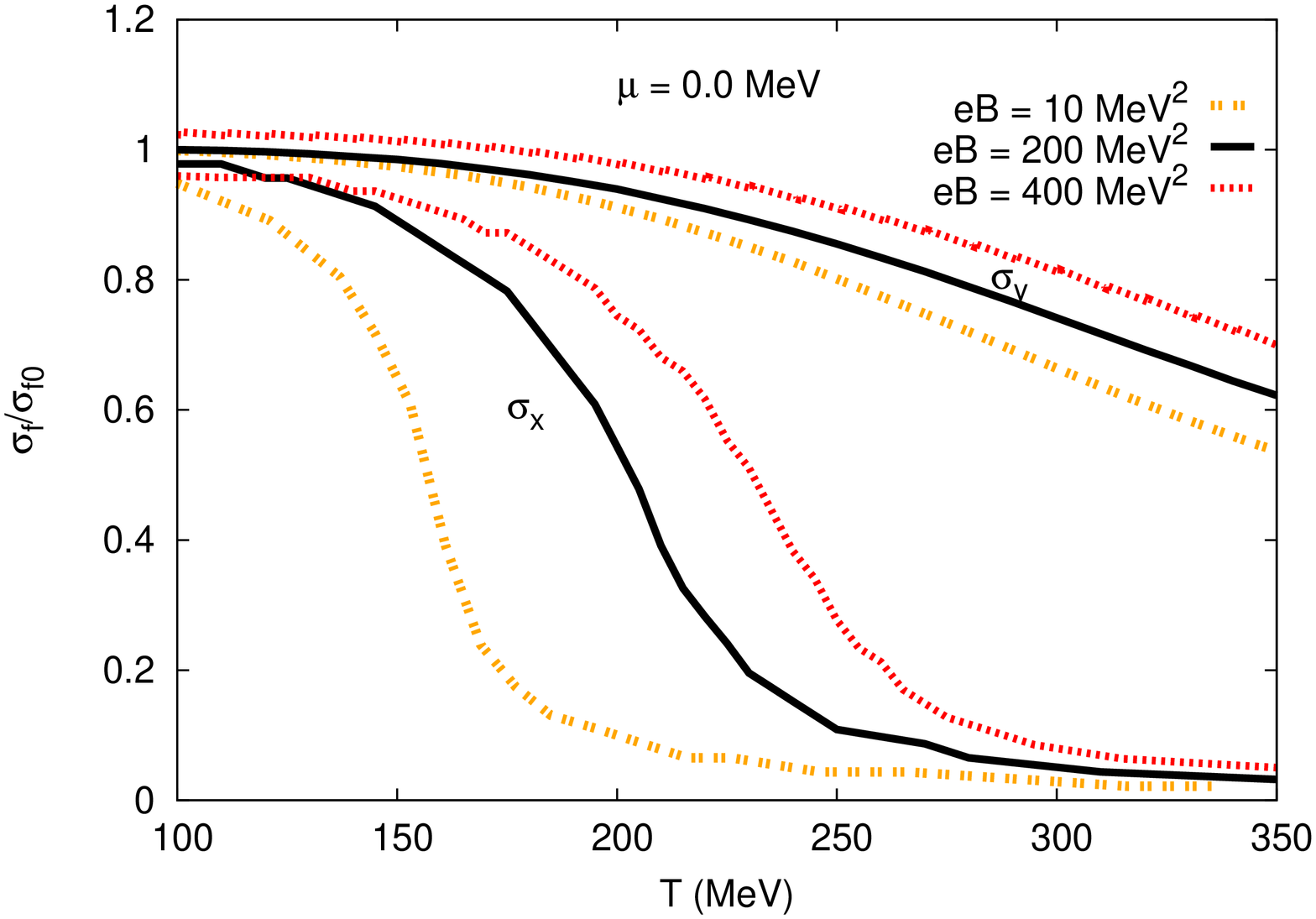}
\includegraphics[width=5.cm,angle=-90]{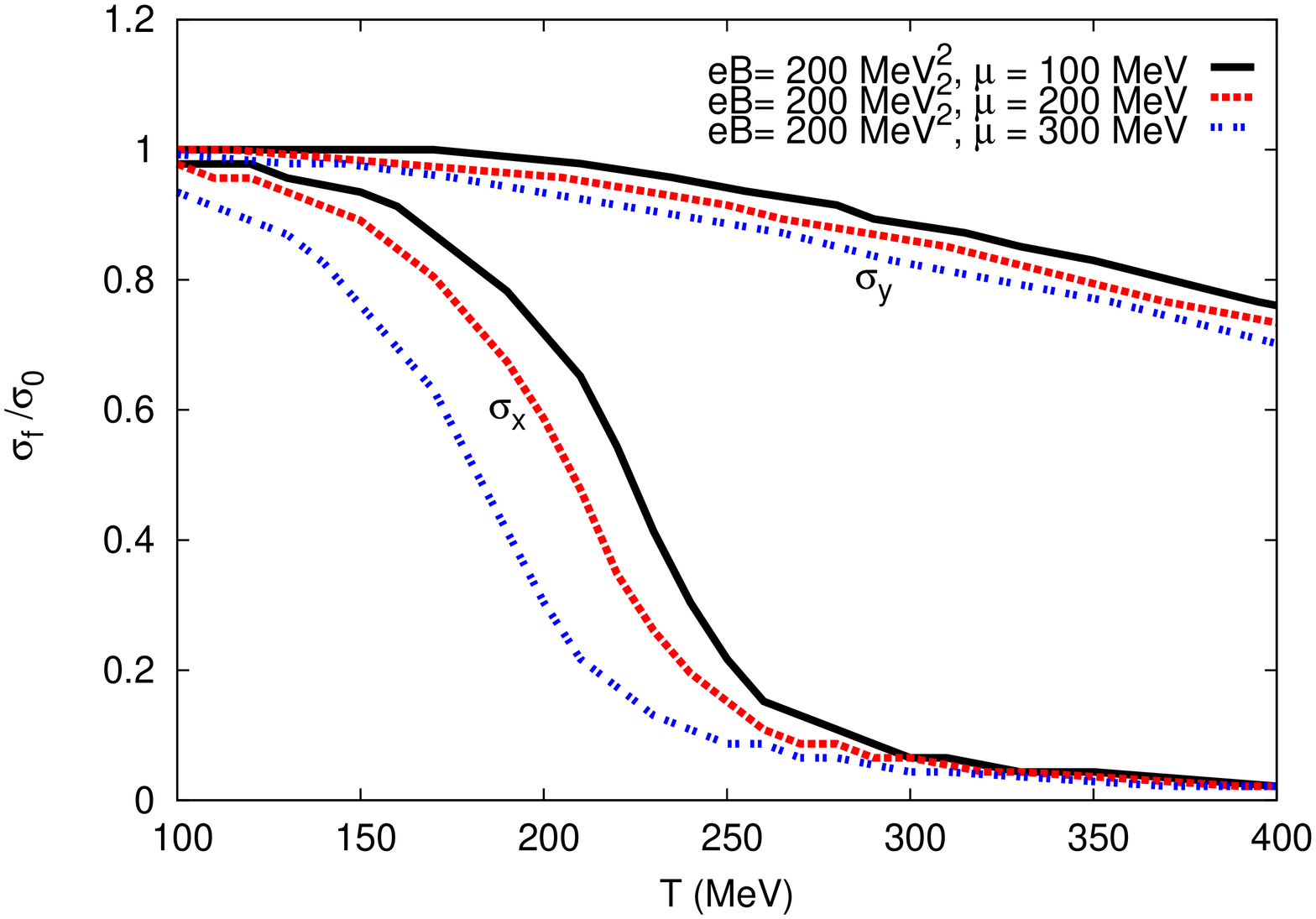}
\caption{(Color online) Left-hand panel (a): the normalized chiral-condensates, $\sigma_x$ (lower curves) and $\sigma_y$ (upper curves), are given as function of temperature at vanishing chemical potential and different magnetic field values, $eB=10~$MeV$^2$ (double-dotted curve),  $200~$MeV$^2$ (solid curve) and  $400~$MeV$^2$ (dotted curve). Right-hand panel (b): the same as in left-hand panel but at a constant magnetic field $eB=200~$MeV$^2$ and different quark chemical potentials, $\mu=100~$MeV (solid curve),  $200~$MeV (dotted curve) and  $300~$MeV (double-dotted curve).   
\label{fig:sig-with-t}}
}
\end{figure}

In left-hand panel (a) of Fig. \ref{fig:sig-with-t}, the normalized chiral condensates, $\sigma_x$ and $\sigma_y$, are given as function of temperature at  vanishing chemical potential and different magnetic field values, $eB=10~$MeV$^2$ (double-dotted curve),  $200~$MeV$^2$ (solid curve) and  $400~$MeV$^2$ (dotted curve). We notice that both condensates increase with increasing the magnetic field, $eB$. This dependence seems to explain the increase in the chiral critical temperature $T_c$ with the magnetic field. This - in turn - agrees with various studies using PLSM and PNJL \cite{Marco2010,Marco2011,Marco2011-2,Fraga2013,Jens13,Skokov2011}.  The condensates become moderated (smoother) with increasing magnetic field.

The right-hand panel (b) of Fig. \ref{fig:sig-with-t} shows the chiral condensates, $\sigma_x$ and $\sigma_y$, as function of temperature at constant magnetic field $eB=200~$MeV$^2$, and finite chemical potentials, $\mu=100~$MeV (solid curve), $200~$MeV (dotted curve) and $300~$MeV (double-dotted curve). Both condensates  decrease with increasing the chemical potentials. This dependence gives a signature for the decreasing behavior of the chiral critical temperature $T_c$  with increasing the chemical potential, which obviously agrees with our previous calculations \cite{Tawfik:2014uka}. The condensates become rowdy (sharper) with increasing chemical potential.

\begin{figure}[htb]
\centering{
\includegraphics[width=5.cm,angle=-90]{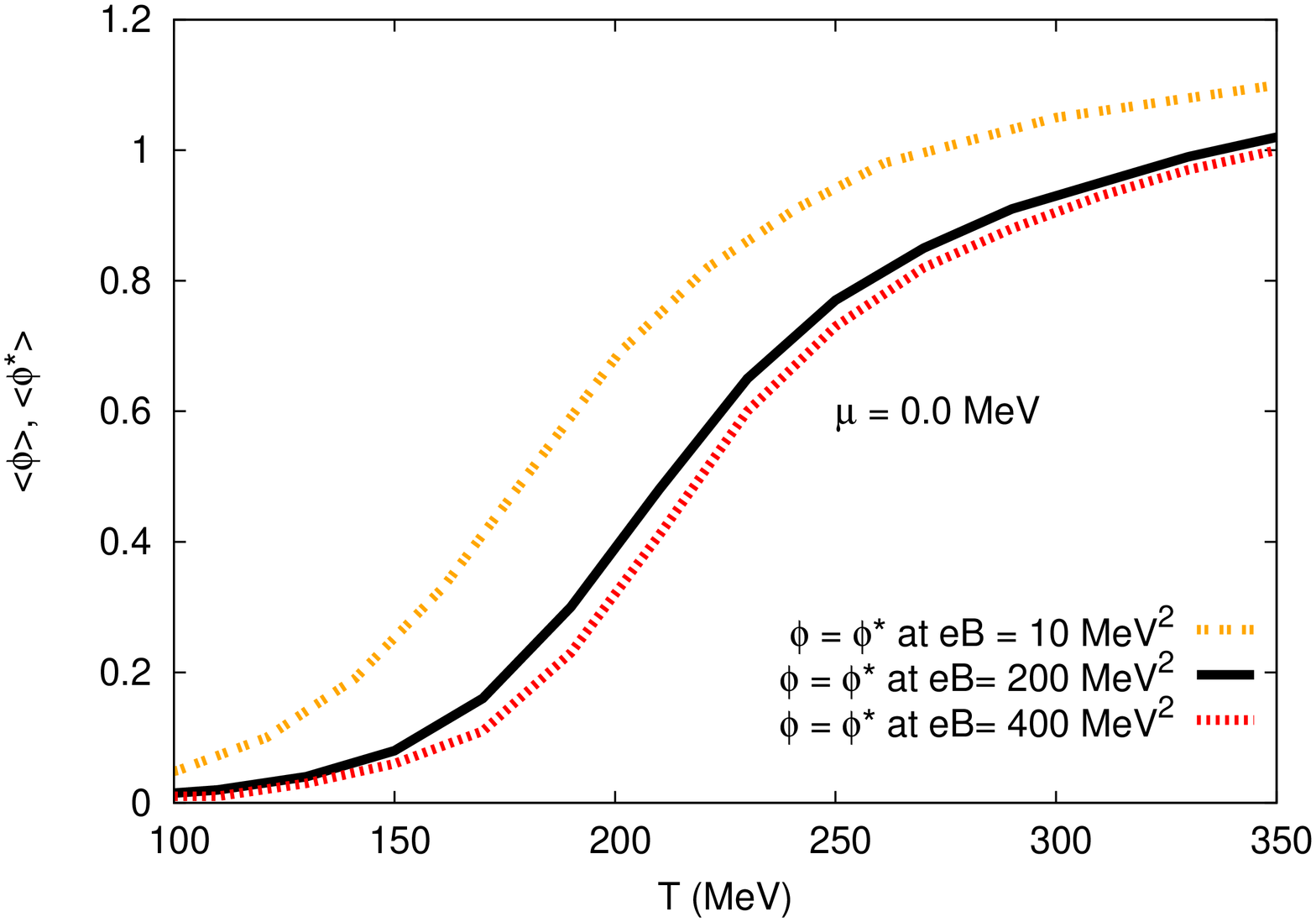}
\includegraphics[width=5.cm,angle=-90]{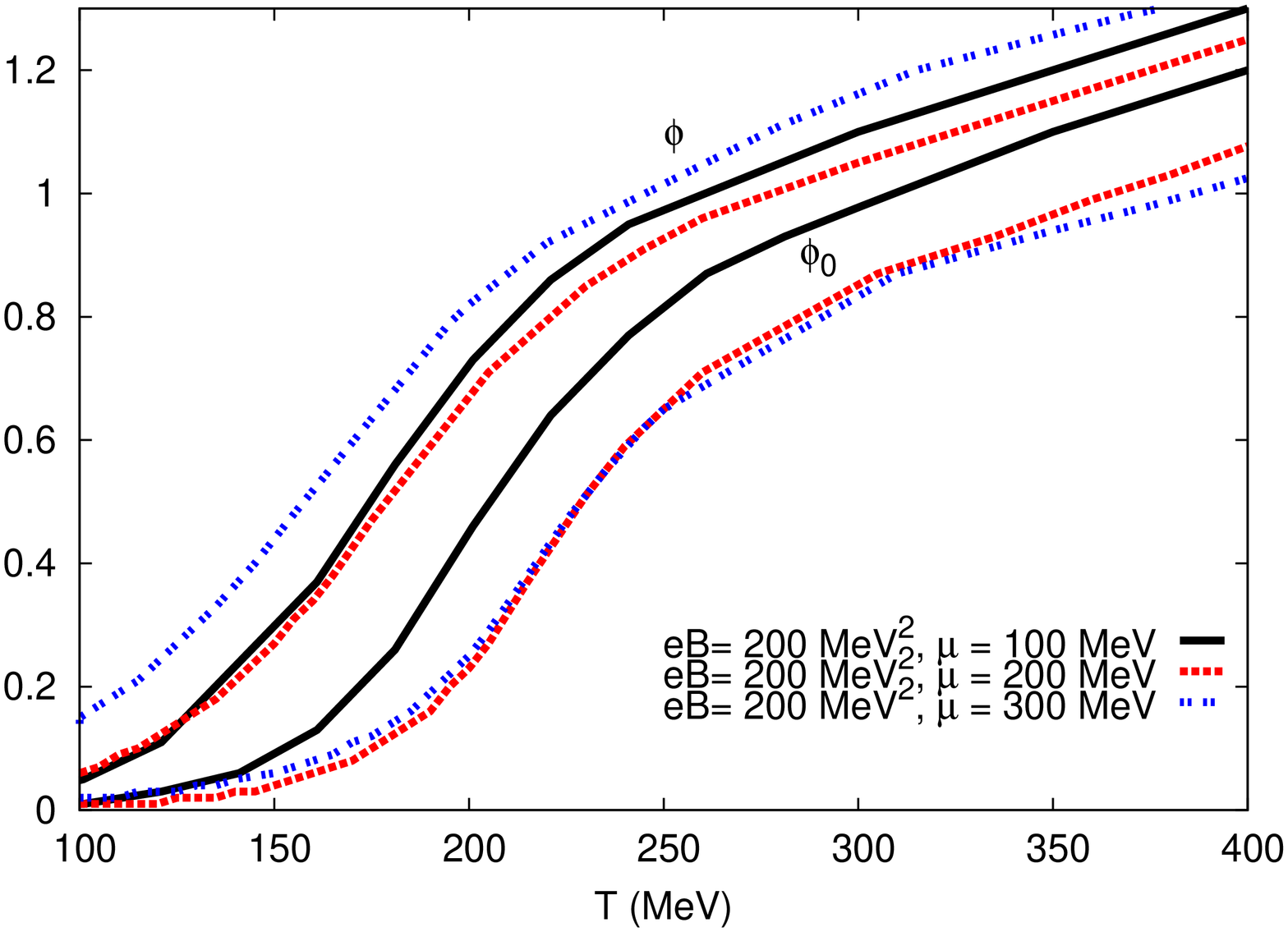}
\caption{(Color online) Left-hand panel (a): the Polyakov-loop field and its conjugation, $\phi$ and $\phi^*$, are given as function of temperature at a vanishing constant chemical potential and different magnetic field values, $eB=10~$MeV$^2$ (double-dotted curve),  $0.2~$GeV$^2$ (solid curve) and  $400~$MeV$^2$ (dotted curve). Right-hand panel (b): the same as in the left-hand panel but at a constant magnetic field, $eB=0.2~$GeV$^2$ and different quark chemical potential values, $\mu=0.1~$GeV (solid curve),  $0.2~$GeV (dashed curve) and  $0.3~$GeV (double-dotted curve).  \label{fig:fi-with-t}}
}
\end{figure}

The left-hand panel (a) of Fig. \ref{fig:fi-with-t} shows the Polyakov-loop field and it is conjugation, $\phi$ (upper curves) and $\phi^*$ (lower curves), as function of temperature at a vanishing chemical potential and different magnetic field values, $eB=0.01~$GeV$^2$ (double-dotted curve),  $0.2~$GeV$^2$ (solid curve) and  $0.4~$GeV$^2$ (dotted curve). Both fields decrease with increasing the magnetic field. This behavior explains the dependence of the confinement critical temperature on the magnetic field. At vanishing chemical potential, $\phi=\phi^*$. Both Polyakov-loop fields become smoother with increasing magnetic field.

The right-hand panel (b) draws the same as in left-hand panel but at a constant magnetic field $eB=0.2~$GeV$^2$ and different quark chemical potential values, $\mu=0.1~$GeV (solid curve),  $0.2~$GeV (dashed curve) and  $0.3~$GeV (double-dotted curve). We find that $\phi$ increases with increasing the chemical potential values but $\phi^*$ decreases. This behavior seems to agree with our previous calculations \cite{Tawfik:2014uka}.  At finite chemical potential, $\phi>\phi^*$.

We conclude that the Polyakov-loop fields, $\phi$ and $\phi^*$, increase with $T$, Fig. \ref{fig:fi-with-t}. At vanishing $\mu$, both $\phi$ and $\phi^*$ decrease with increasing $eB$. At finite $\mu$, we find that $\phi$ increases, while $\phi^*$ decreases with $eB$.

\subsection{Thermodynamic quantities}
\label{subsec:thermo}

In this section, we introduce some thermal quantities like energy density and trace anomaly.  As we discussed in Ref. \cite{Tawfik:2014uka}, the purely mesonic potential, Eq. (\ref{Upotio}) gets infinity at very low temperature and entirely vanishes at high temperature. From this numerical estimation, we concluded that this part of potential is only effective at very low temperatures. Its dependence on the external magnetic field has been checked and was found that finite $eB$ comes up with very tiny contribution to this potential part. As the present study is performed at temperatures around the critical one, this potential part can be removed from the effective potentials given in Eq. (\ref{potential}).  In Eq. (\ref{Upotio}), the chiral condensates, $\sigma$'s, are small at finite temperature, Fig. \ref{fig:sig-with-t}. Therefore,  much smaller values are expected for their higher orders and multiplications. Opposite situation is likely at very small temperatures.

\subsubsection{Energy density}
\label{subsubsec:presser}

The energy density, $\epsilon/T^4$, at finite quark chemical potential, $\mu_f$, can be obtained as
\begin{eqnarray}
\epsilon(T, \mu_f, eB) &=& -\frac{\partial}{\partial (1/T)} \ln Z(T, \mu_f, eB). 
\label{e1} 
\end{eqnarray} 
In section \ref{subsec:condensates}, we have estimated the parameters, the two chiral condensates, $\sigma_x$ and $\sigma_y$ and  the two order parameters of the Polyakov-loop and it's conjugation, $\phi$ and $\phi^*$, respectively. Thus, we can substitute all these into Eq. (\ref{e1}).

\begin{figure}[htb]
\centering{
\includegraphics[width=5.cm,angle=-90]{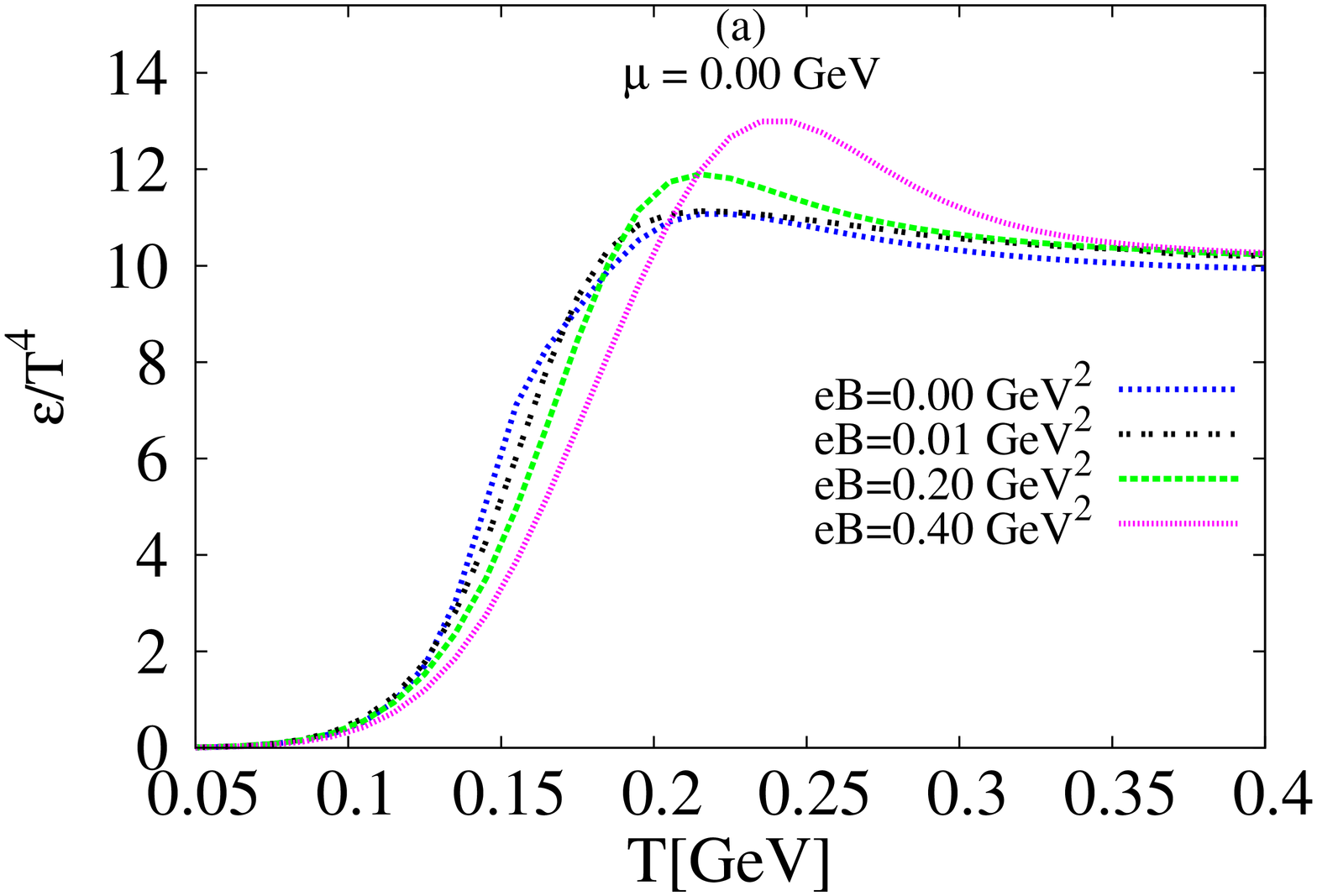}
\includegraphics[width=5.cm,angle=-90]{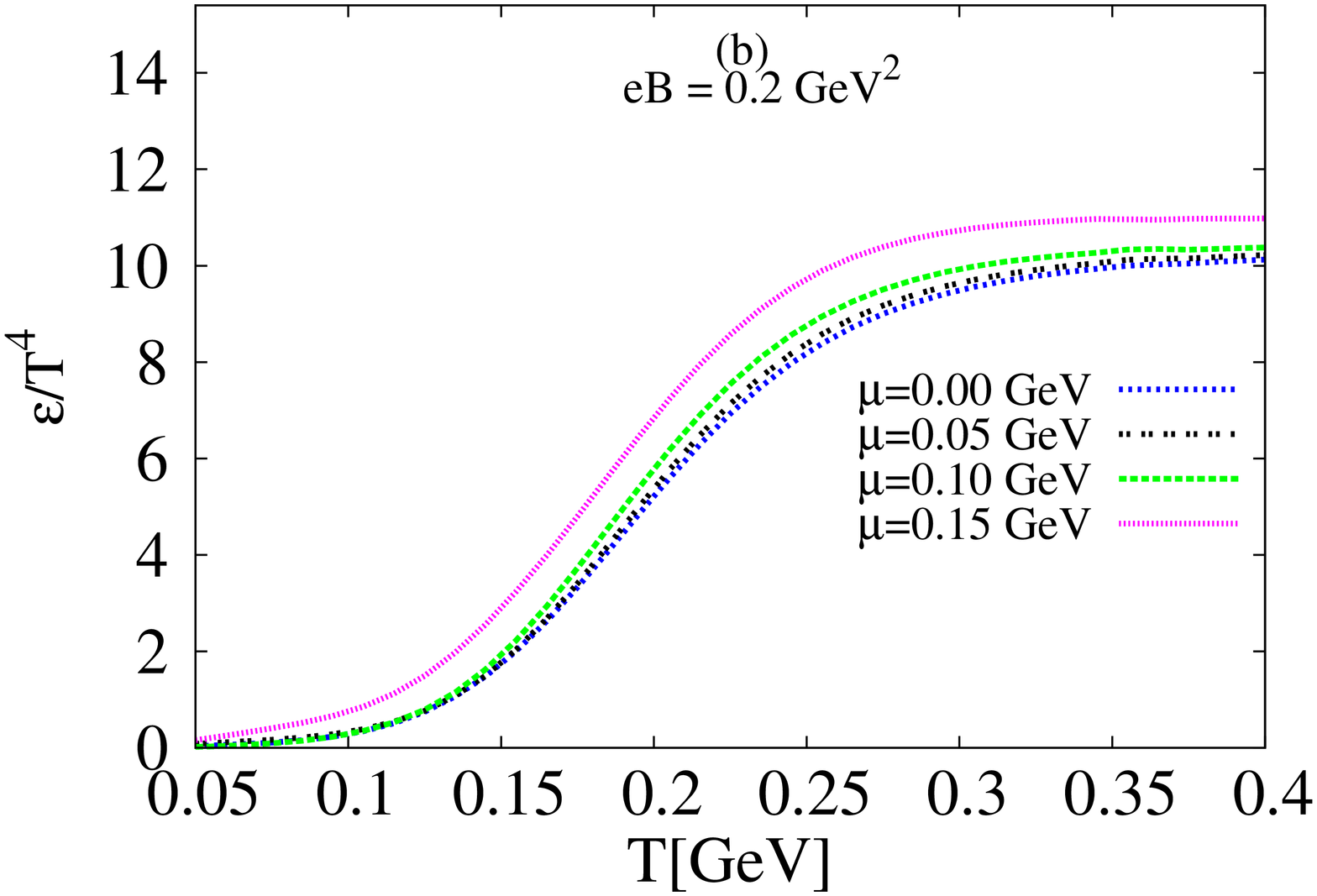}
\caption{(Color online) Left-hand panel (a): the normalized energy density, $\epsilon/T^4$, is given as function of temperature at a vanishing chemical potential and different magnetic fields, without magnetic field (solid curve) \cite{Tawfik:2014uka}, $eB=0.01~$GeV$^2$ (long-dashed curve),  $eB=0.2~$GeV$^2$ (dashed curve) and  $eB=0.4~$GeV$^2$ (double-dotted curve). The right-hand panel (b) shows the same as in the left-hand panel but at a constant magnetic field $eB=0.2~$GeV$^2$ and different chemical potential values, $\mu=0.1~$GeV (long-dashed curve), $0.2~$GeV (dash-dotted curve) and  $0.3~$GeV (double-dotted curve). The upper curves represent results from Eq. \ref{new-qqpotio} plus Eq. \ref{thermalOMG}, while lower curves are based on thermodynamic derivatives from the thermal potential, Eq. \ref{new-qqpotio}.
\label{fig:pr}}
}
\end{figure}

The left-hand panel (a) of Fig. \ref{fig:pr} presents the normalized energy density, $\epsilon/T^4$, as function of temperature at vanishing chemical potential.  In calculating the results, Eqs. (\ref{new-qqpotio}) and Eq. (\ref{thermalOMG}) are implemented as given in Eq. (\ref{potentialNEW}). The general temperature-dependence is not absent. Also, we notice that $\epsilon/T^4$  is sensitive to the change in $eB$ \cite{Tawfik:2014hwa}.  Increasing $eB$ seems to increase the critical temperature, at which the system undergoes phase transition. As the chiral condensates become smoother with increasing $eB$, the thermodynamic quantities, such as energy density, behave accordingly, i.e. the phase transition becomes smoother as well.

The right-hand panel (b) shows $\epsilon/T^4$ as function of temperature at a constant magnetic field $eB=0.2~$GeV$^2$ and varying quark chemical potentials, $\mu=0.1~$GeV (long-dashed curve), $0.2~$GeV (dash-dotted curve) and  $0.3~$GeV (double-dotted curve). The solid curve represents the results in absence of an external magnetic field but at $\mu=0.1~$GeV. We note that $\epsilon/T^4$  is not as sensitive to the change in $\mu$ \cite{Tawfik:2014uka} as to the external magnetic field.  Despite the lack of chemical potential dependency, which can be understood due to the large magnetic field applied, it is believed to affect contrary to the chemical potential. To this indirect dependency of $\mu$ and $e B$, we shall devote a separate work. Again, it seems that increasing $\mu_f$, decreasing $T_c$.

\subsubsection{Trace anomaly}
\label{subsubsec:trace}

\begin{figure}[htb]
\centering{
\includegraphics[width=5.cm,angle=-90]{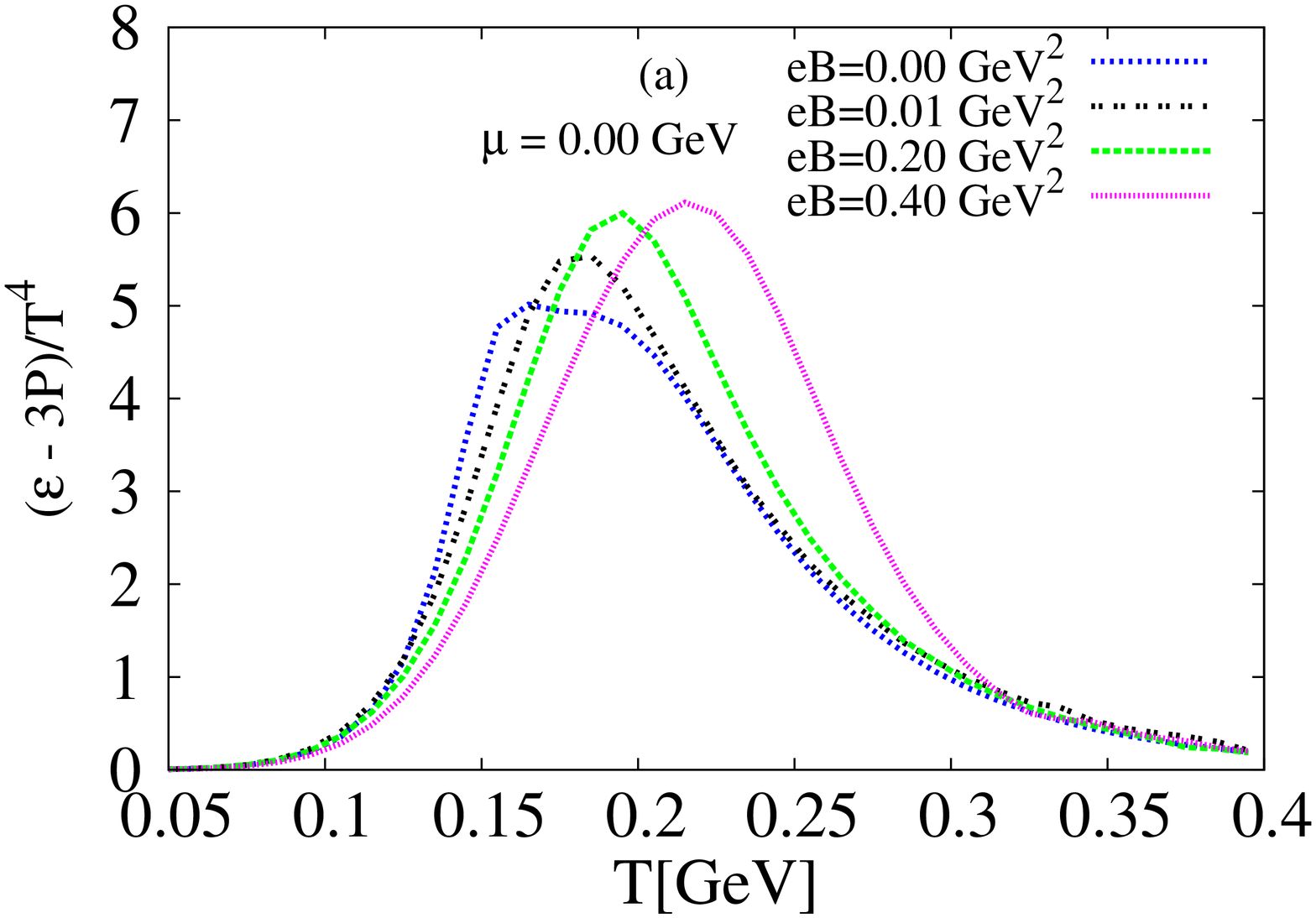}
\includegraphics[width=5.cm,angle=-90]{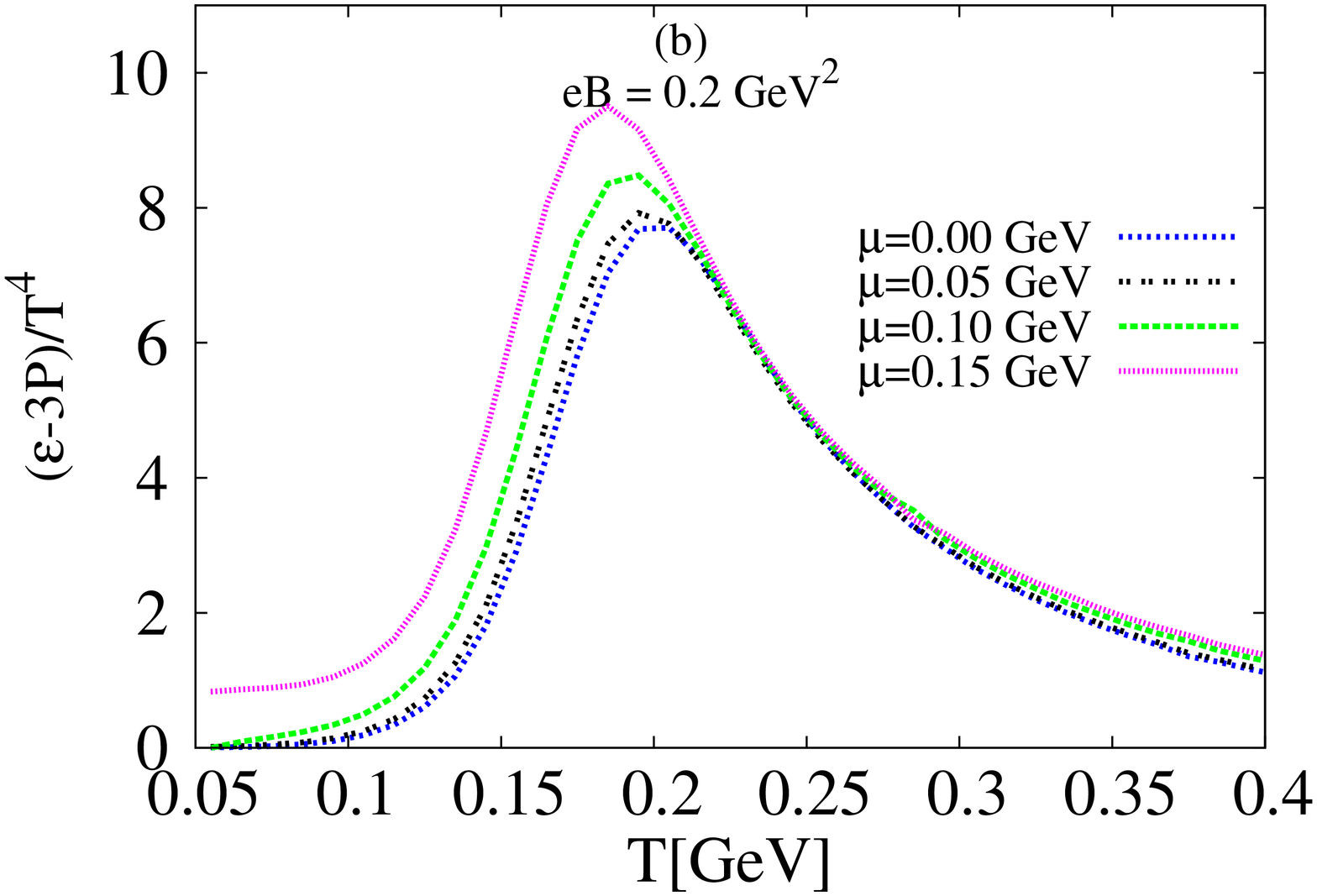}
\caption{(Color online) The same as in Fig. \ref{fig:pr} but for the the trace anomaly $(\epsilon - 3 p)/T^4$. 
\label{fig:tr}}
}
\end{figure}

At finite quark chemical potential, the trace anomaly known as interaction measure reads
\begin{eqnarray} 
\dfrac{\epsilon(T, \mu_f, eB) - 3 p(T, \mu_f, eB) }{T^4} =\,T \frac{\partial}{\partial T} \; \frac{p(T, \mu_f, eB) }{T^4}. \label{tr1} 
\end{eqnarray}
In Fig. \ref{fig:tr}, we notice that the normalized trace-anomaly under the effect of an external magnetic field becomes smaller than the corresponding quantity in absence of magnetic field \cite{Tawfik:2014uka} at high temperature. This can be explained due the restrictions added to the quark energy by the Landau quantization through the magnetic field. We find that increasing $eB$ increases the critical temperature. This behavior can be understood because of the dependence of the chiral condensates, Fig. \ref{fig:sig-with-t} and the Polyakov-loop potential, Fig. \ref{fig:fi-with-t} on $eB$. 

In the left-hand panel (a), the trace anomaly $(\epsilon - 3 p)/T^4$, is given as function of temperature $T$ at vanishing chemical potential but different values of the magnetic fields, vanishing \cite{Tawfik:2014uka} (solid curve), $eB=0.01~$GeV$^2$ (long-dashed curve),  $eB=0.2~$GeV$^2$ (dash-dotted curve) and  $eB=0.4~$GeV$^2$ (double-dotted curve). We notice that the trace anomaly increases with $T$ until the chiral symmetry is restored. Then, increasing $T$ reduces the normalized trace anomaly. The peak represents the critical temperature $T_c$ corresponding to a certain magnetic field. We find that $T_c$ increases with increasing $eB$. 

The right-hand panel (b) of Fig. \ref{fig:tr} shows the same as in the left-hand panel but at a constant magnetic field $eB=0.2~$GeV$^2$ and different chemical potentials, $\mu=0.0~$GeV (solid curve), $0.1~$GeV  (long-dashed curve),  $200~$MeV  (dash-dotted curve) and $300~$GeV (double-dotted curve). The trace anomaly increases with $T$ until the chiral symmetry is fully restored. The peaks are positioned at $T_c$ of the certain value for chemical potential. Here, we find that $T_c$ decreases with increasing $\mu$.  The sensitivity to $\mu$ is not as strong as to $eB$. This might be interpreted as the high magnetic field applied seems to contradict the effects of the chemical potential. In other words, should the magnetic field adds energy to the system, the chemical potential requires energy in order to produce new particles. We notice that the dependence on the quark chemical potential is more obvious that that shown in Fig. \ref{fig:pr}.

\subsubsection{Magnetic catalysis effect}
\label{subsubsec:M_F_C}

In App. \ref{appnd:1}, we discuss the magnetic catalysis, Eq. (\ref{DR}), and the so-called dimension reduction concepts, Eq.(\ref{DR}). Due to the effects of the magnetic field, the latter would mean modifying the sum over the three-dimensional momentum space to a one-dimensional one. According to Ref. \cite{Tawfik:2014uka}, the effect of this reduction reduces also the value of the quantity by almost two third from the expected value. This would explain the difference between results at vanishing and that at finite $eB$, left-hand panels (a) of Figs. \ref{fig:pr} and \ref{fig:tr}, for instance.   In the present work, we distinguish between two types of systems. In the first one, the Landau quantization should be implemented, i.e. taking into account the magnetic effects, while in the other system, the external magnetic field is not taken into consideration, i.e. no magnetic contribution to the thermal system.

\subsection{Higher-order moment of particle multiplicity}
\label{sec:H-M}

The higher-order moment of the particle multiplicity is defined  \cite{Tawfik:2014uka,Tawfik:2012si} as
\bea
m_i &=& \frac{\partial^i}{\partial\, \mu^i} \frac{p(T,\mu,B)}{T^4},
\eea
where the pressure $p(T,\mu,B)=-\,T\, \partial\, \ln {\cal Z}(T,\mu,B)/\partial V$ is related to the partition function, which in tern is related to the potential, $\ln {\cal Z}(T,\mu,B)=-V\, \Omega(T,\mu,B)/T$.

In this section, we introduce the first four non-normalized moments of the particle multiplicity calculated in PLSM under the effects of an external magnetic field. The thermal evolution is studied at a constant chemical potential but different magnetic fields and also at a constant magnetic field but different chemical potentials. Doing this, it is possible to map out the chiral phase-diagram, for which we determine the irregular behavior in the higher-order moments as function of $T$ and $\mu$. 

\subsubsection{Non-normalized higher-order moments}
\label{sec:non-H-M}

\begin{figure}[htb]
\centering{
\includegraphics[width=5.cm,angle=-90]{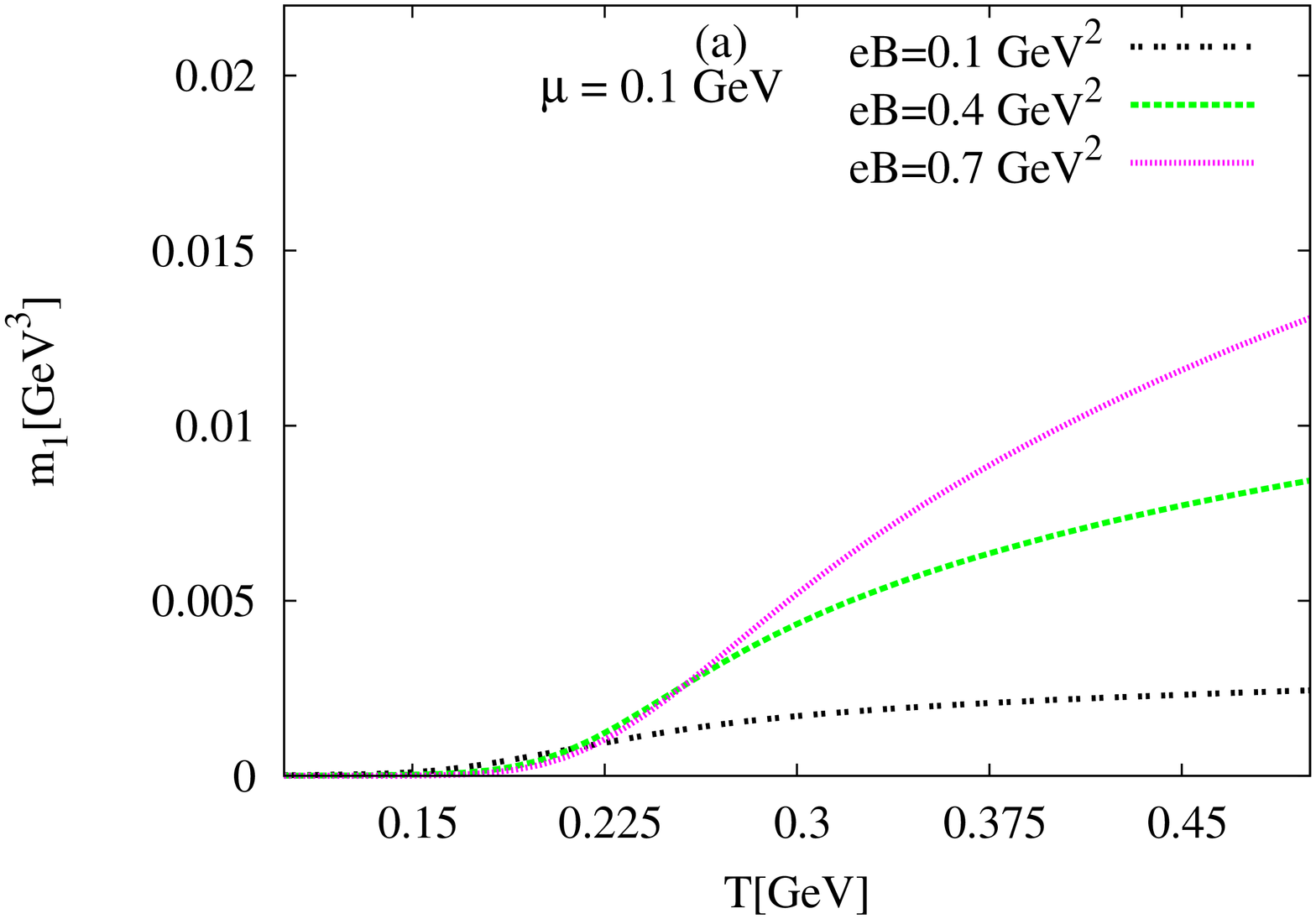}
\includegraphics[width=5.cm,angle=-90]{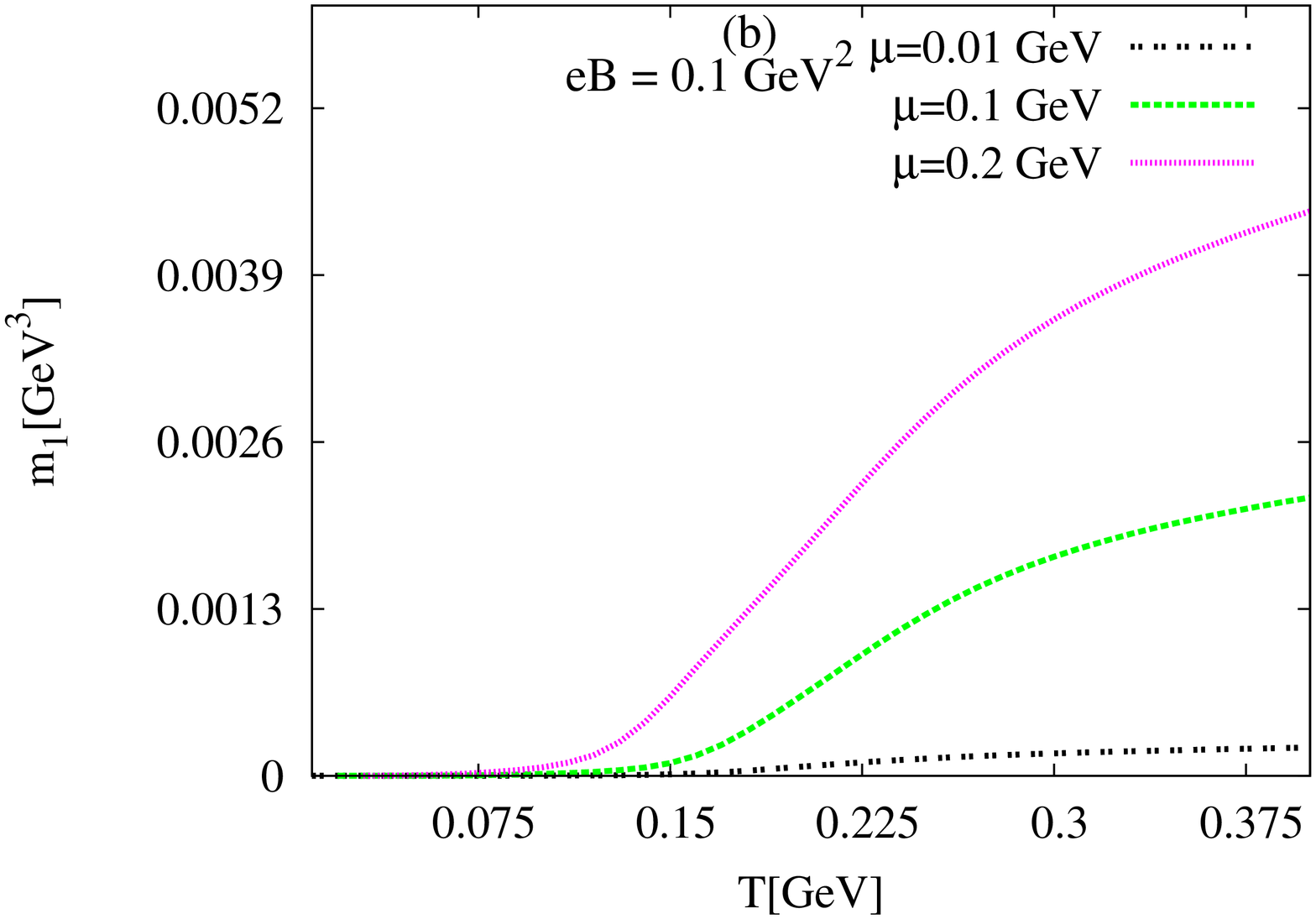}
\caption{(Color online) Left-hand panel (a): non-normalized quark number density, $m_1$, is given as function of temperature at a constant  chemical potential $\mu=0.1~$GeV and different values of the magnetic field,  $eB=0.1~$GeV$^2$ (double-dotted curve),  $0.4~$GeV$^2$ (dashed curve) and  $0.7~$GeV$^2$ (dotted curve). Right-hand panel (b) shows the same as in the left-hand panel but at a constant magnetic field $eB=0.1~$GeV$^2$ and different chemical potentials,  $\mu=0.01~$GeV (double-dotted curve),  $0.1~$GeV  (dashed curve) and  $0.2~$GeV (dotted curve).  \label{fig:m1}}
}
\end{figure}

\begin{figure}[htb]
\centering{
\includegraphics[width=5.cm,angle=-90]{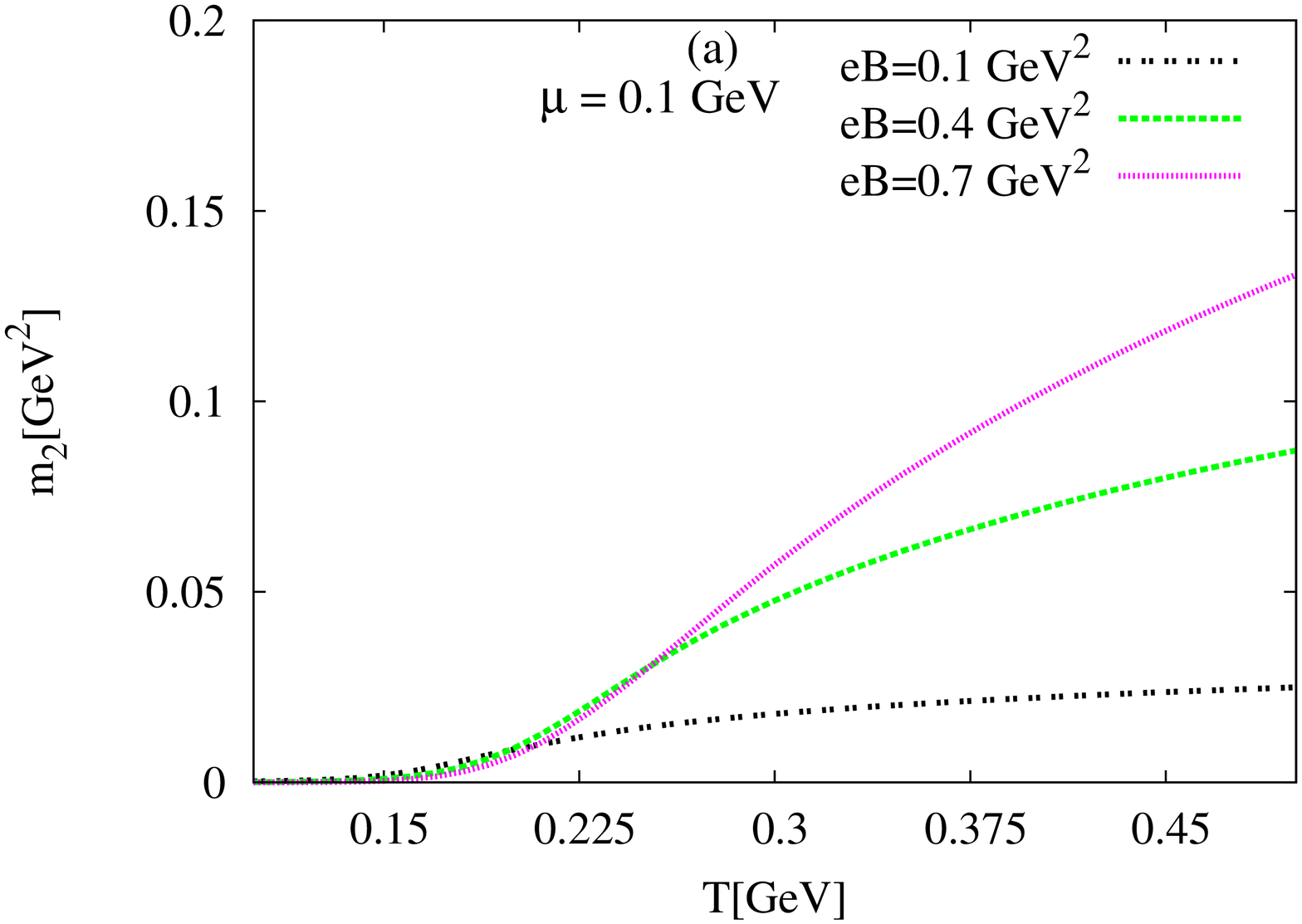}
\includegraphics[width=5.cm,angle=-90]{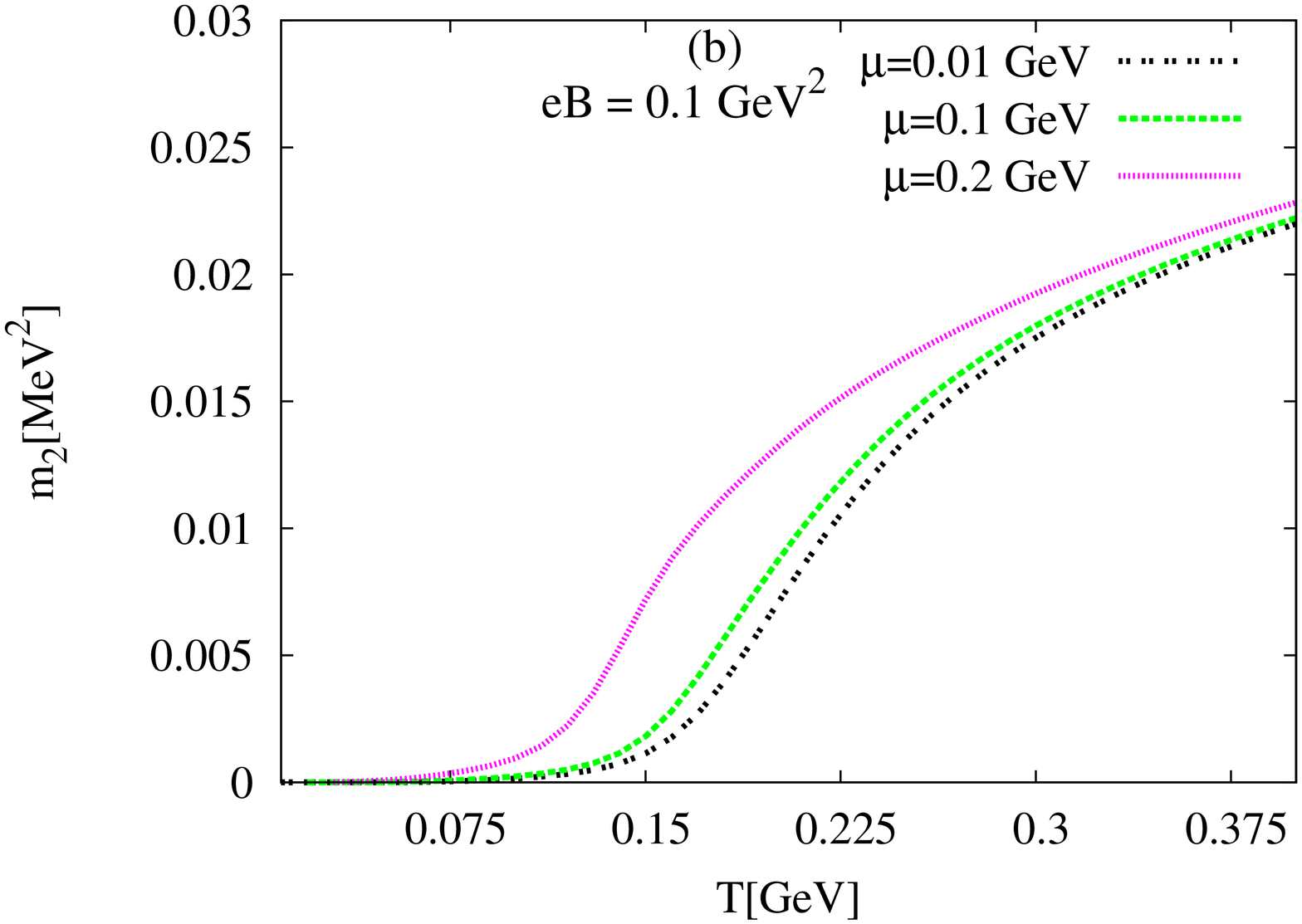}
\caption{(Color online) The same as in Fig. \ref{fig:m1} but for quark number susceptibility $m_2$. \label{fig:m2}}
}
\end{figure}

\begin{figure}[htb]
\centering{
\includegraphics[width=5.cm,angle=-90]{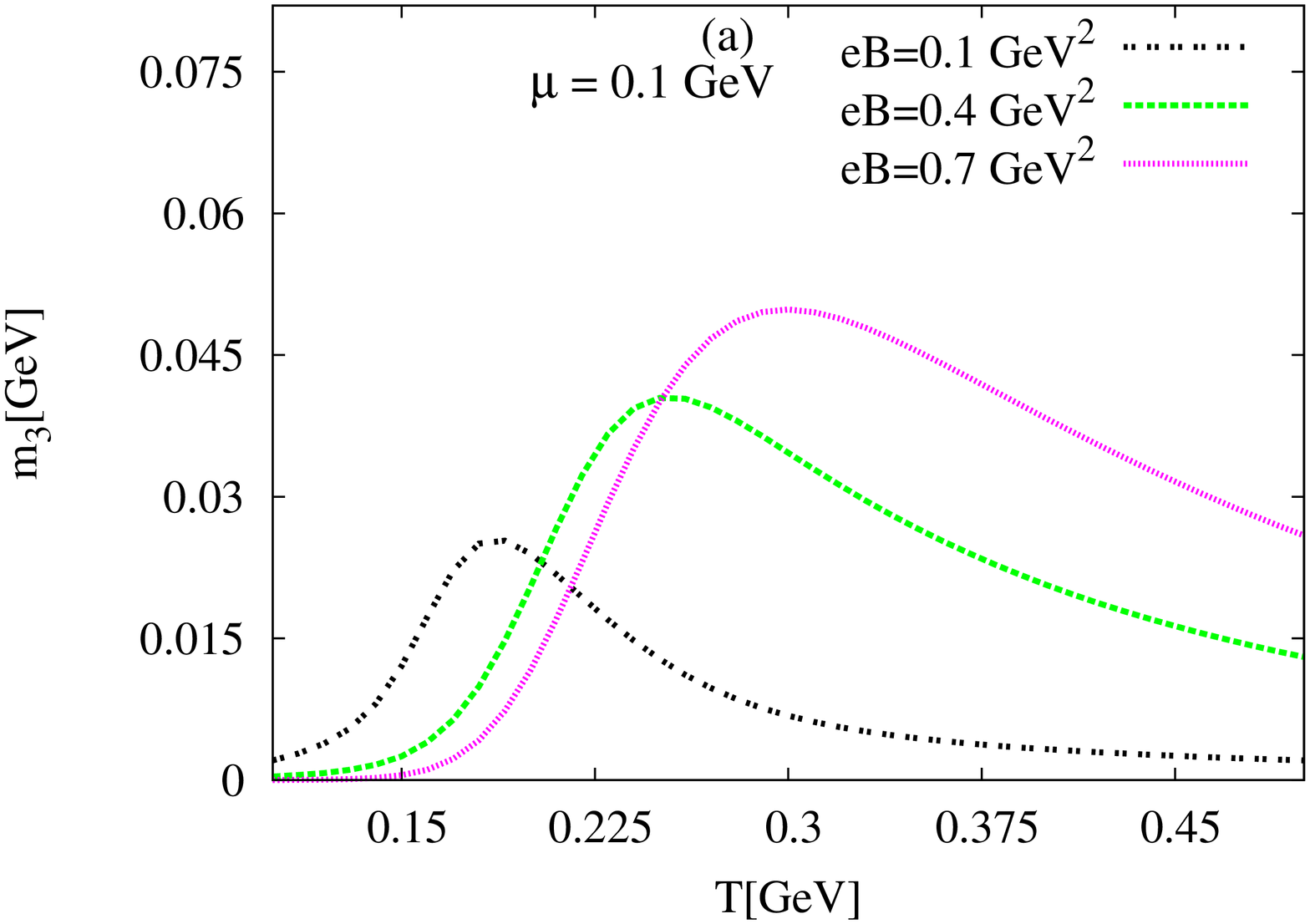}
\includegraphics[width=5.cm,angle=-90]{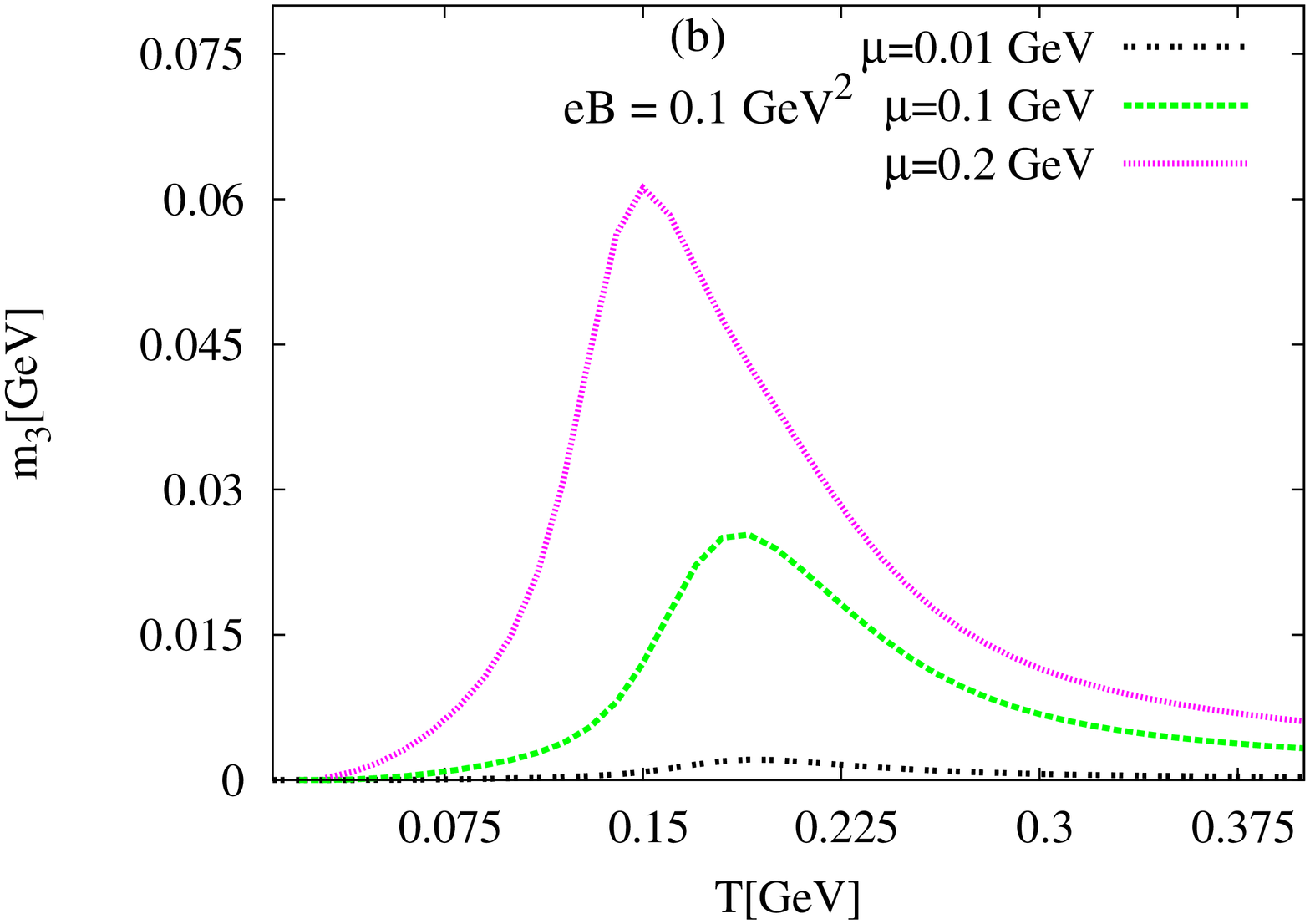}
\caption{(Color online) The same as in Fig. \ref{fig:m1} but for the third-order moment of quark number density $m_3$. \label{fig:m3}}
}
\end{figure}

\begin{figure}[htb]
\centering{
\includegraphics[width=5.cm,angle=-90]{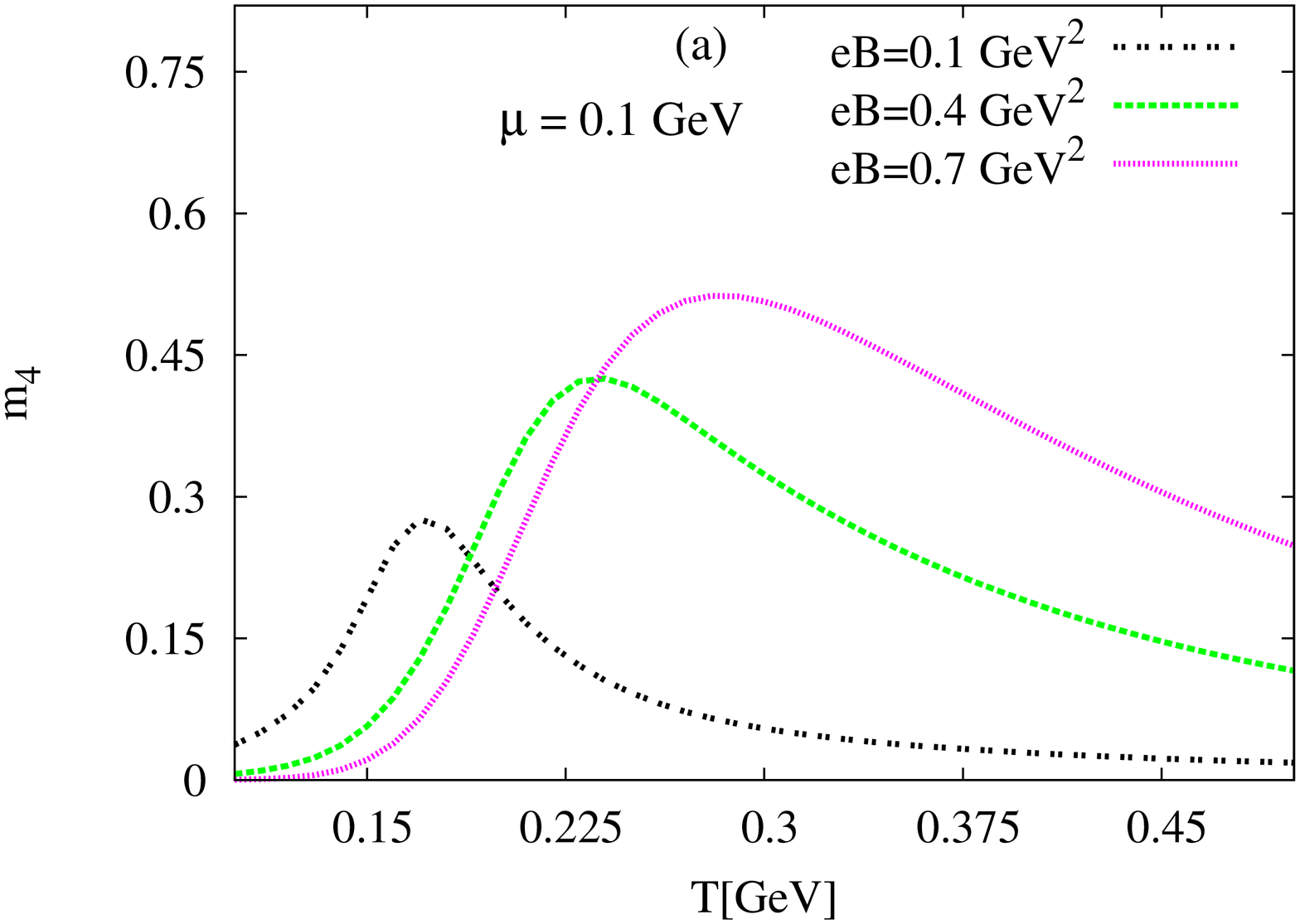}
\includegraphics[width=5.cm,angle=-90]{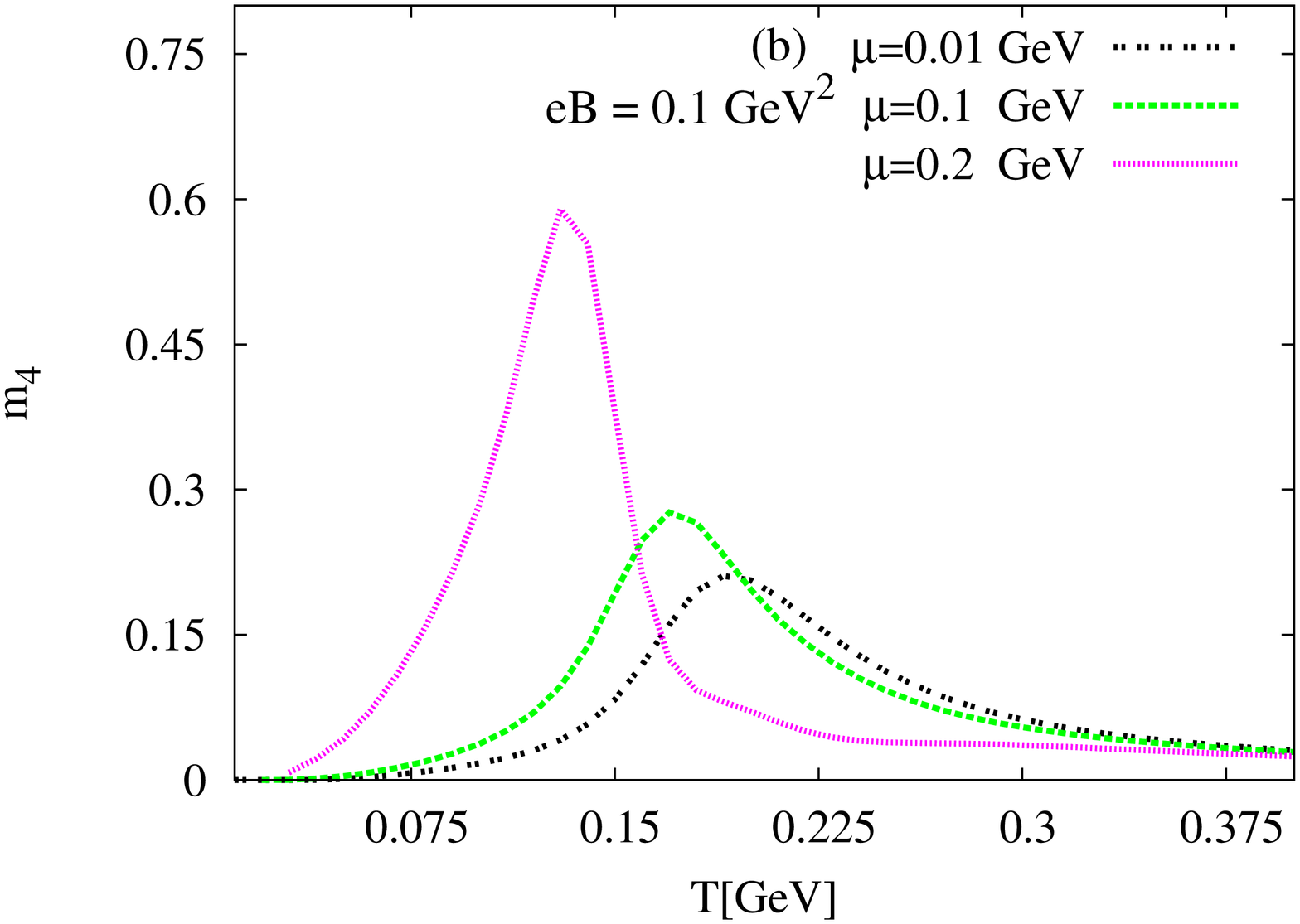}
\caption{(Color online) The same as in Fig. \ref{fig:m1} but for the fourth-order moment of quark number density $m_4$. \label{fig:m4}}
}
\end{figure}

Here, we introduce the non-normalized higher-order moments of the particle multiplicity \cite{Tawfik:2014uka}.  The left-hand panels (a) of Figs. \ref{fig:m1}, \ref{fig:m2}, \ref{fig:m3} and \ref{fig:m4} show the first four non-normalized moments of the quark distributions. These quantities are given as function of temperature at a constant chemical potential $\mu=0.1~$GeV and different magnetic fields,  $eB=0.1~$GeV$^2$ (double-dotted curve),  $0.4~$GeV$^2$ (dashed curve) and  $0.7~$GeV$^2$ (dotted curve). We find that increasing temperature rapidly increases the four moments. Furthermore, the thermal dependence is obviously enhanced, when moving from lower to higher orders. The values of the moment are increasing as we increase the magnetic field. The fluctuation in the third- and fourth-order moments reflect the increase of the critical temperature $T_c$ with increasing the magnetic field. The critical temperature can, for instance, be defined where the peaks are positioned.

The right-hand panels (b) of Figs. \ref{fig:m1}, \ref{fig:m2}, \ref{fig:m3} and \ref{fig:m4}, present the same as in the left-hand panels but at a constant magnetic field $eB=0.1~$GeV$^2$ and different chemical potentials,  $\mu=0.01~$GeV (double-dotted curve),  $0.1~$GeV  (dashed curve) and  $0.2~$GeV (dotted curve).  It is apparent that increasing temperature rapidly increases the four moments of quark number density. Furthermore, the thermal dependence is obviously enhanced, when moving from lower to higher orders. The values of the moment are increasing as we increase the chemical potential. But the critical temperature $T_c$ decrease with  $\mu$. The peaks are positioned at the critical temperature.

\subsubsection{Normalized higher-order moments}
\label{sec:NH-M}

The statistical normalization of the higher-order moments requires a scaling of the non-normalized quantities, section \ref{sec:non-H-M}, with respect to the standard deviation $\sigma$, which is related to the susceptibility $\chi$ or the fluctuations \cite{Tawfik_rev2014,Tawfik:2012si} in the particle multiplicity. It is conjectured that the dynamical phenomena could be indicated by large fluctuations in these dimensionless moments and therefore, the chiral phase-transition can be mapped out \cite{Tawfik_rev2014}. Due to the sophisticated derivations, we restrict the discussion here to dimensionless higher-order moments \cite{Tawfik:2014uka}. This can be done when the normalization is done with respect to the temperature or chemical potential. 

The higher-order moments of the particle multiplicity normalized with respect to temperature are studied in dependence on the temperature at a constant chemical potential  and different magnetic fields. Also they are studied at different  chemical potentials and a constant magnetic field. The corresponding expressions were deduced in Ref. \cite{Tawfik:2014uka}. 

\begin{figure}[htb]
\centering{
\includegraphics[width=5.cm,angle=-90]{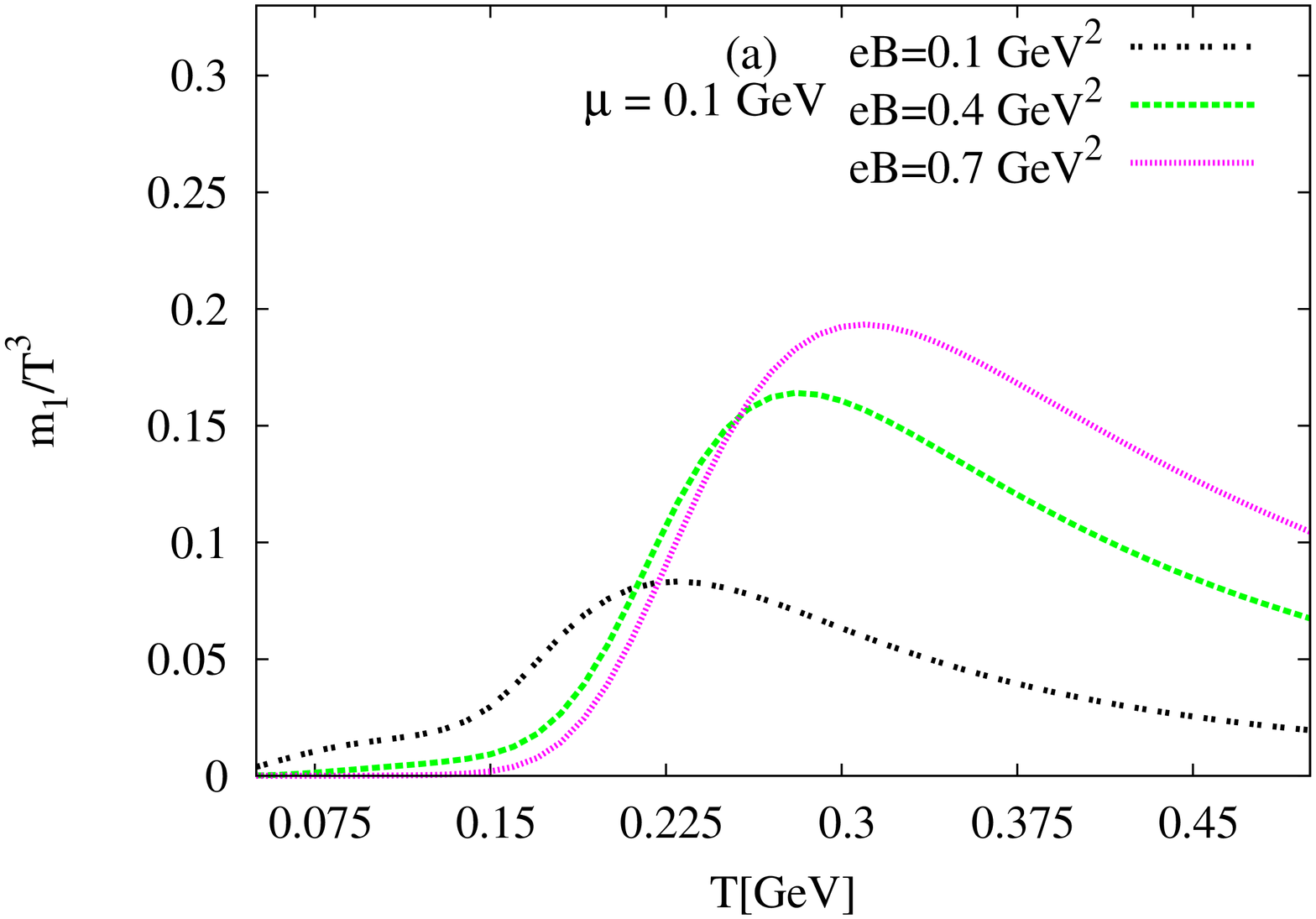}
\includegraphics[width=5.cm,angle=-90]{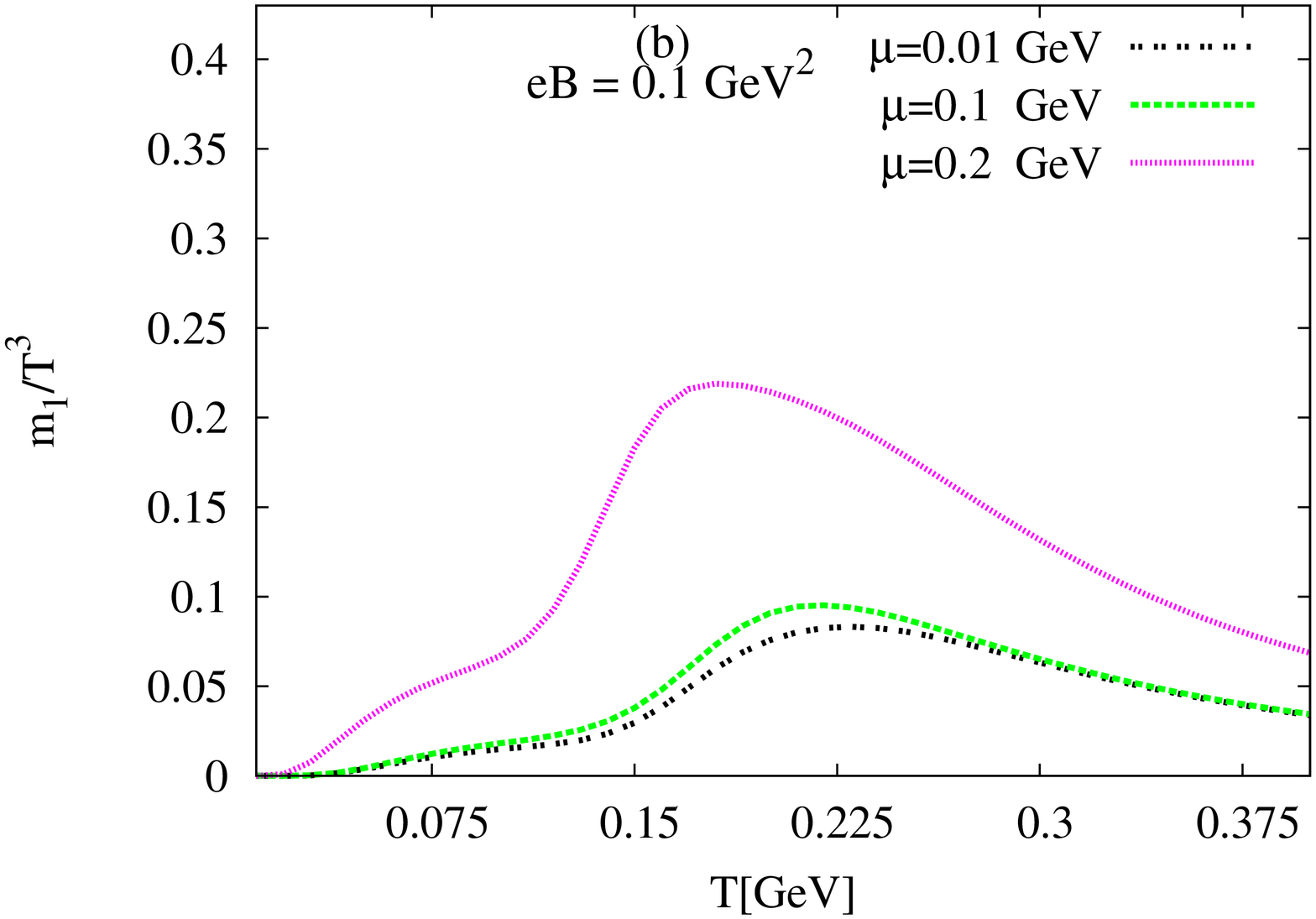}
\caption{(Color online) Left-hand panel (a): the dimensionless quark number density $ m_1/T^3$, is given as function of temperature at constant chemical potential $\mu=0.1~$GeV and different magnetic fields,  $eB=0.1~$GeV$^2$ (double-dotted curve),  $eB=0.4~$GeV$^2$ (dashed curve) and $eB=0.7~$GeV$^2$ (dotted curve). Right-hand panel (b) shows the same as in the left-hand panel but at a constant magnetic field $eB=0.1~$GeV$^2$ and different chemical potential values,  $\mu=0.01~$GeV (double-dotted curve), $\mu=0.1~$GeV (dashed curve) and  $\mu=0.2~$GeV (dotted curve).  
\label{fig:mn1}}
}
\end{figure}

\begin{figure}[htb]
\centering{
\includegraphics[width=5.cm,angle=-90]{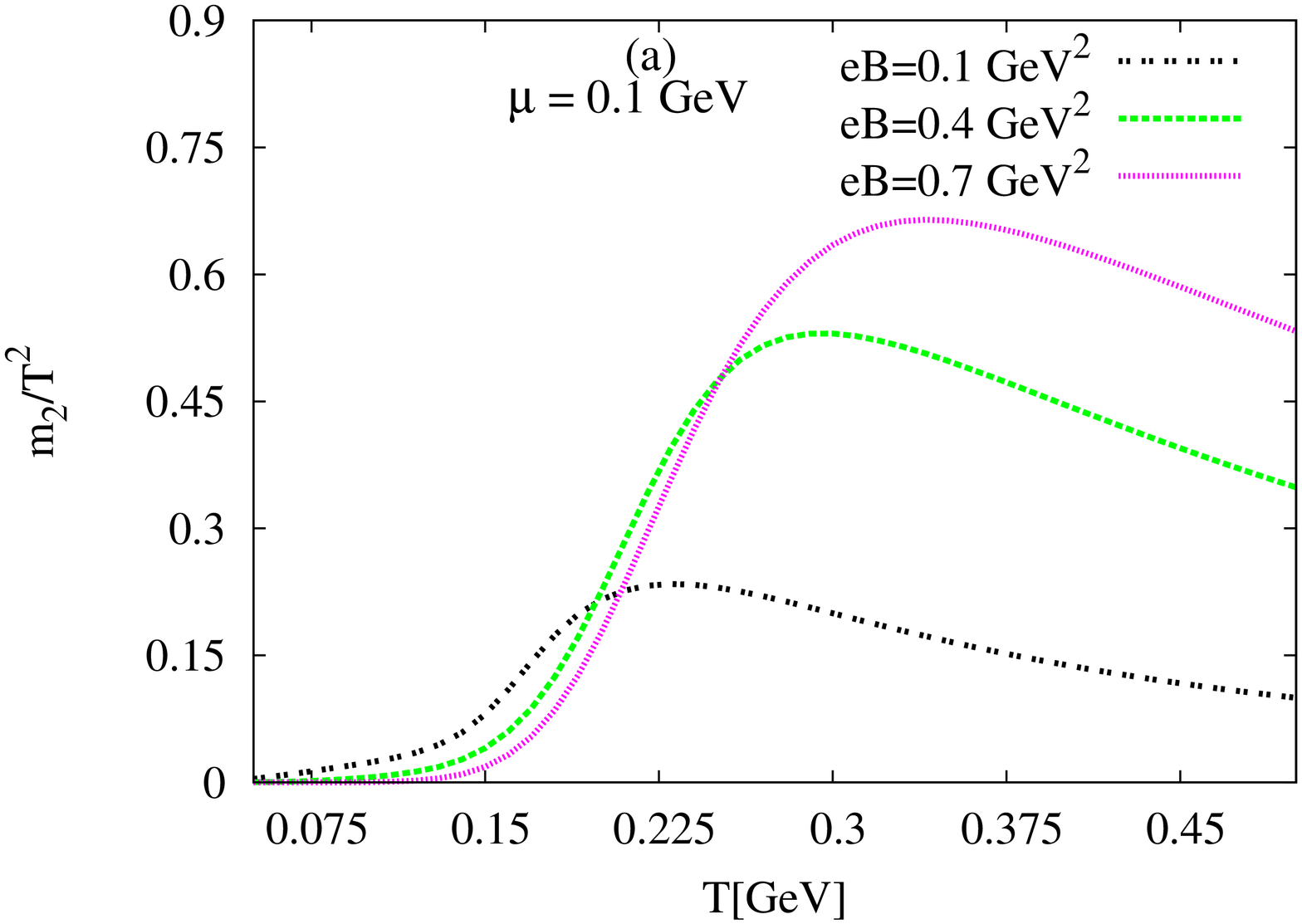}
\includegraphics[width=5.cm,angle=-90]{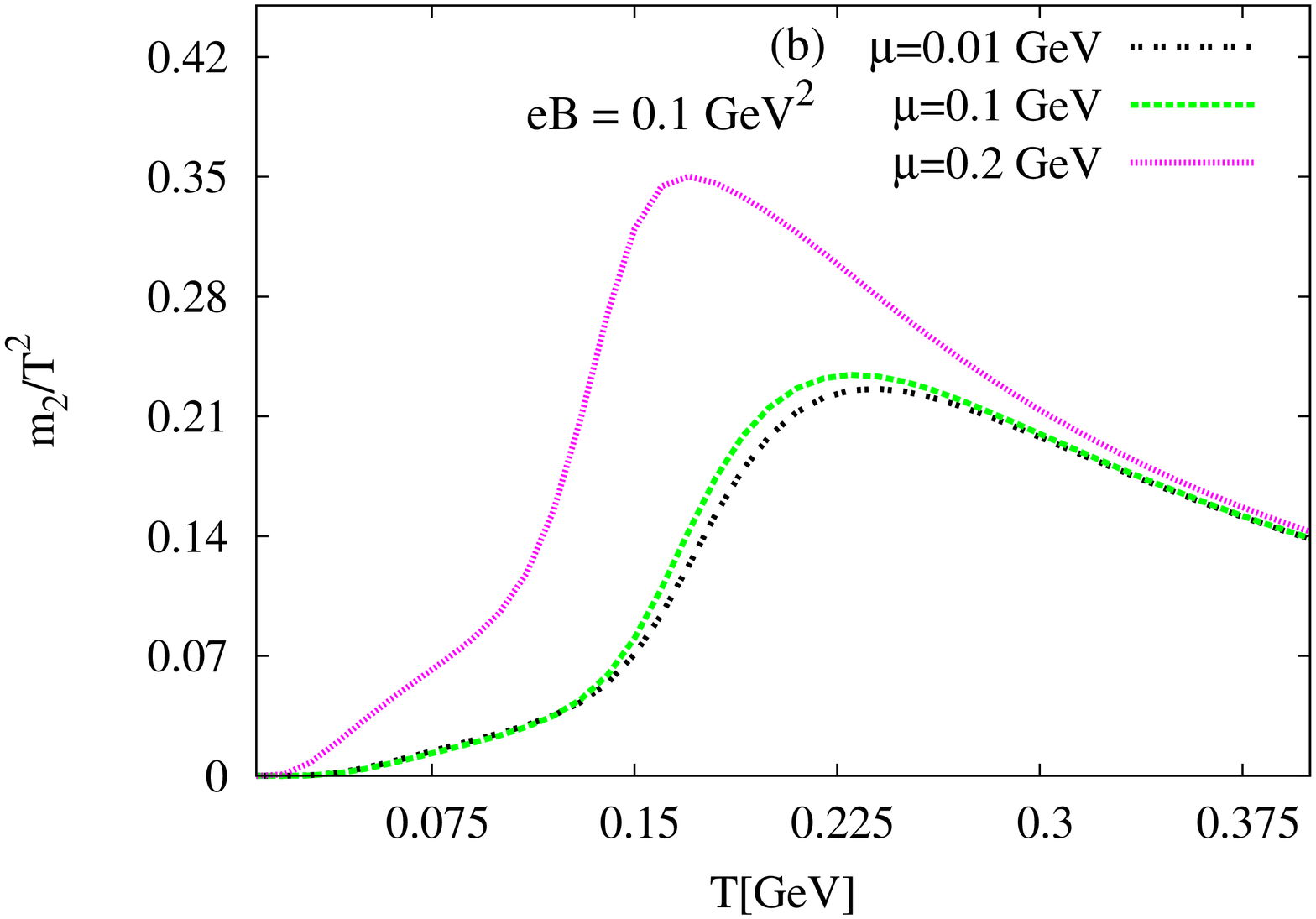}
\caption{(Color online) The same as in Fig. \ref{fig:mn1} but for the dimensionless quark number susceptibility $m_2/T^2$.  \label{fig:mn2}}
}
\end{figure}

\begin{figure}[htb]
\centering{
\includegraphics[width=5.cm,angle=-90]{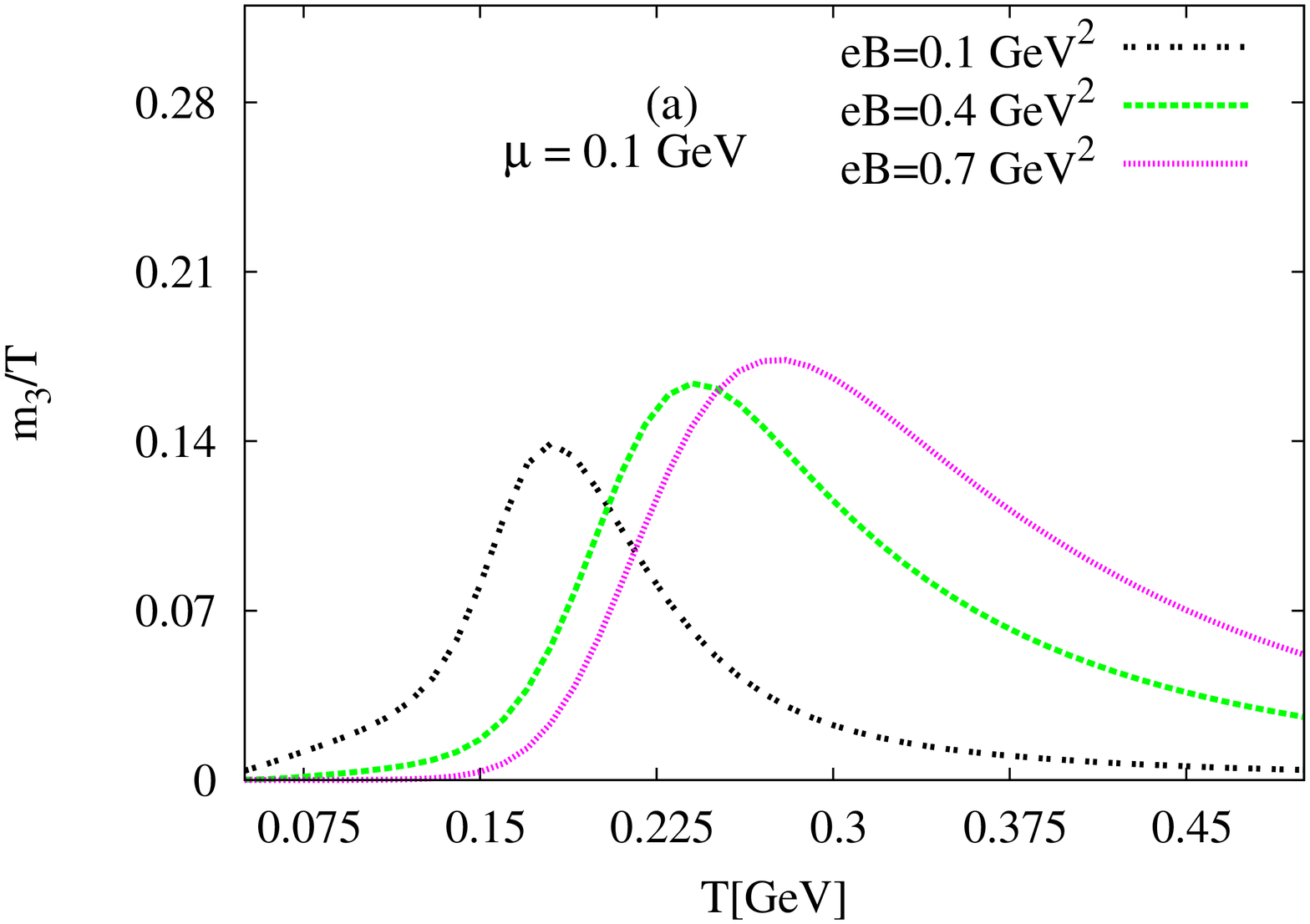}
\includegraphics[width=5.cm,angle=-90]{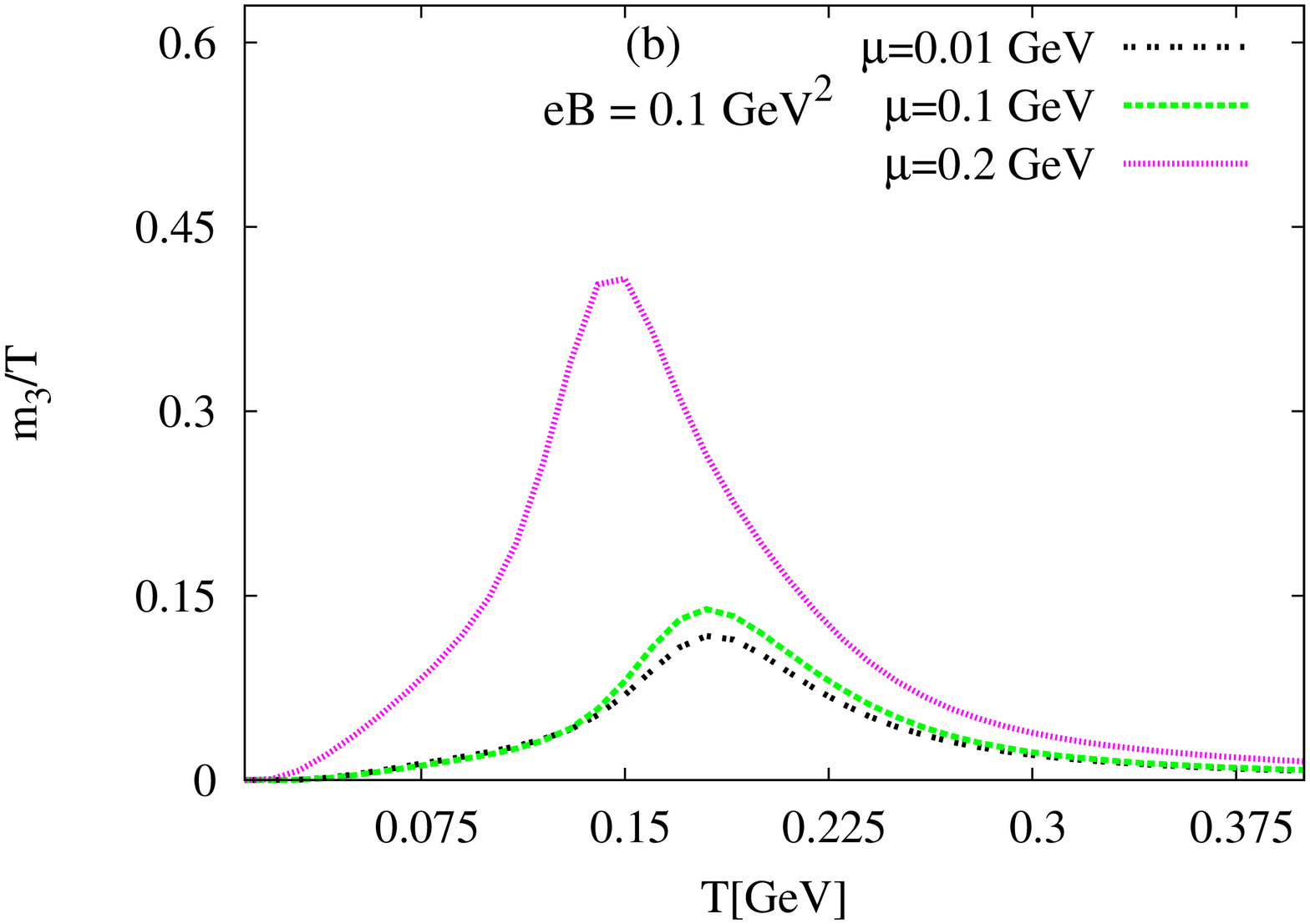}
\caption{(Color online) The same as in Fig. \ref{fig:mn1} but for the dimensionless third-order moment of the quark number density $m_3/T$.    \label{fig:mn3}}
}
\end{figure}

In left-hand panel (a) of Figs. \ref{fig:mn1}, \ref{fig:mn2} and \ref{fig:mn3} the first three normalized moments are given as function of temperature at a constant chemical potential $\mu=0.1~$GeV and different magnetic fields, $eB=0.1~$GeV$^2$ (double-dotted curve),  $0.4~$GeV$^2$ (dashed curve) and $0.7~$GeV$^2$ (dotted curve). We find that the values of the moments are increasing as the magnetic field increases. The fluctuations in the normalized  moments would define the dependence of the critical temperature $T_c$ on the magnetic field.  

The right-hand panels (b) of Figs. \ref{fig:mn1}, \ref{fig:mn2} and \ref{fig:mn3} present the first three normalized moments as function of temperature but at a constant magnetic field $eB=0.1~$GeV$^2$ and different chemical potentials, $\mu=0.01~$GeV (double-dotted curve), $0.1~$GeV (dashed curve) and  $0.2~$GeV (dotted curve). We notice that the moments of quark multiplicity increase with the chemical potentials. That the peaks at corresponding critical temperatures can be used to map out the chiral phase-diagrams, $T$ vs. $eB$ and $T$ vs. $\mu$.

\subsection{Chiral phase-transition}
\label{subsec:phase-T}

Now we can study the effects of the magnetic field on the chiral phase-transition. In a previous work \cite{Tawfik:2014uka}, we have introduced and summarized different methods to calculate the critical temperature and chemical potential, $\mu_c$, by using the fluctuations in the normalized  higher-order moments of the quark multiplicity or by using the order parameters. The latter is  implemented in the present work. The PLSM has two order-parameters. The first one presents the chiral phase-transition. This is related to strange and non-strange chiral condensates, $\sigma_x$ and $\sigma_y$. The second one gives hints for the confinement-deconfinement phase-transition, the Polyakov-loop fields, $\phi$ and $\phi^*$. Therefore, for the models having Polyakov-loop potential, we can follow a procedure as follows.  We start with a constant value of the magnetic field. By using strange and non-strange chiral-condensates, a dimensionless quantity reflecting the difference between the non-strange and strange condensates $\Delta_{q,s}(T)$ as a function of temperature at fixed chemical potentials will be implemented. This procedure give {\it one point} in the $T$-$\mu$-chart, at which the chiral phase transition takes place. At the same chemical potential as in previous step, we deduce the other order-parameter related to the Polyakov-loop fields as a function of temperature. These calculations give another point ({\it in $T$ and $\mu$ chart}), at which the deconfinement phase-transition takes place. By varying the chemical potential, we repeat these steps. Then, we find a region (or point), in (at) which the two order-parameters, chiral and deconfinement, cross each other,  i.e. equal each other. It is assumed that such a point represents phase transition(s) at the given chemical potential. In doing this, we get a set of points in a two-dimensional chart, the QCD phase-diagram.

\begin{figure}[htb]
\centering{
\includegraphics[width=8.cm,angle=-90]{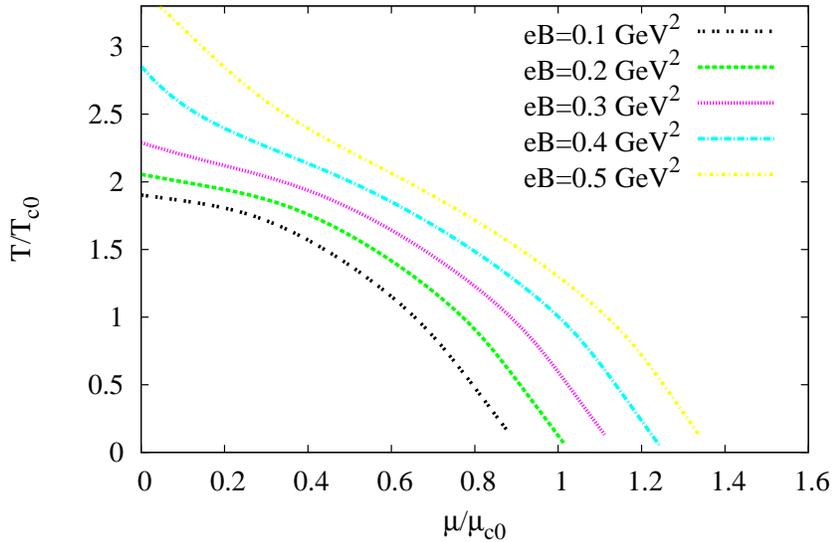}
\caption{(Color online) The chiral phase-diagram, $T/T_{c0}$ vs. $\mu/\mu_{c0}$, at different values of the magnetic fields, $eB=0.1$, $0.2$, $0.3$, $0.4$ and $0.5~$GeV$^2$ from top to bottom. The normalization quantities $T_{c0}=0.15~$MeV and $\mu_{c0}=0.3~$GeV were deduced in Ref. \citep{Tawfik:2014uka}. \label{fig:P-D}}
}
\end{figure}

In Fig.  \ref{fig:P-D}, we compare five chiral phase-diagrams, $T/T_{c0}$ vs. $\mu/\mu_{c0}$, with each others at $eB=0.1$, $0.2$, $0.3$, $0.4$ and $0.5~$GeV$^2$ from top to bottom. $T/T_{c0}$ are plotted against $\mu/\mu_{c0}$, where the two normalization quantities $T_{c0}=0.15~$GeV and $\mu_{c0}=0.3~$GeV were deduced from Ref. \citep{Tawfik:2014uka}. This should give an indication about the behavior of the critical temperature and the critical chemical potential of the system under the effect of the magnetic field. Apparently, we conclude that both critical temperature and critical chemical potential increase with increasing the magnetic field.

\subsection{Meson masses}
\label{sec:masses}

The masses can be deduced from the second derivative of the grand potential, Eq. (\ref{potential}) with respect to the corresponding fields, evaluated at its minimum, which is estimated at vanishing expectation values of all scalar and pseudoscalar fields
\begin{eqnarray}
m_{i,ab}^{2}&=& \dfrac{\partial^{2}\Omega(T,\mu_{f})}{\partial \xi_{i,a} \partial \xi_{i,b}} |_{min},
\end{eqnarray}
where $a$ and $b$ range from $0, \cdots, 8$ and  $\xi_{i,a}$ and $\xi_{i,b}$ are scalar and pseudoscalar mesonic fields, respectively. Obviously, $i$ stands for scalar and pseudoscalar mesons.

The scalar meson masses \cite{Vivek} 
\begin{eqnarray}
m^{2}_{\sigma} &=& m^2_{s,{00}}\cos^{2}\theta_{s}+m^2_{s,88}\sin^{2}\theta_{s} + 2 m^{2}_{s,08}\sin\theta_{s}\cos\theta_{s}, \label{ms1} \\
m^{2}_{f_0} &=& m^2_{s,00}\sin^{2}\theta_{s}+m^{2}_{s,88}\cos^{2}\theta_{s} - 2 m^2_{s,08}\sin\theta_{s}\cos\theta_{s}, \label{ms2}\\
m^{2}_{\sigma_{NS}} &=& \frac{1}{3}(2m^{2}_{s,00}+m^{2}_{s,88}+2\sqrt{2}m^{2}_{s,08}), \label{ms3} \\
m^{2}_{\sigma_{S}} &=& \frac{1}{3}(m^{2}_{s,00}+2m^{2}_{s,88}-2\sqrt{2}m^{2}_{s,08}), \label{ms4}
\end{eqnarray}
where $\theta_{s}$ is the scalar mixing angle \cite{Vivek}
\bea
\theta_s &=& \frac{1}{2} \text{ArcTan}\left[\frac{2 (m_s^2)_{08}}{(m_s^2)_{00}-(m_s^2)_{88}}\right] ,\nn
\eea
with $ (m_s^2)_{a b} = m^2\, \delta_{a\, b} - 6 {\cal G}_{a b c} \bar{\sigma}_c + 4\, {\cal F}_{a b c d}\, \bar{\sigma}_c\, \bar{\sigma}_d$.
The expressions for ${\cal G}_{a b c}$ and ${\cal F}_{a b c d}$ can be found in Ref. \cite{Vivek}. On the tree level, $\bar{\sigma}_c$ can be determined according to $\partial U(\bar{\sigma})/\partial \bar{\sigma}_a=0=m^2\,  \bar{\sigma}_a - 3 {\cal G}_{a b c} \bar{\sigma}_b\,  \bar{\sigma}_c + (4/3)\, {\cal F}_{a b c d}\, \bar{\sigma}_b\, \bar{\sigma}_c\,  \bar{\sigma}_d - h_a$.

\begin{figure}[htb]
\centering{
\includegraphics[width=5.cm,angle=-90]{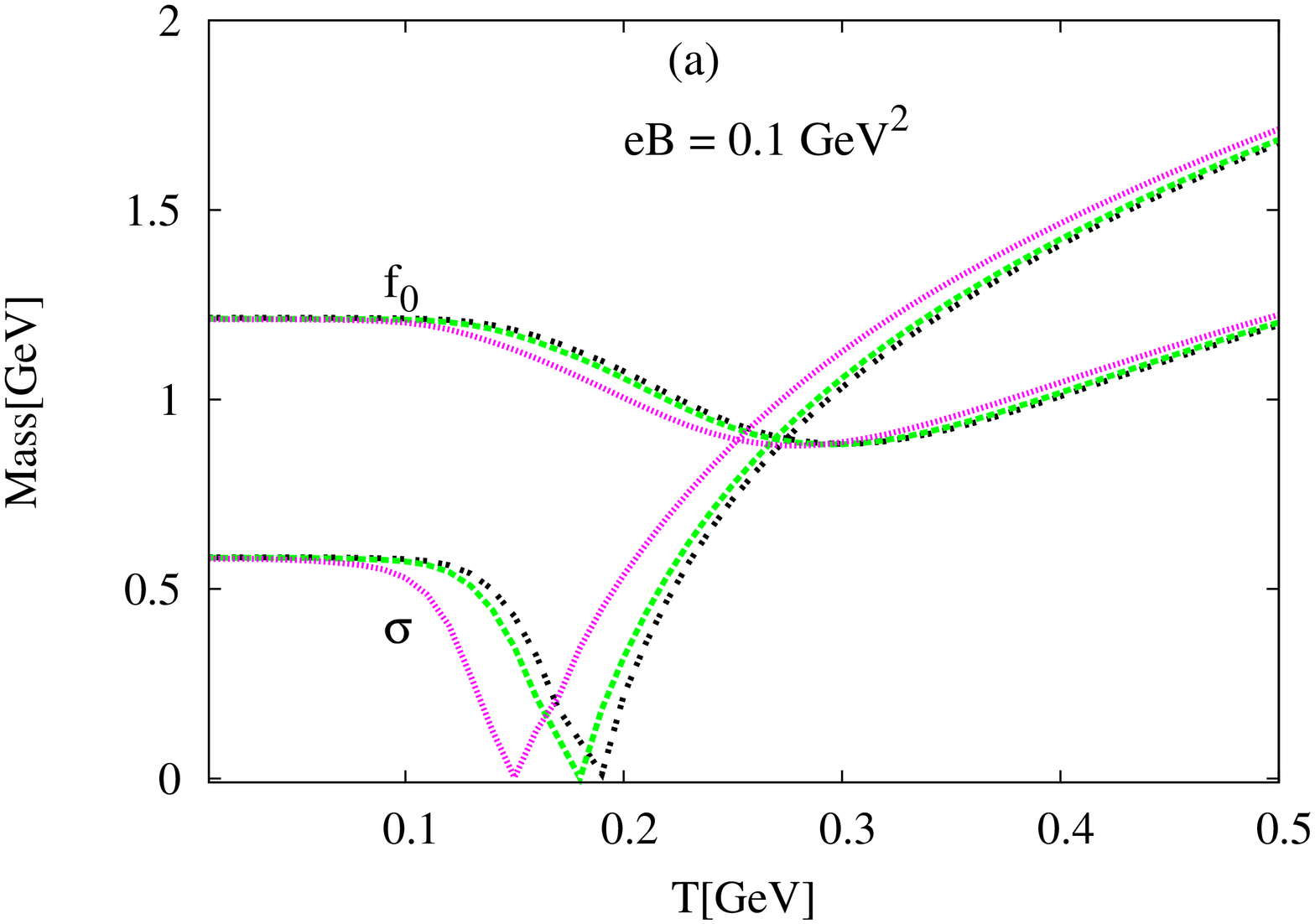}
\includegraphics[width=5.cm,angle=-90]{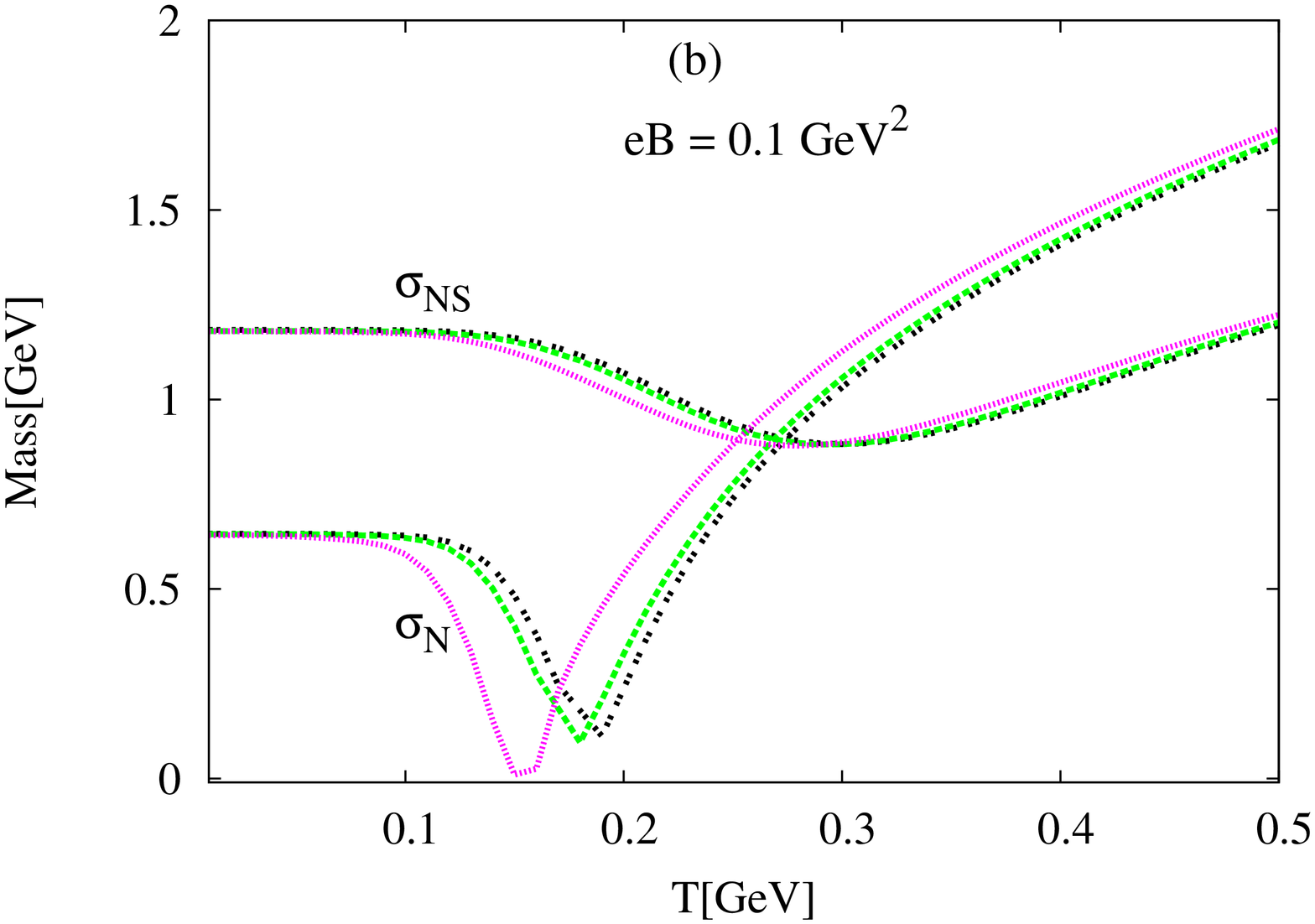}
\caption{(Color online) Left-hand panel (a): the scalar meson masses, $m_{\sigma}$ from Eq. (\ref{ms1}), $m_{f_0}$ from Eq. (\ref{ms2}) are given as function of temperature at a constant  magnetic field $eB=0.1~$GeV$^2$ and different chemical potentials, $\mu=0.0~$GeV (dotted curve),  $0.1~$GeV (dashed curve) and  $0.2~$GeV (double-dotted curve). Right-hand panel (b): the same as in left-hand panel (a) but for $m_{\sigma_{NS}}$ from Eq. (\ref{ms3}) and $m_{\sigma_{S}}$ from Eq. (\ref{ms4}).  
\label{fig:sif_f0_Mu}}
}
\end{figure}

In Fig. \ref{fig:sif_f0_Mu}, the  scalar meson masses, $m_{\sigma}$ from Eq. (\ref{ms1}), and $m_{f_0}$ from Eq. (\ref{ms2}) are given as function of temperature at a constant  magnetic field $eB=0.1~$GeV$^2$ and different chemical potentials $\mu=0.0~$GeV (dotted curve),  $0.1~$GeV (dashed curve) and  $0.2~$GeV (double-dotted curve). We conclude that the scalar meson masses decrease as the temperature increases. This remains until $T$ reaches the critical value. Then, the vacuum effect becomes dominant and rapidly increases with the temperature. The effect of the chemical potential is very obvious. The masses decrease with the increase in chemical potential. This explains the phase diagram of temperatures and chemical potentials at a certain magnetic field. The decrease of the critical temperature with increasing chemical potential is represented by the bottoms (minima) in thermal behavior of meson masses before switching on the vacuum effect.

\begin{figure}[htb]
\centering{
\includegraphics[width=4.cm,angle=-90]{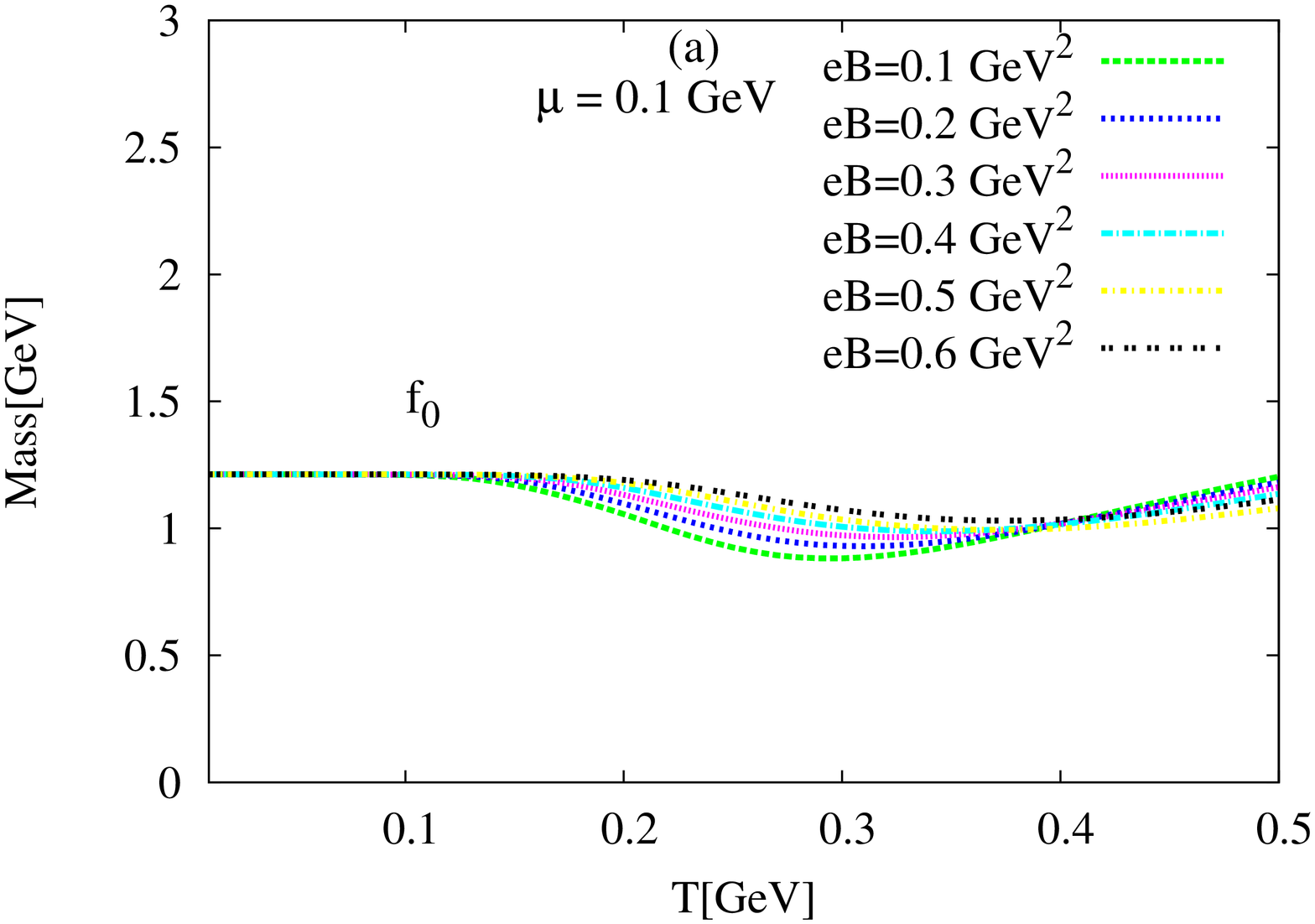}
\includegraphics[width=4.cm,angle=-90]{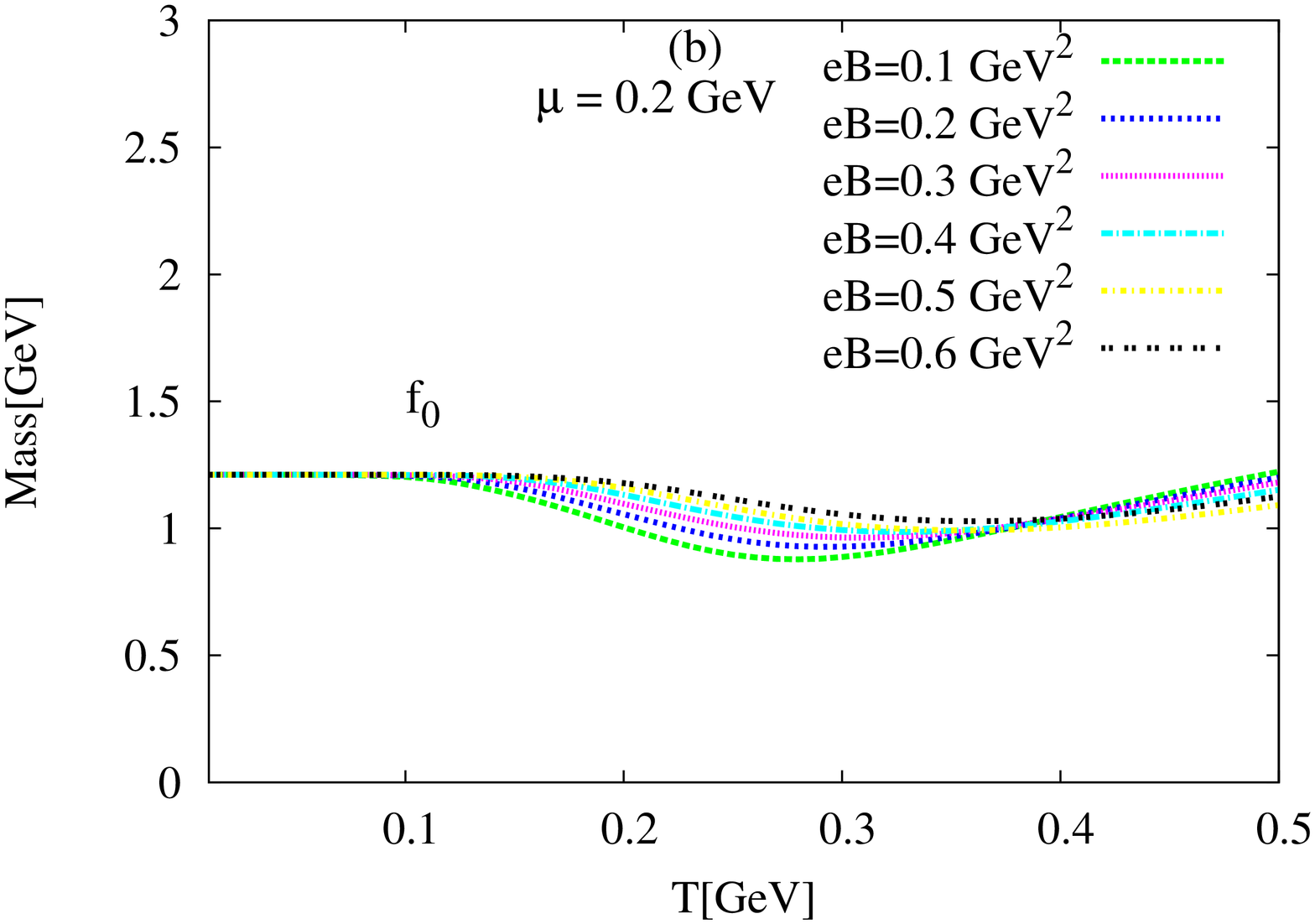}\\
\includegraphics[width=4.cm,angle=-90]{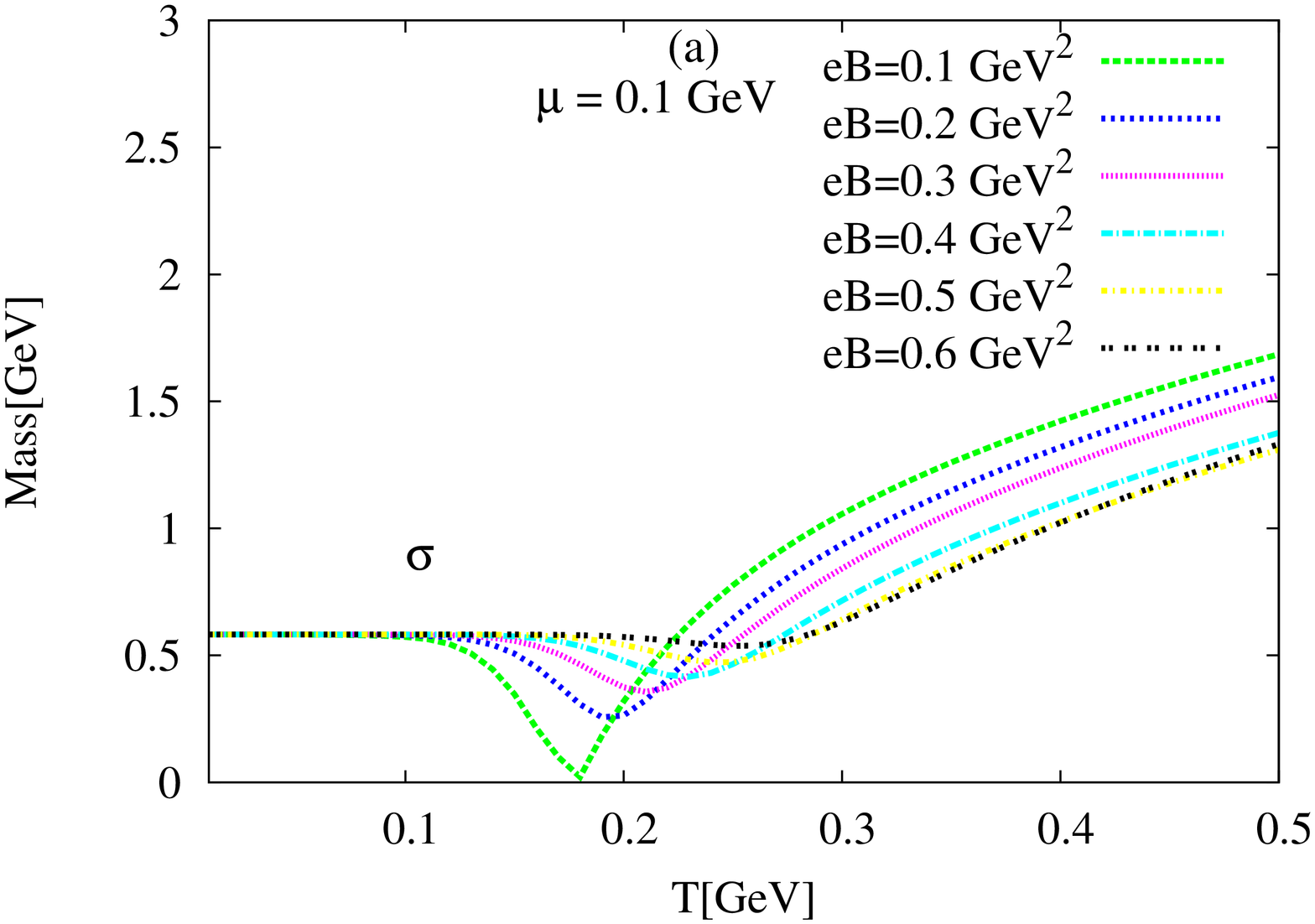}
\includegraphics[width=4.cm,angle=-90]{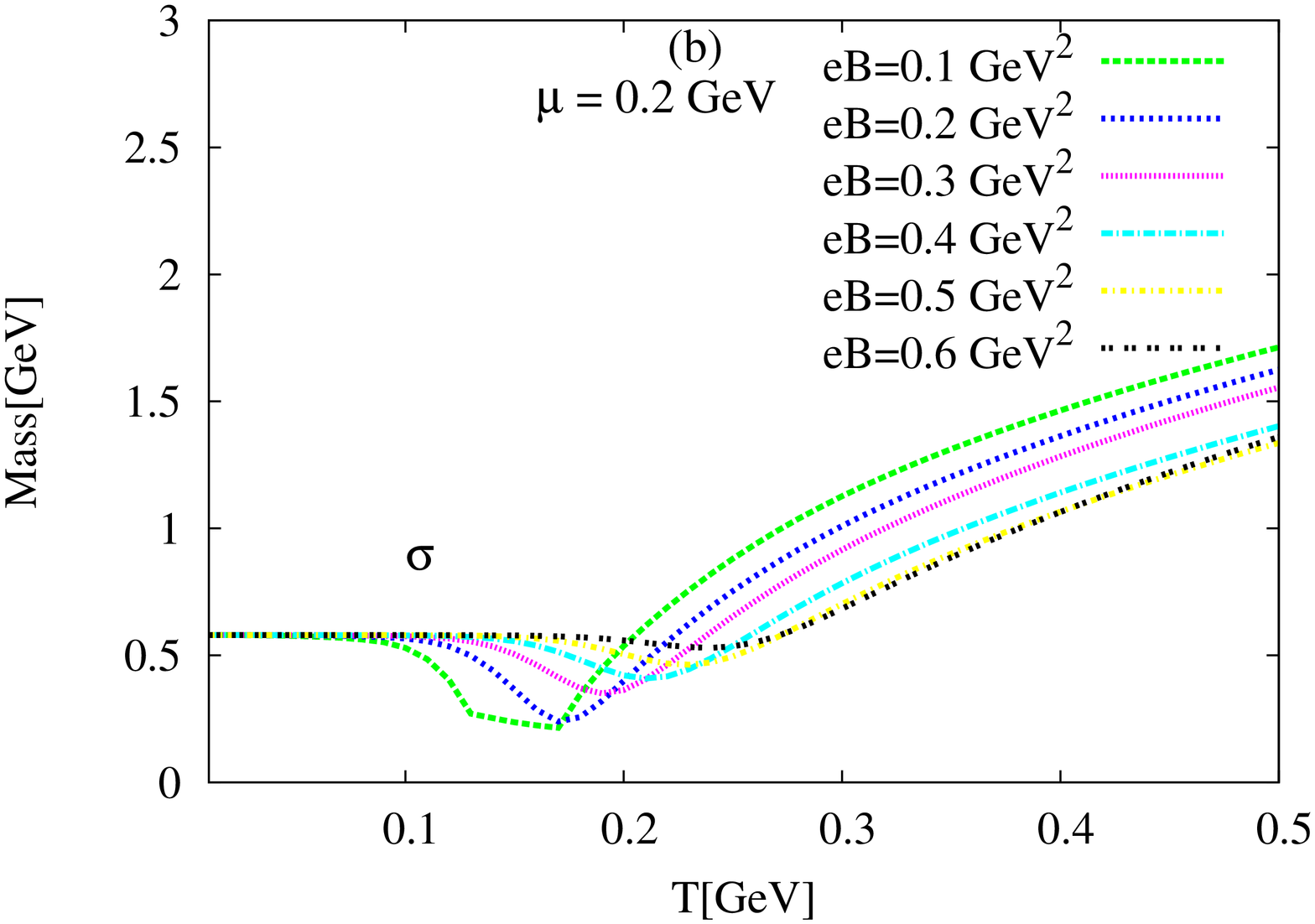}\\
\includegraphics[width=4.cm,angle=-90]{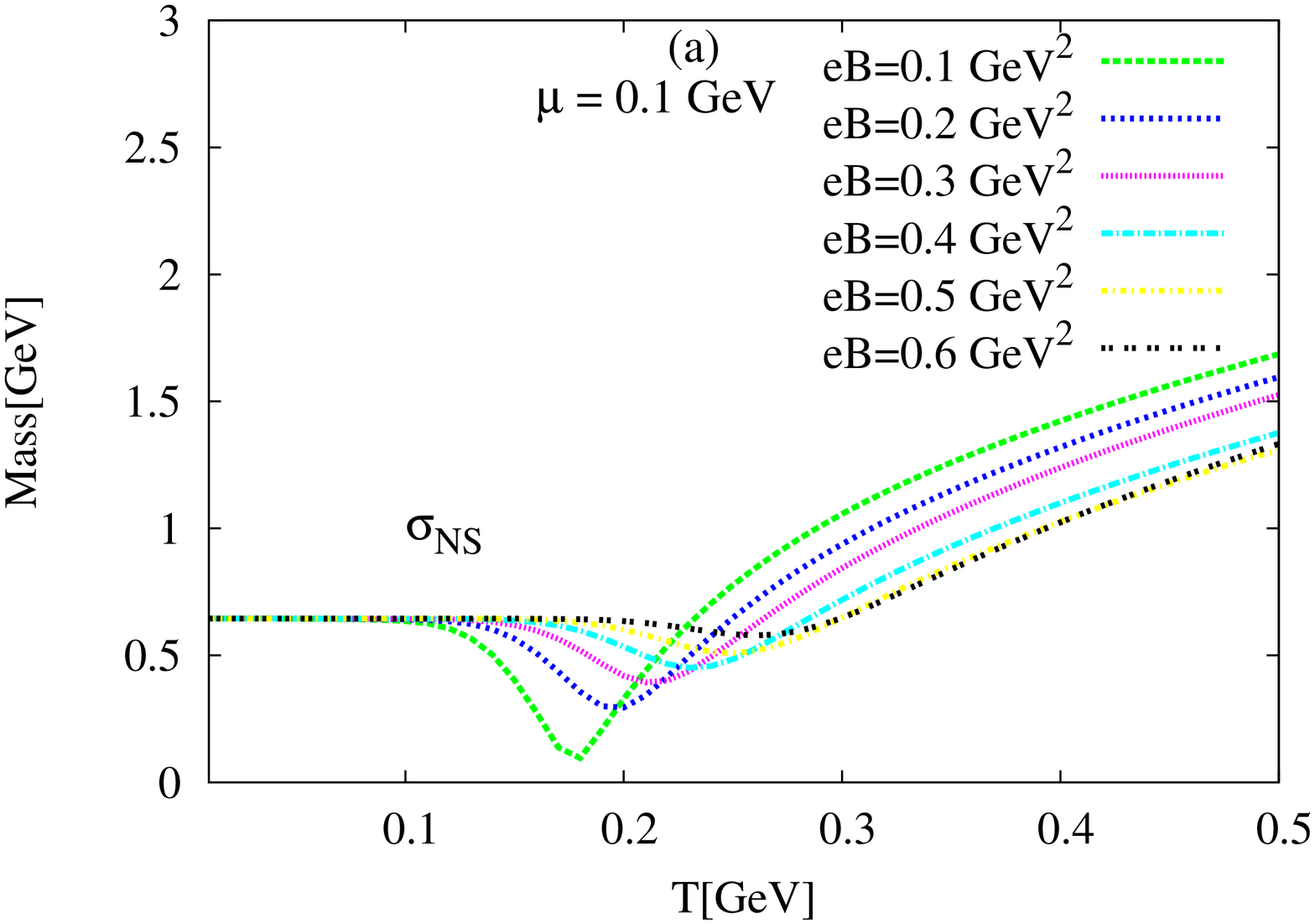}
\includegraphics[width=4.cm,angle=-90]{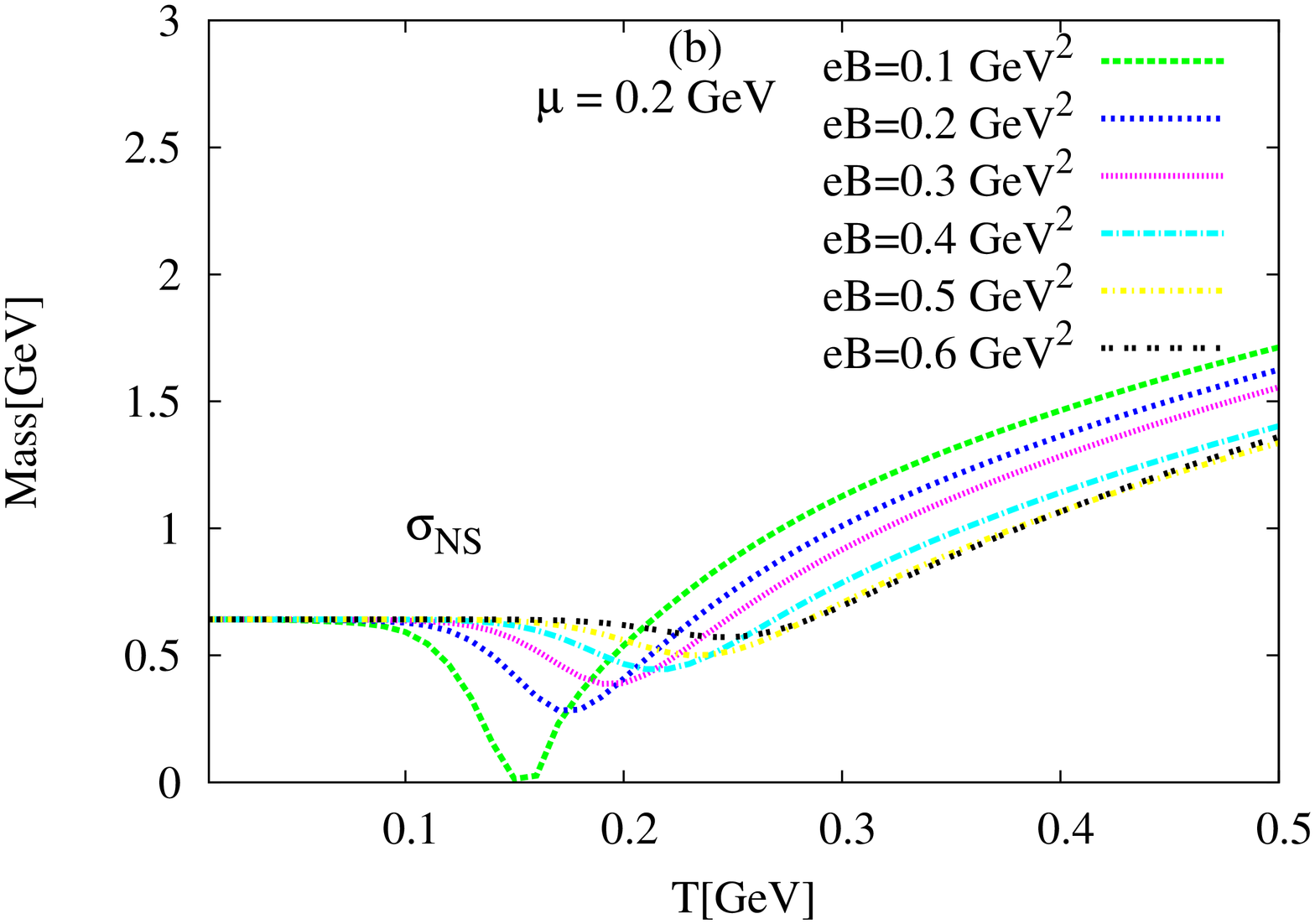}\\
\includegraphics[width=4.cm,angle=-90]{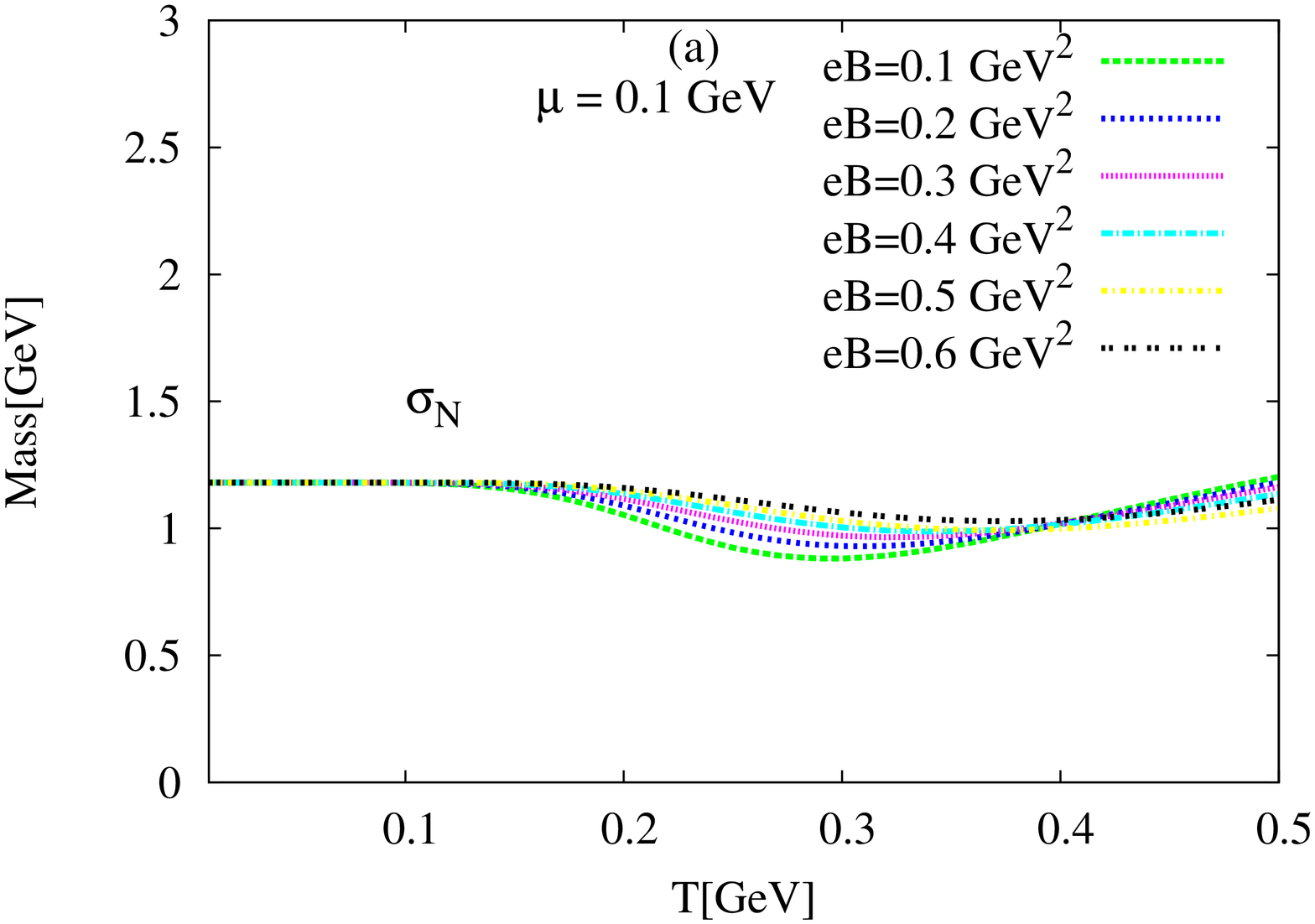}
\includegraphics[width=4.cm,angle=-90]{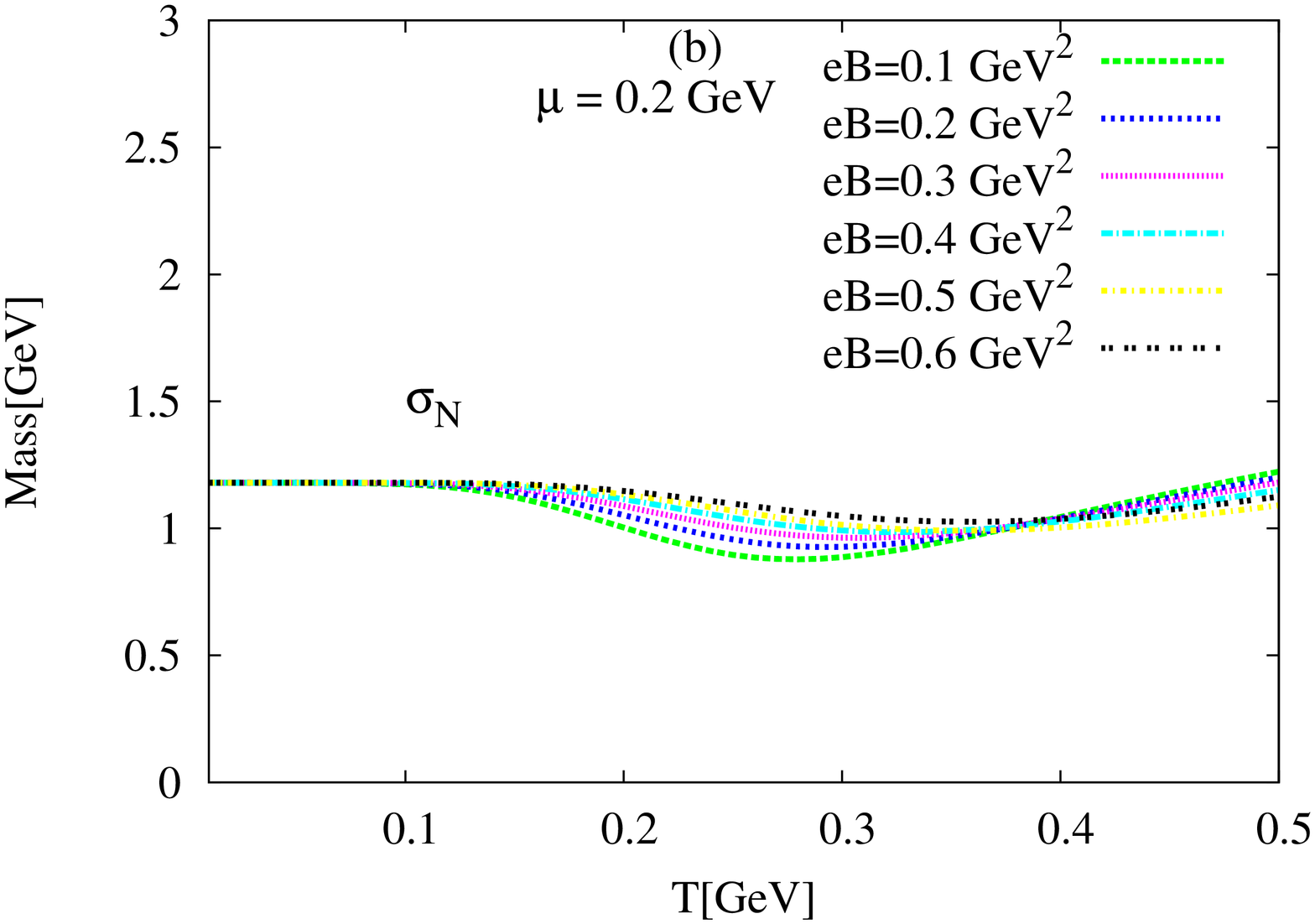}
\caption{(Color online) Left-hand panel (a): the four scalar meson masses are given as function of temperature at a constant chemical potential $\mu=0.1~$GeV and different magnetic fields, $eB=0.1$, $0.2$, $0.3$, $0.4$, $0.5$ and $0.6~$GeV$^2$ from top to bottom. Right-hand panel (b): the same as in left-hand panel but at chemical potential $\mu=0.2~$GeV.
\label{fig:sif_f0}}
}
\end{figure}

In Fig. \ref{fig:sif_f0}, the four scalar meson masses, $m_{\sigma}$ from Eq. (\ref{ms1}), $m_{f_0}$ from Eq. (\ref{ms2}), $m_{\sigma_{NS}}$ from  Eq. (\ref{ms3}) and $m_{\sigma_{S}}$ from  Eq. (\ref{ms4}) are given as function of temperature at two values of chemical potential, $\mu=0.1~$GeV left-hand panel (a) and $\mu=0.2~$GeV right-hand panel (b) and different magnetic fields, $eB=0.1$, $0.2$, $0.3$, $0.4$, $0.5$ and $0.6~$GeV$^2$ from top to bottom. We notice that the scalar meson masses decrease as the temperature increases, until it reaches the critical temperature. Then, the vacuum effect gets dominant and apparently increases with the temperature. The effect of magnetic field is very obvious. The masses increase as the magnetic field increases. This explains the phase diagram of temperatures and magnetic field at a certain chemical potential.

\begin{figure}[htb]
\centering{
\includegraphics[width=4.cm,angle=-90]{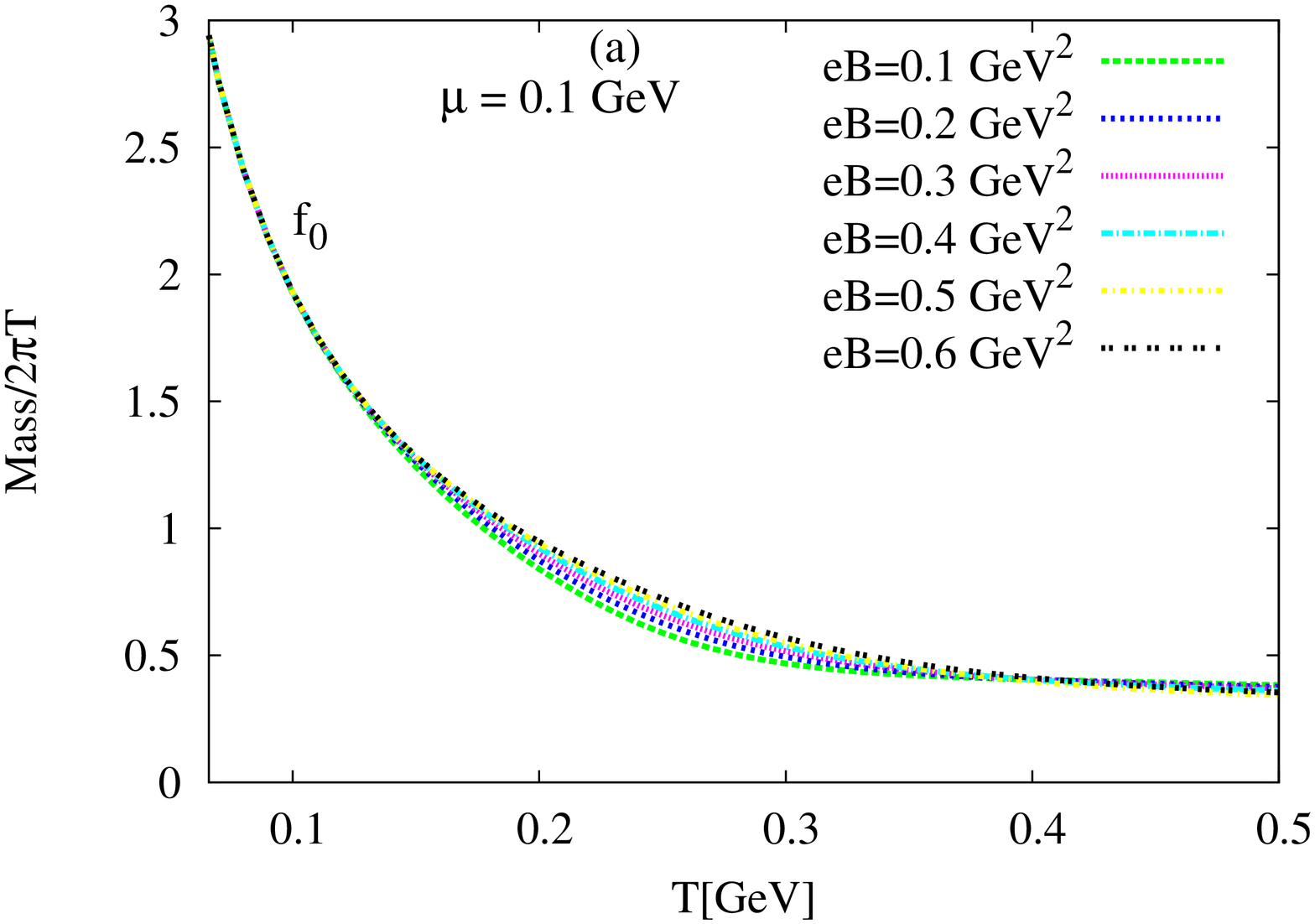}
\includegraphics[width=4.cm,angle=-90]{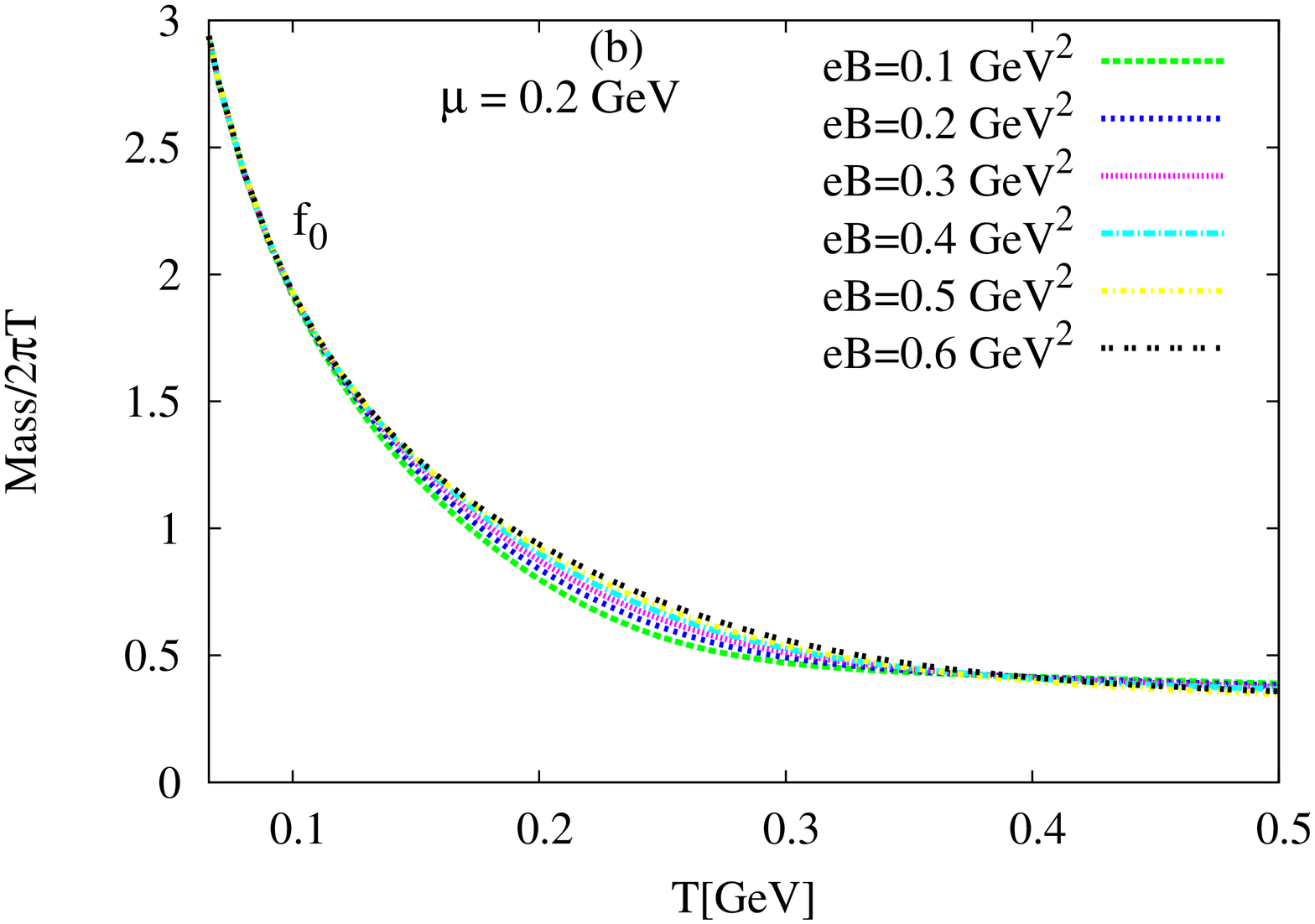}\\
\includegraphics[width=4.cm,angle=-90]{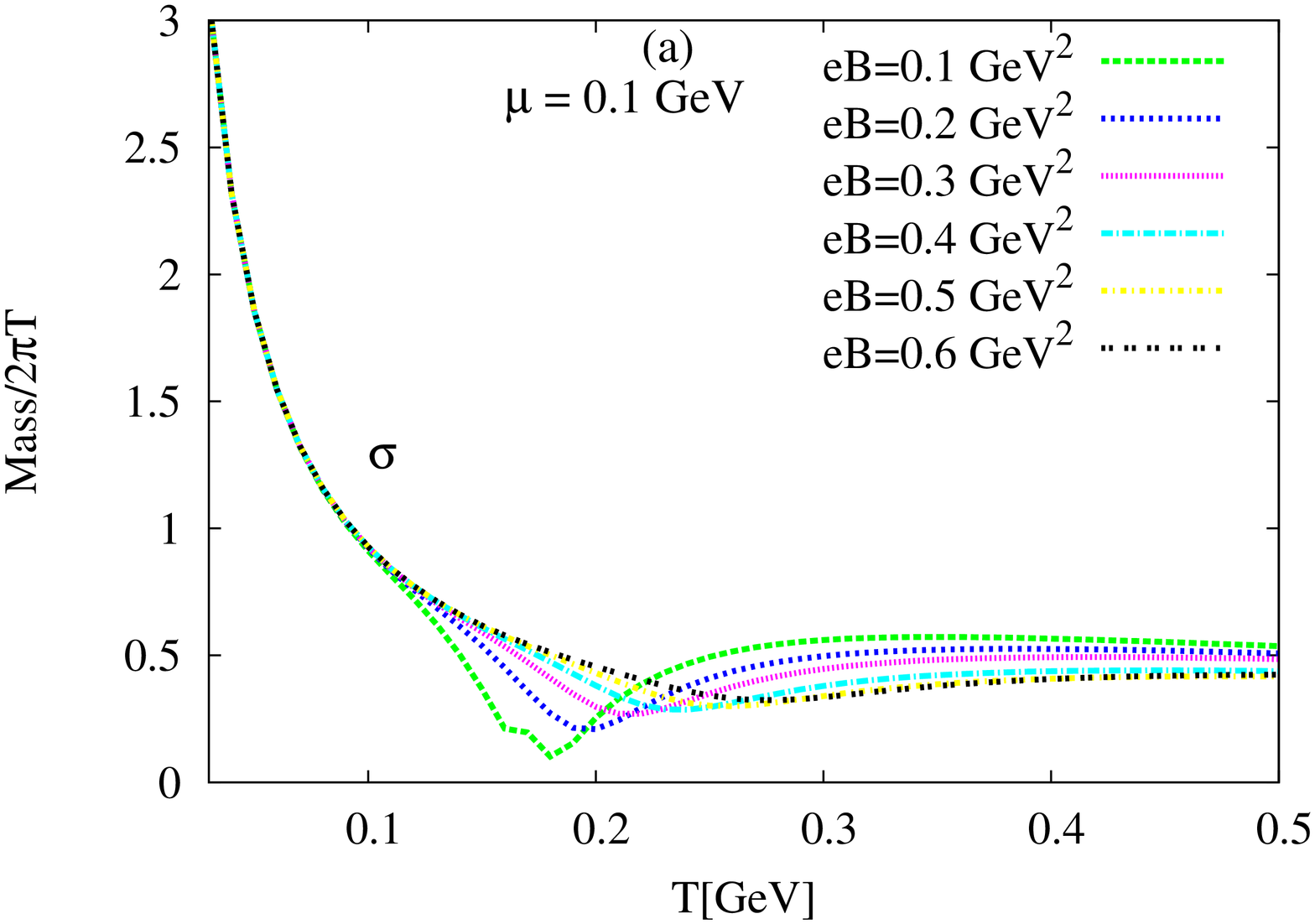}
\includegraphics[width=4.cm,angle=-90]{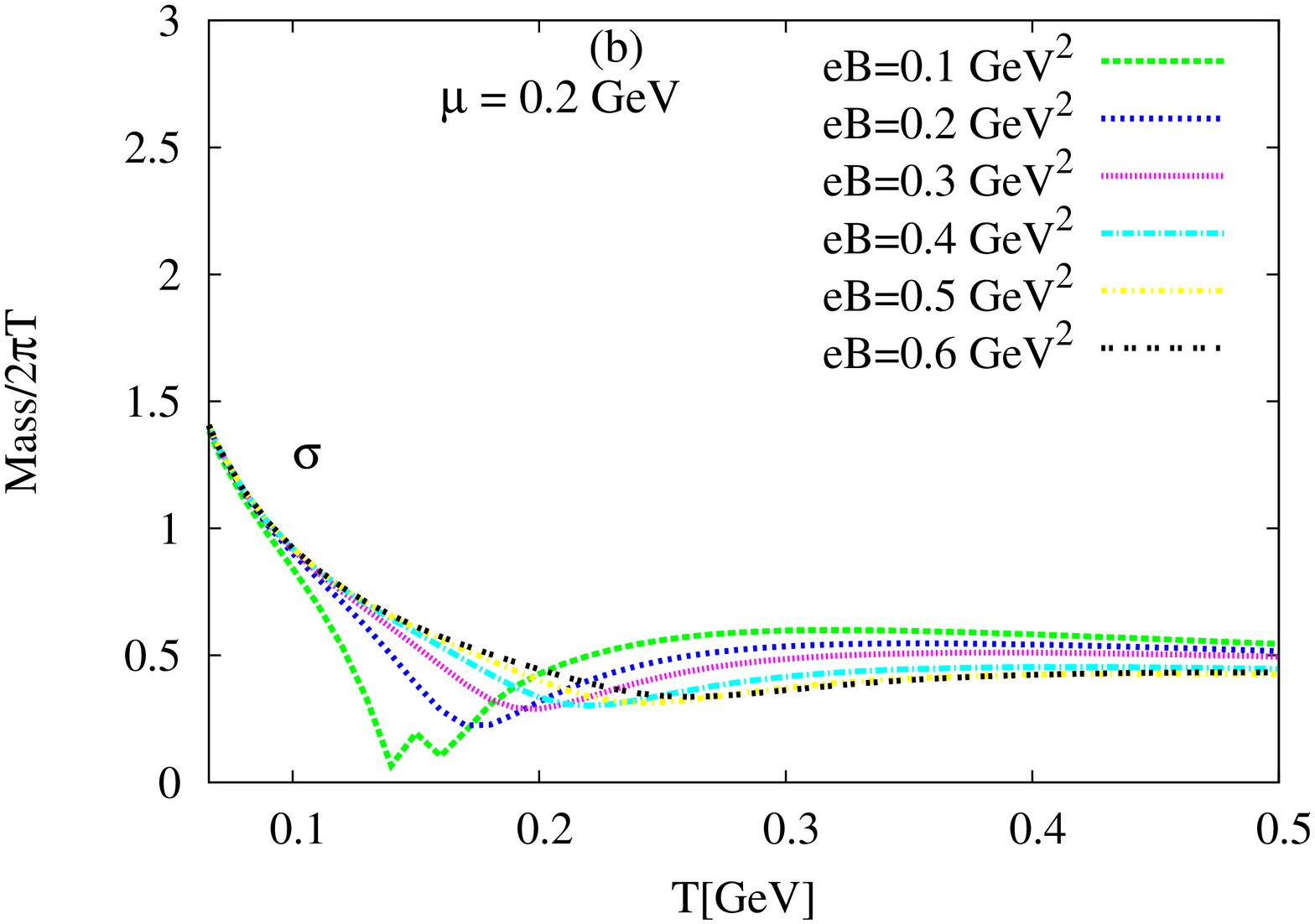}\\
\includegraphics[width=4.cm,angle=-90]{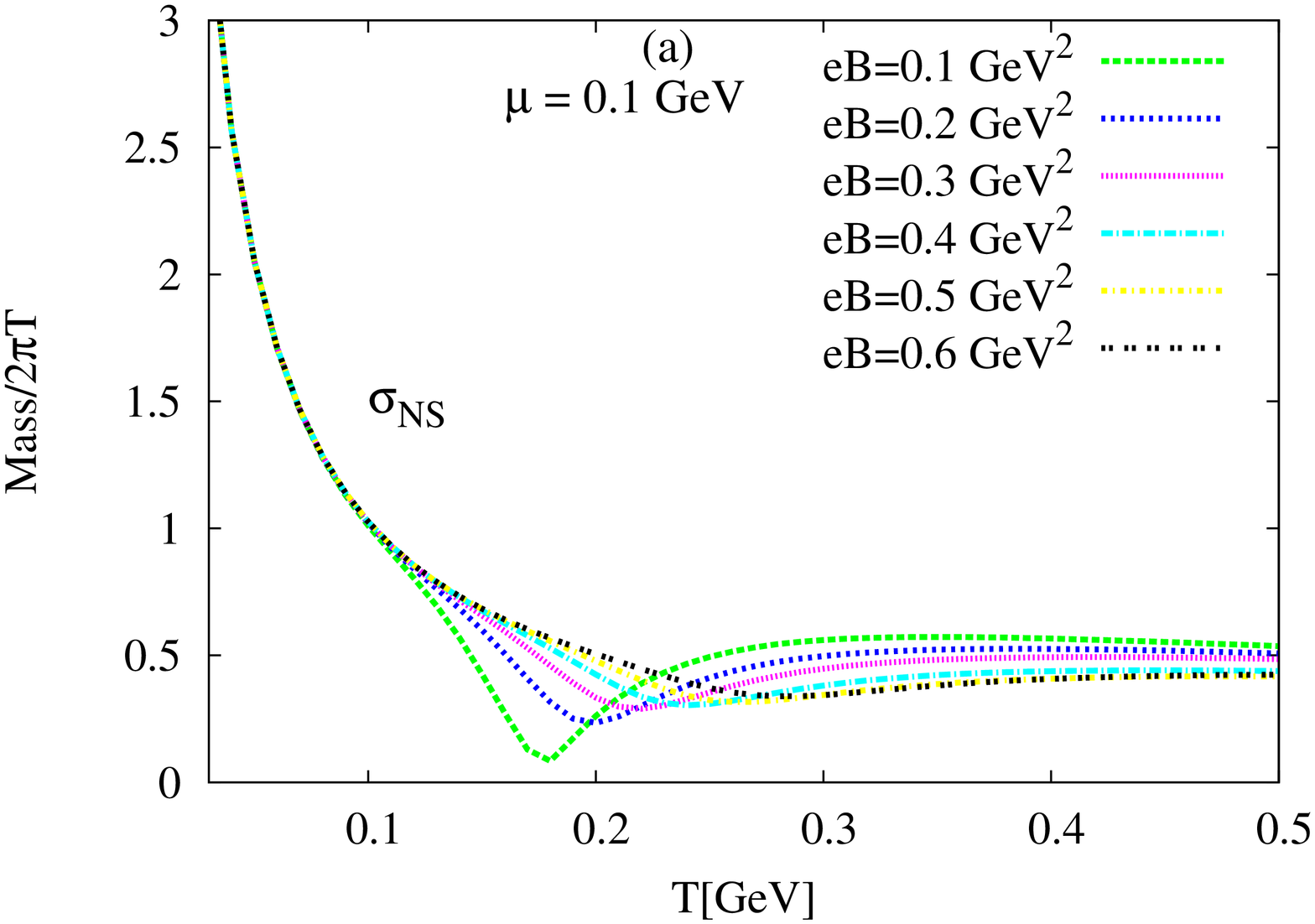}
\includegraphics[width=4.cm,angle=-90]{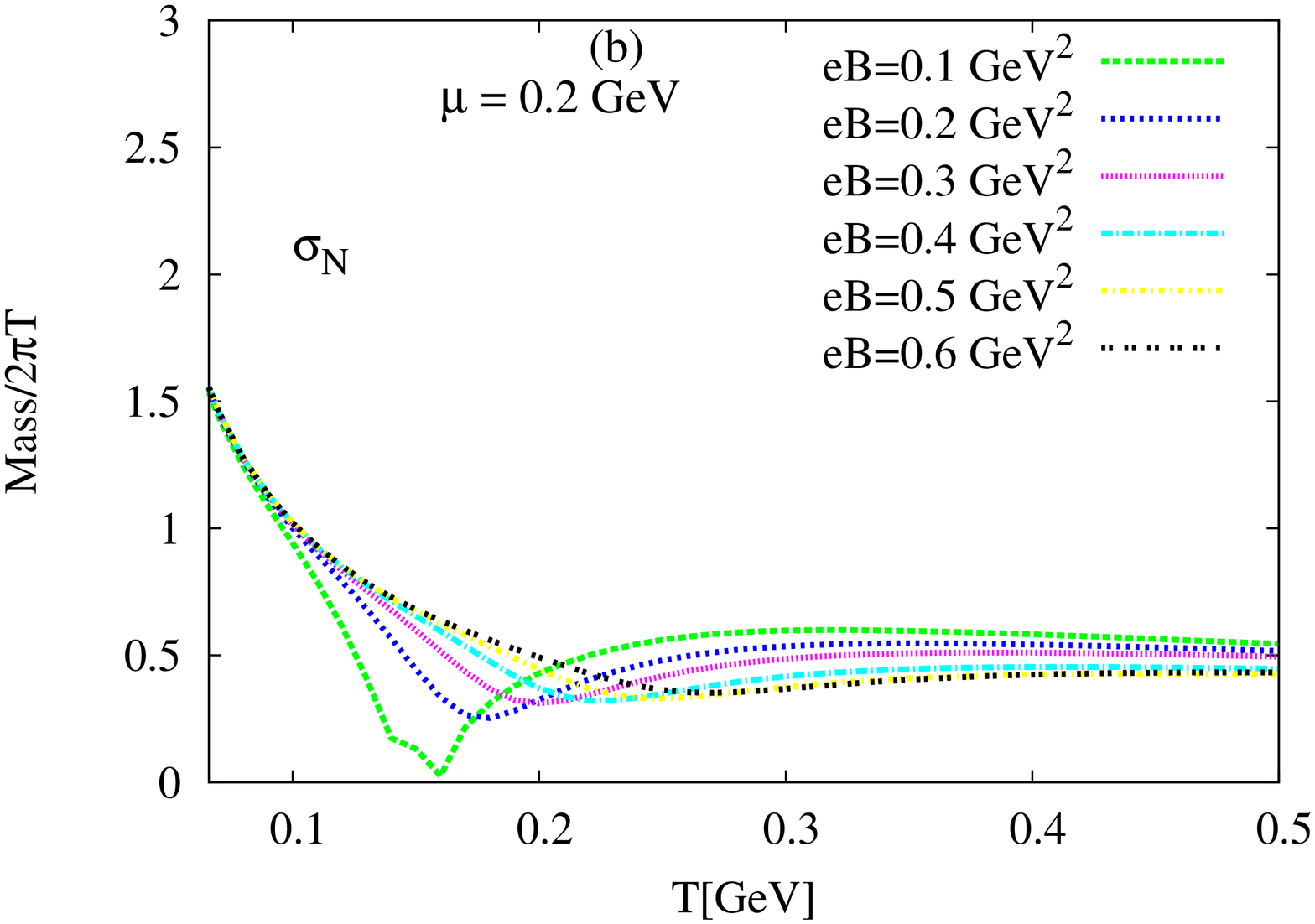}\\
\includegraphics[width=4.cm,angle=-90]{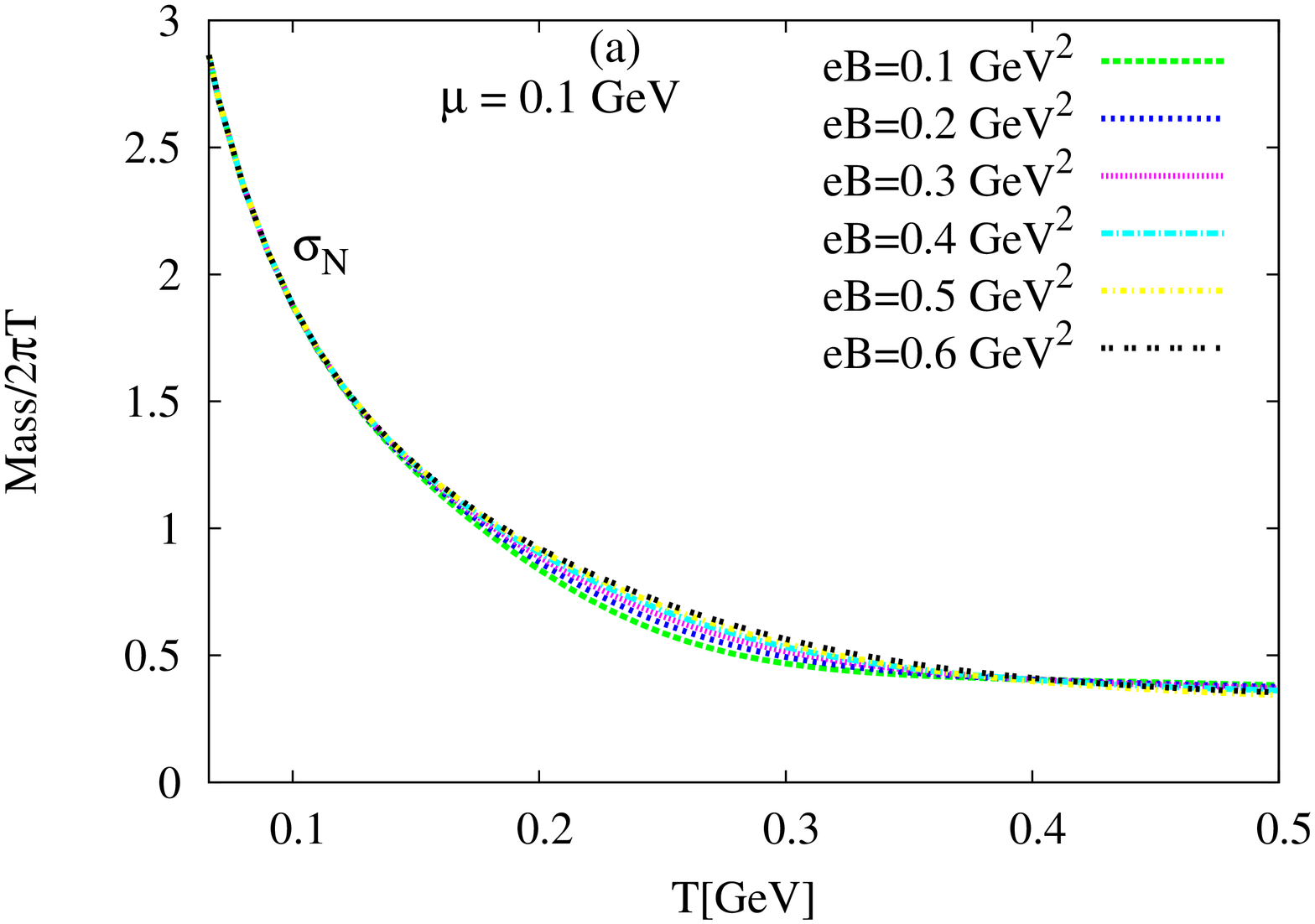}
\includegraphics[width=4.cm,angle=-90]{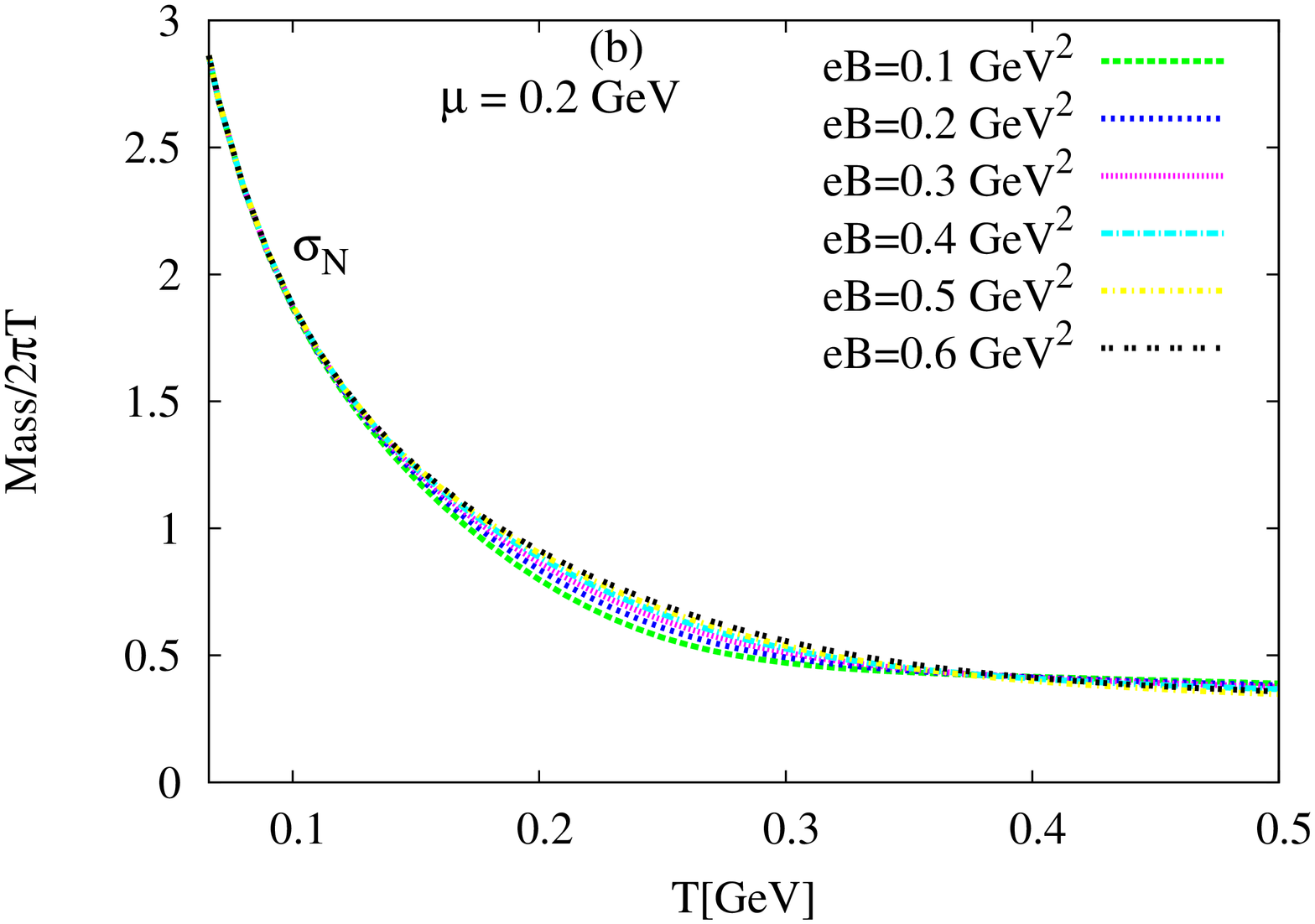}
\caption{(Color online) Left-hand panel (a): the scalar meson masses normalized with respect to the lowest Matsubara frequency are given as function of temperature at a constant chemical potential $\mu=0.1~$GeV and different magnetic fields, $eB=0.1$, $0.2$, $0.3$, $0.4$, $0.5$ and $0.6~$GeV$^2$ from top to bottom. Right-hand panel (b): the same as in left-hand panel but at chemical potential $\mu=0.2~$GeV. 
\label{fig:sif_f0_norm}}
}
\end{figure}

In Fig. \ref{fig:sif_f0_norm}, the  normalized scalar meson masses, $m_{\sigma}$ from Eq. (\ref{ms1}), $m_{f_0}$ from Eq. (\ref{ms2}), $m_{\sigma_{NS}}$  from Eq. (\ref{ms3}) and $m_{\sigma_{S}}$ from Eq. (\ref{ms4}) are given as function of temperature at two values of chemical potential, $\mu=0.1~$GeV in left-hand panel (a) and $\mu=0.2~$GeV in right-hand panel (b) and different magnetic fields, $eB=0.1$, $0.2$, $0.3$, $0.4$, $0.5$ and $0.6~$GeV$^2$ from top to bottom. The normalization is done due to the lowest Matsubara frequencies, $2 \pi T$, App. \ref{sec:matsub}. At high temperatures, we notice that the masses of almost all meson states become temperature independent, i.e. constructing a kind of a universal bundle. This would be seen as a signature for meson dissociation into quarks. In other words, the meson states undergo deconfimement phase-transition. It is worthwhile to highlight that the various meson states likely have different critical temperatures.

At low temperatures, the scalar meson masses normalized to the lowest Matsubara frequency rapidly decrease as the temperature increases. Then, starting from the critical temperature, we find that the thermal dependence almost vanishes. The magnetic field effect is clear, namely the meson masses increase with increasing magnetic field. This characterizes $T$ vs. $eB$ phase diagram.

The pseudoscalar meson masses \cite{Vivek} 
\begin{eqnarray}
m^{2}_{\eta'} &=& m^2_{p,{00}}\cos^{2}\theta_{p}+m^2_{p,88}\sin^{2}\theta_{p}+2m^{2}_{p,08}\sin\theta_{p}\cos\theta_{p}, \label{mp1} \\
m^{2}_{\eta} &=&  m^2_{p,00}\sin^{2}\theta_{p}+m^{2}_{p,88}\cos^{2}\theta_{p}-2m^2_{p,08}\sin\theta_{p}\cos\theta_{p}, \label{mp2}\\
m^{2}_{\eta_{NS}} &=& \frac{1}{3}(2m^{2}_{p,00} + m^{2}_{p,88} + 2 \sqrt{2}m^{2}_{p,08}), \label{mp3}\\
m^{2}_{\eta_{S}} &=& \frac{1}{3}(m^{2}_{p,00} + 2 m^{2}_{p,88} - 2 \sqrt{2}m^{2}_{p,08}), \label{mp4}
\end{eqnarray}
where $\theta_{p}$ is the pseudoscalar mixing angle \cite{Vivek}
\bea
\theta_p &=& \frac{1}{2} \text{ArcTan}\left[\frac{2 (m_p^2)_{08}}{(m_p^2)_{00}-(m_p^2)_{88}}\right] ,\nn
\eea
with $(m_p^2)_{a b} = m^2\, \delta_{a\, b} + 6 {\cal G}_{a b c} \bar{\sigma}_c + 4\, {\cal H}_{a b c d}\, \bar{\sigma}_c\, \bar{\sigma}_d$. 
The expressions for ${\cal H}_{a b c d}$ are given in Eq. (11c) in Ref. \cite{Vivek}.

\begin{figure}[htb]
\centering{
\includegraphics[width=5.cm,angle=-90]{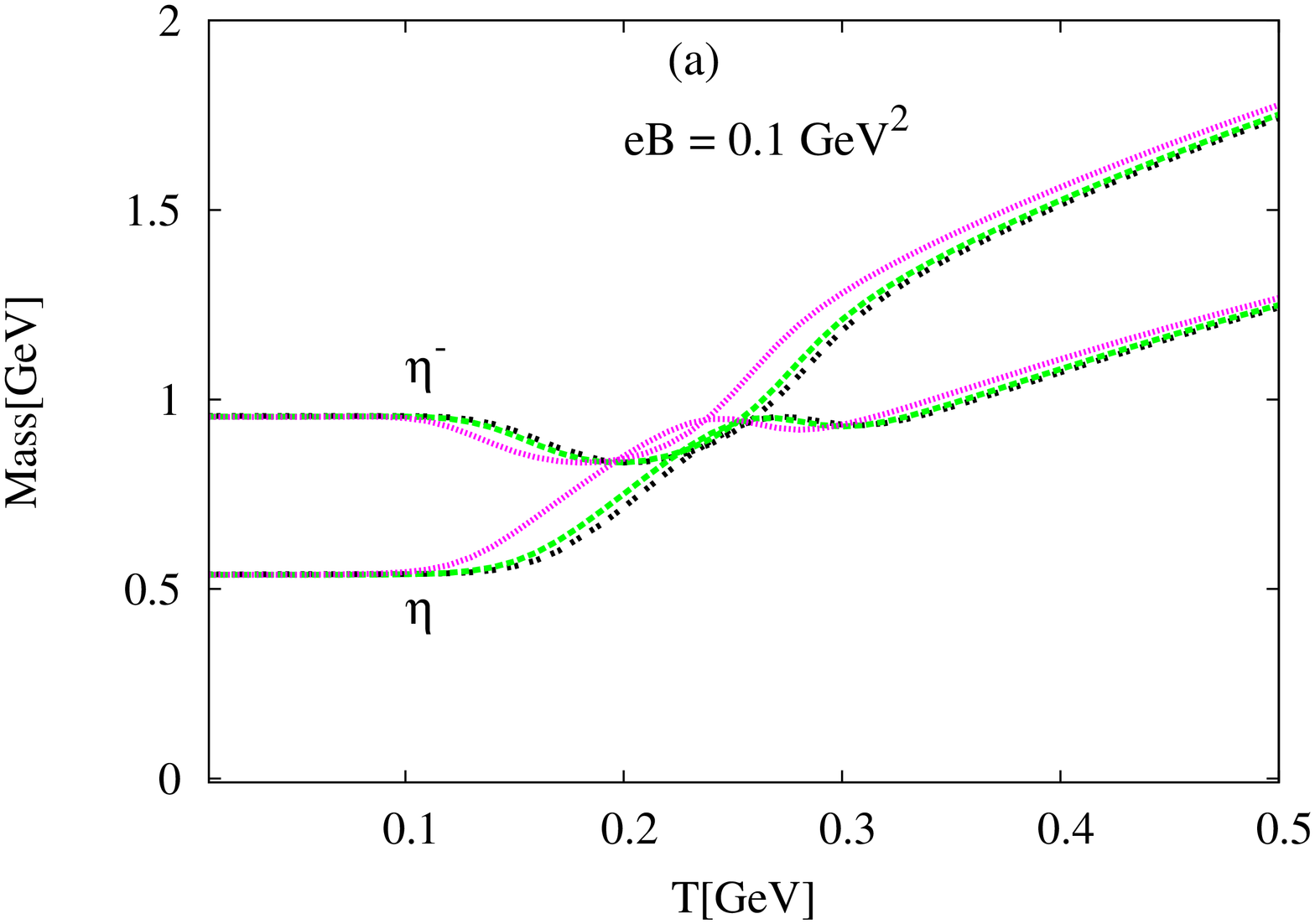}
\includegraphics[width=5.cm,angle=-90]{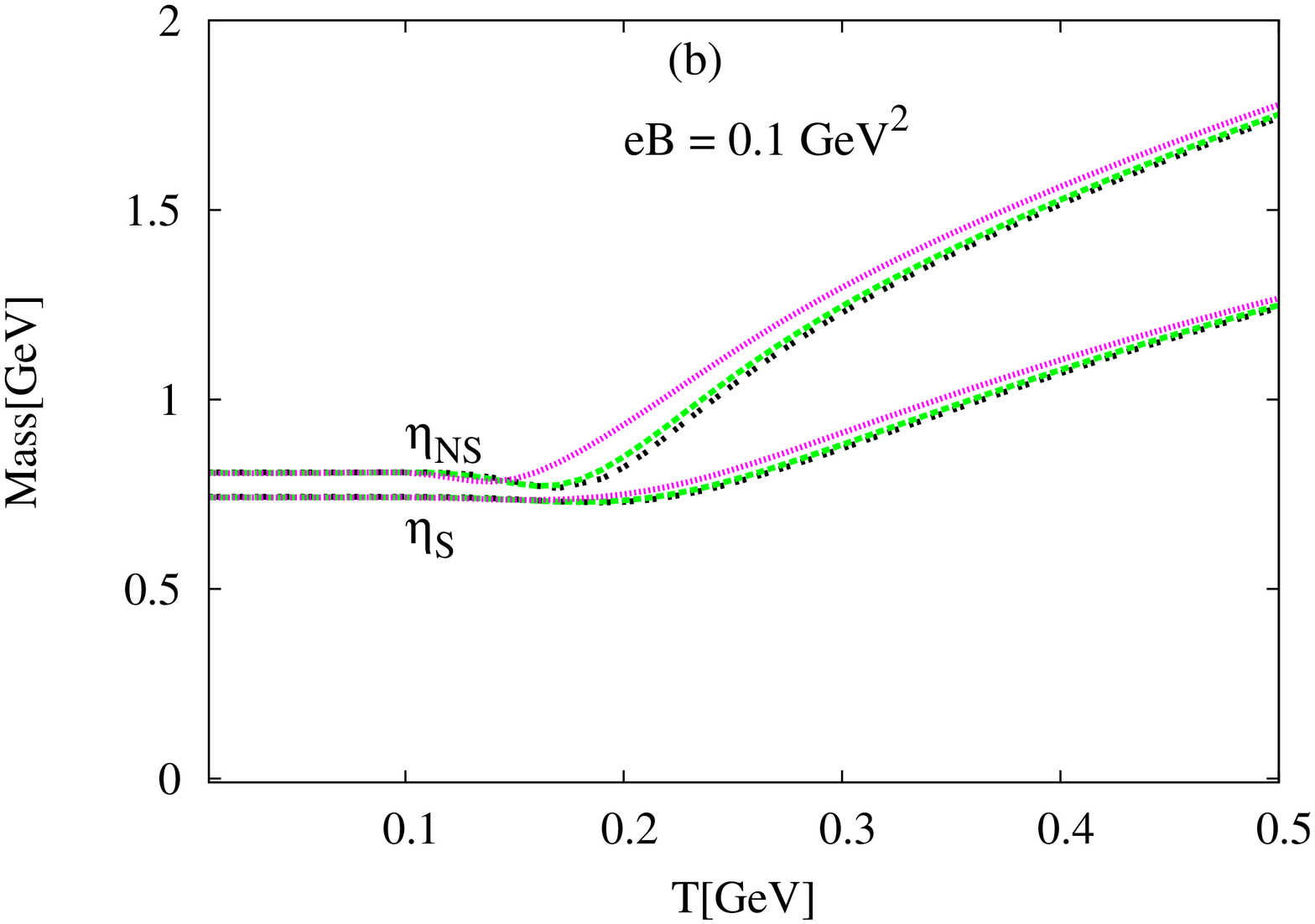}
\caption{(Color online) The same as in Fig. \ref{fig:sif_f0_Mu} but for pseudoscalar meson masses, $m_{\eta'}$ from Eq. (\ref{mp1}), $m_{\eta}$ from Eq. (\ref{mp2}), $m_{\eta_{NS}}$ from Eq. (\ref{mp3}) and $m_{\eta_{S}}$ from Eq. (\ref{mp4}). 
\label{fig:eta_etad_Mu}}
}
\end{figure}

In Fig. \ref{fig:eta_etad_Mu}, the  pseudoscalar meson masses, $m_{\eta'}$ from Eq. (\ref{mp1}), $m_{\eta}$ from Eq. (\ref{mp2}), $m_{\eta_{NS}}$ from Eq. (\ref{mp3}) and $m_{\eta_{S}}$ from Eq. (\ref{mp4}) are given as function of temperature at a constant  magnetic field $eB=0.1~$GeV$^2$ and different chemical potentials, $\mu=0.0~$GeV (dotted curve),  $0.1~$GeV (dashed curve) and  $0.2~$GeV (double-dotted curve).  It is obvious that the pseudoscalar meson masses remain constant at low temperature. At temperatures $\geq T_c$, the vacuum effect becomes dominant. Accordingly. the pseudoscalar meson masses increase with the temperature. We shall notice that even contribution by the vacuum will be moderated through the normalization with respect to the lowest Matsubara frequency.

\begin{figure}[htb]
\centering{
\includegraphics[width=4.cm,angle=-90]{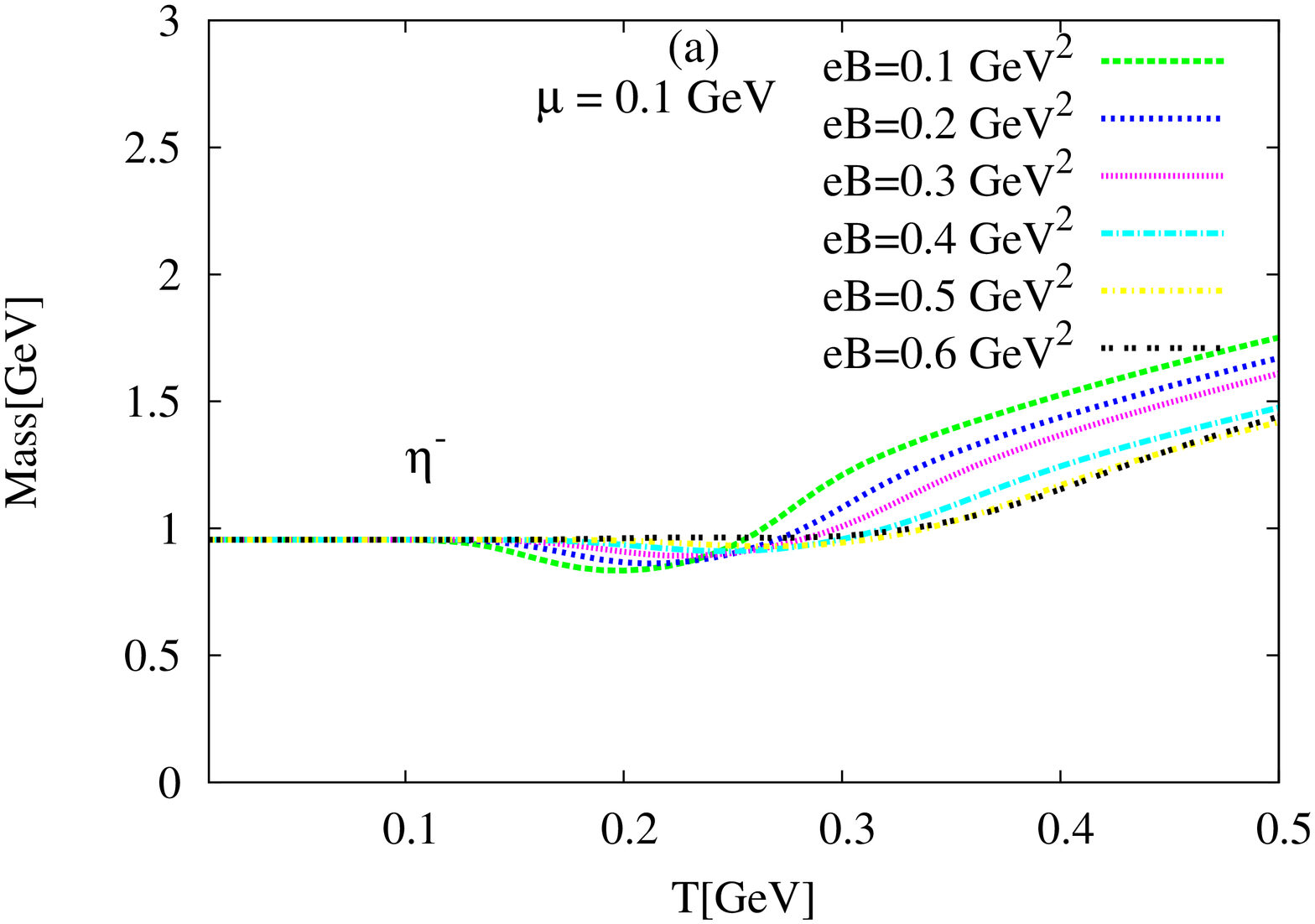}
\includegraphics[width=4.cm,angle=-90]{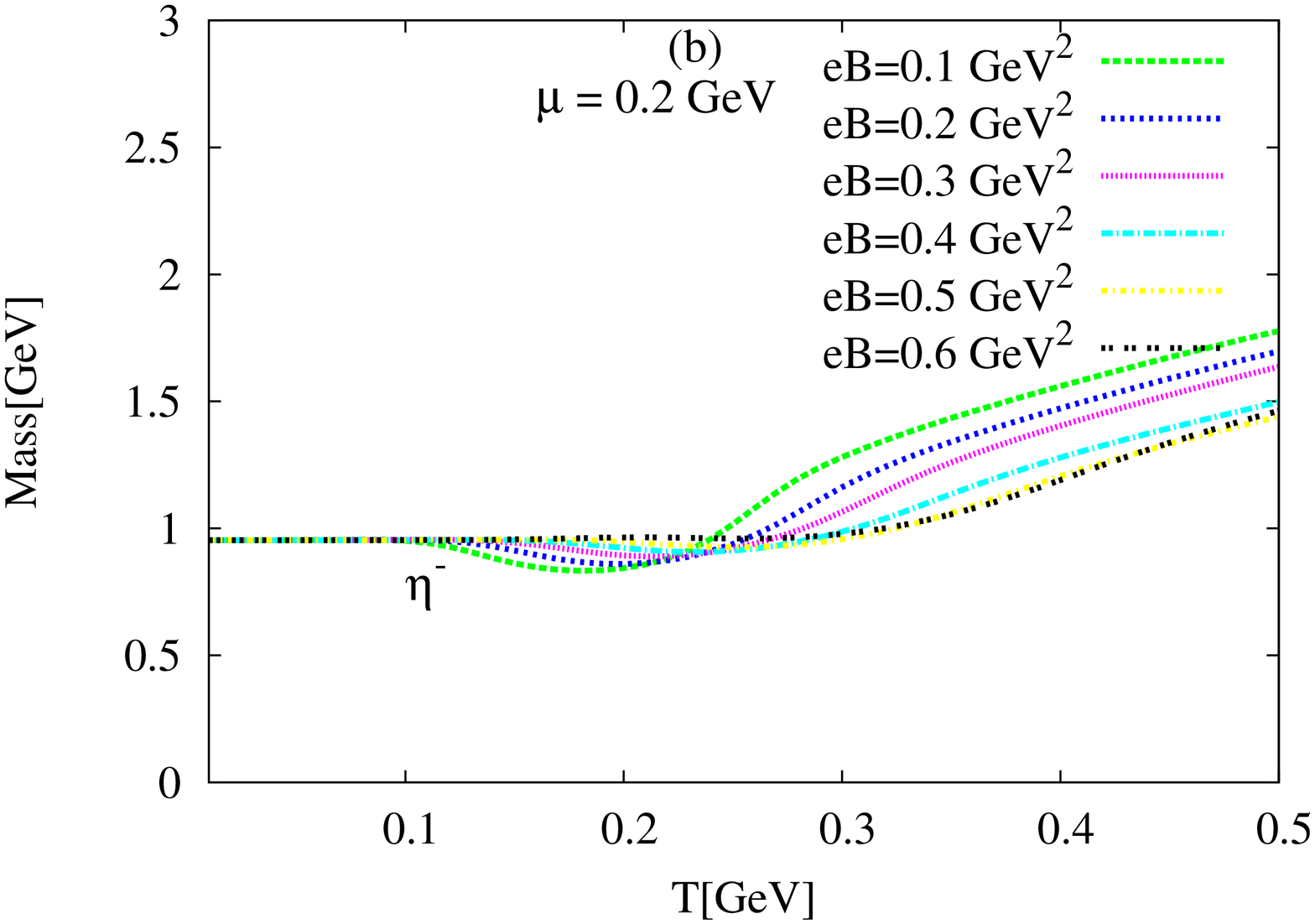}\\
\includegraphics[width=4.cm,angle=-90]{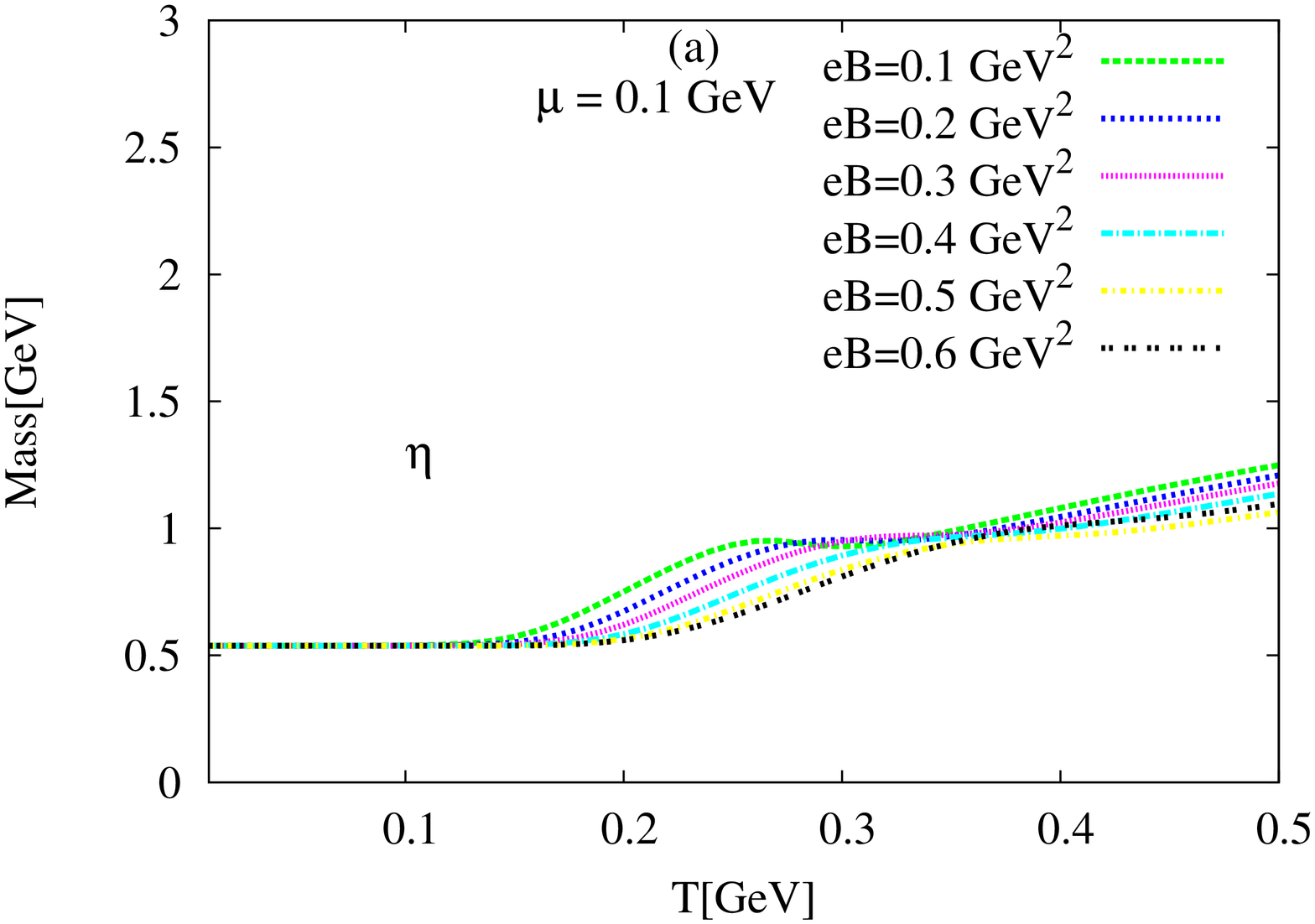}
\includegraphics[width=4.cm,angle=-90]{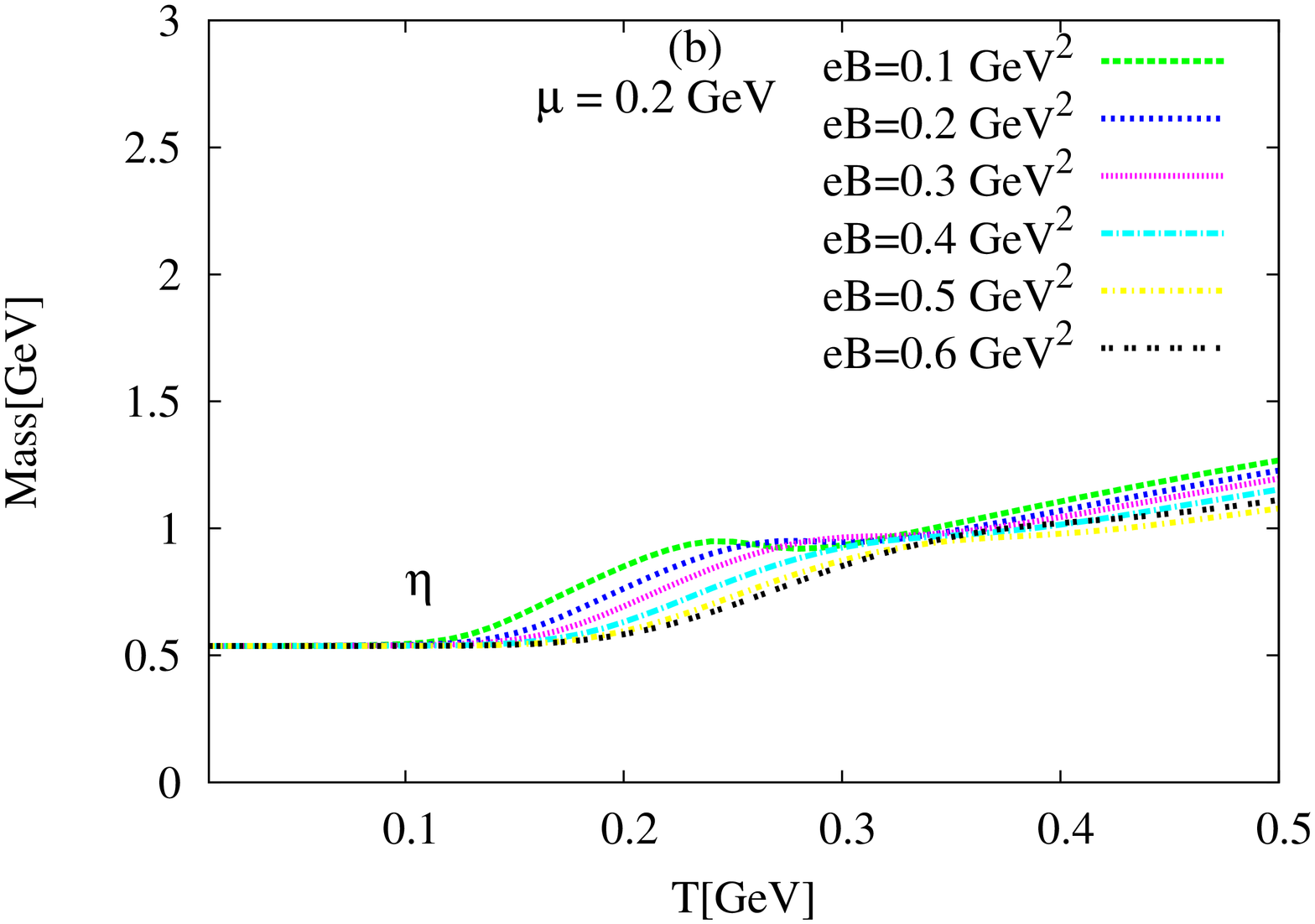}\\
\includegraphics[width=4.cm,angle=-90]{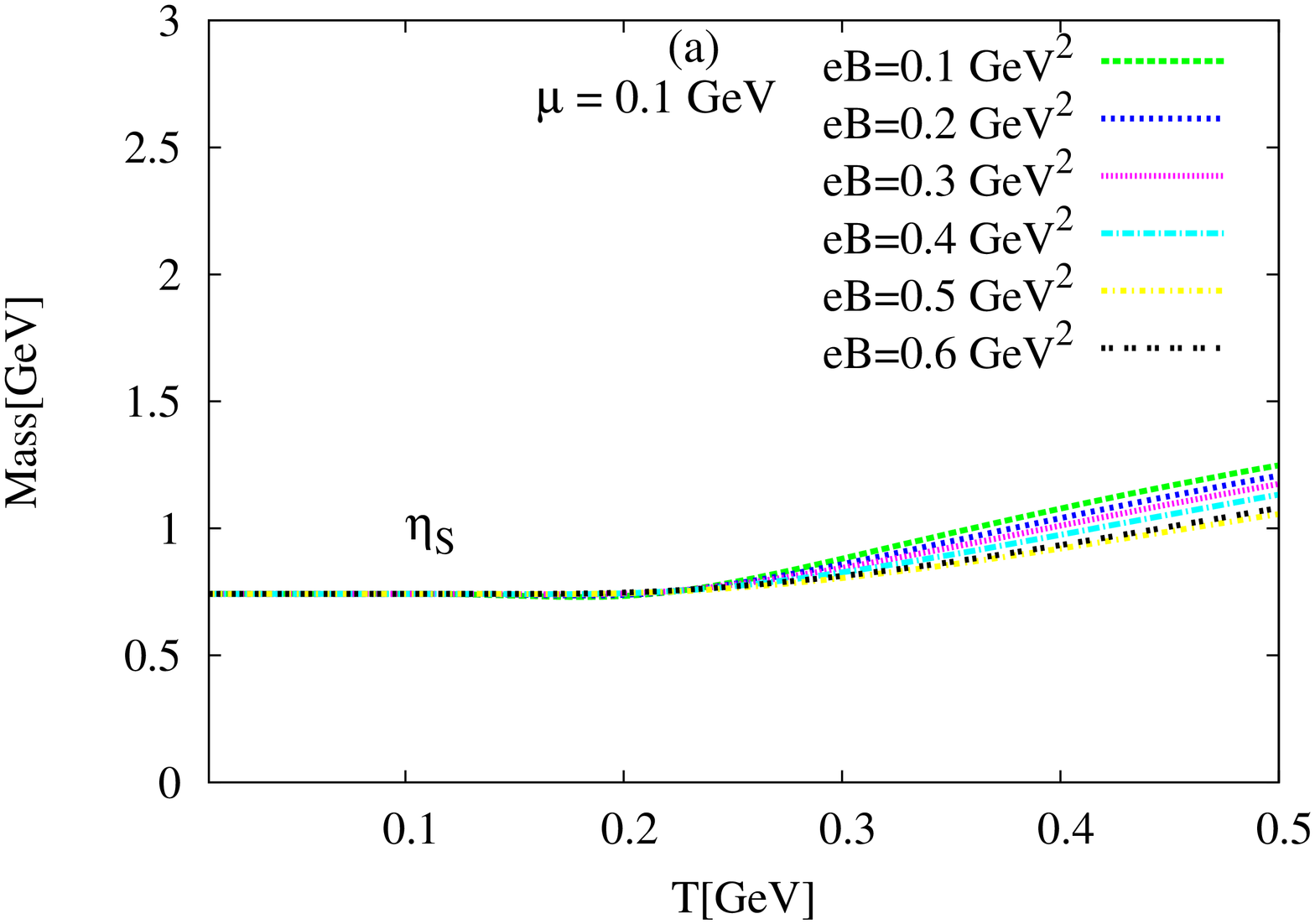}
\includegraphics[width=4.cm,angle=-90]{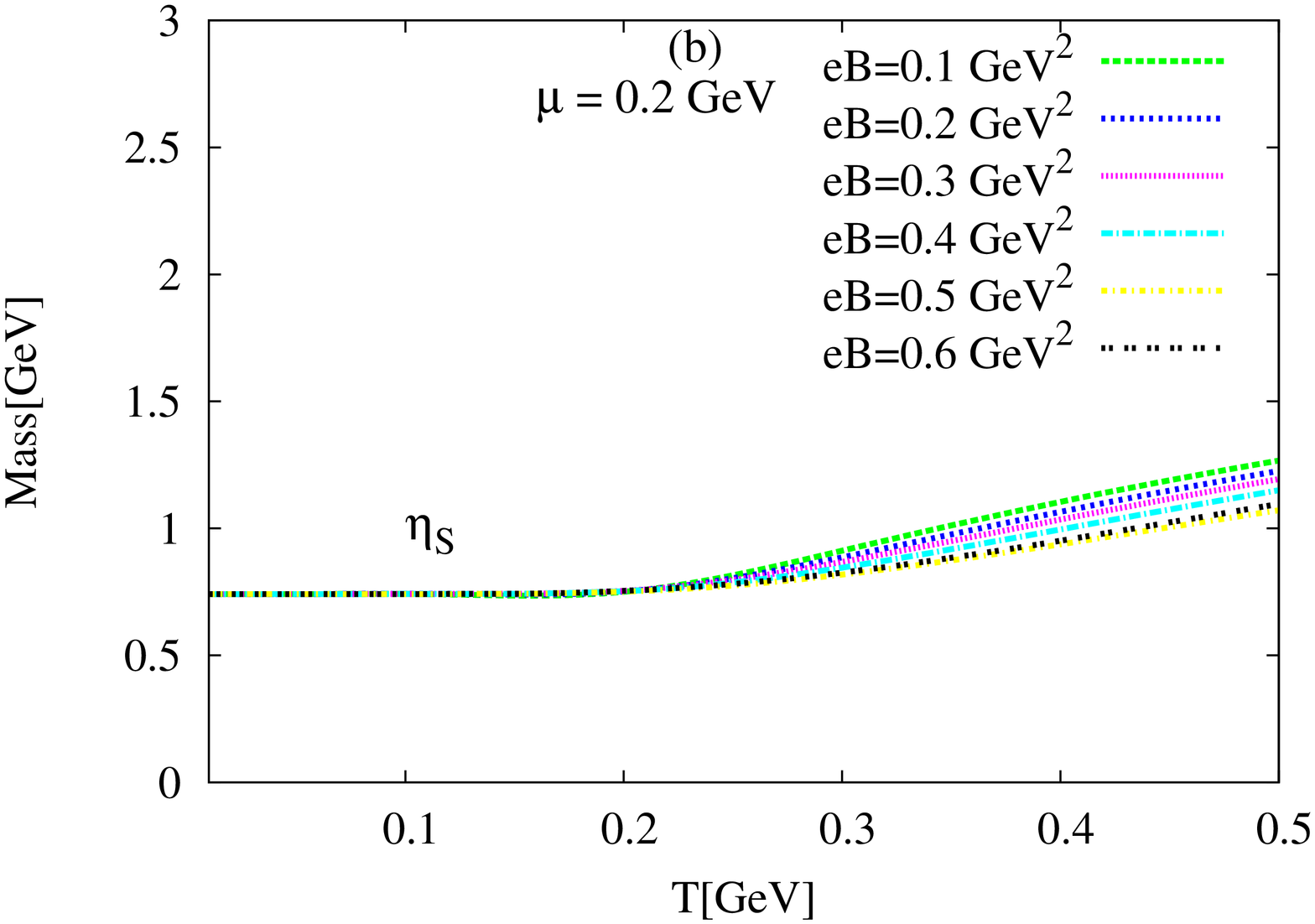}\\
\includegraphics[width=4.cm,angle=-90]{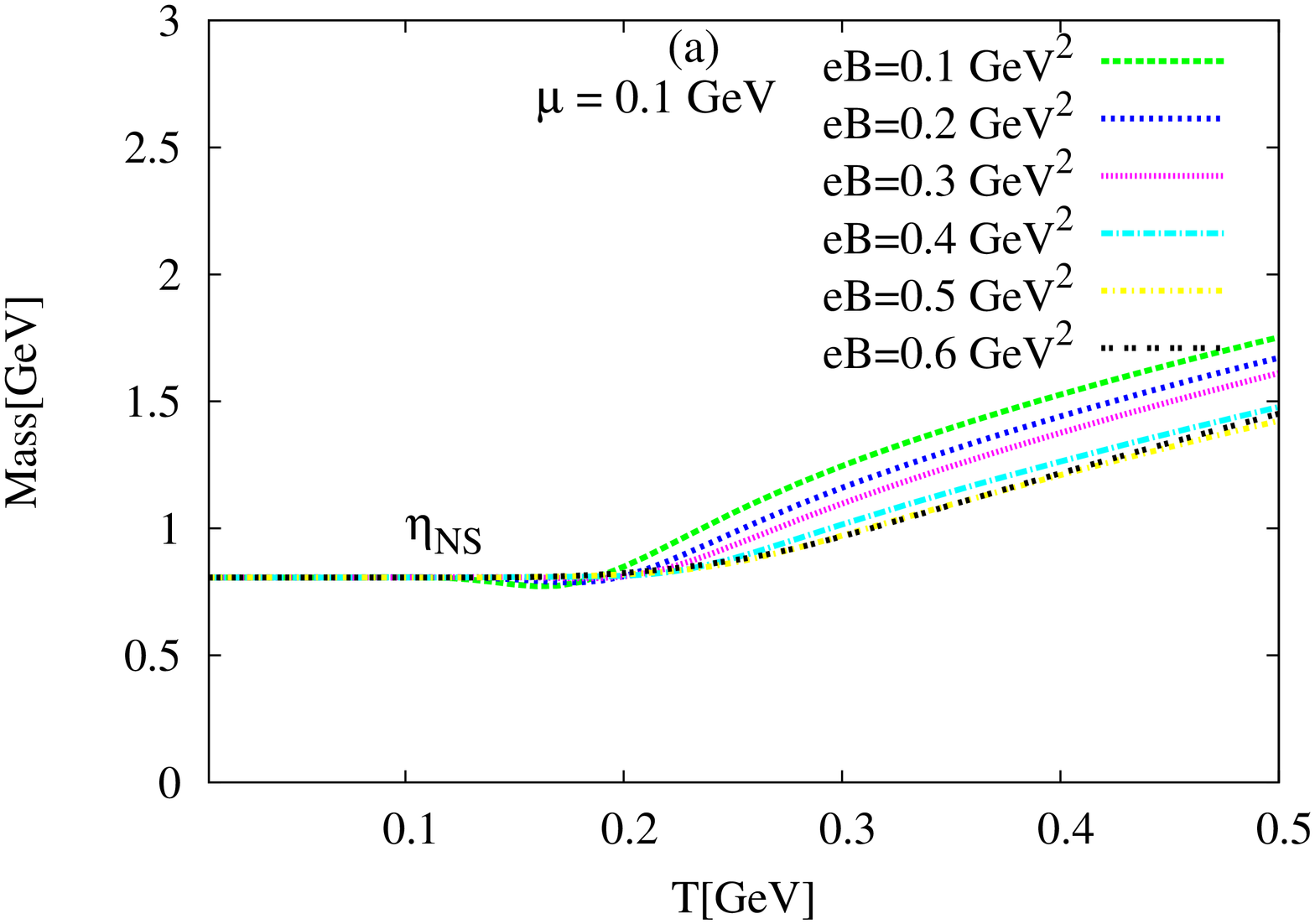}
\includegraphics[width=4.cm,angle=-90]{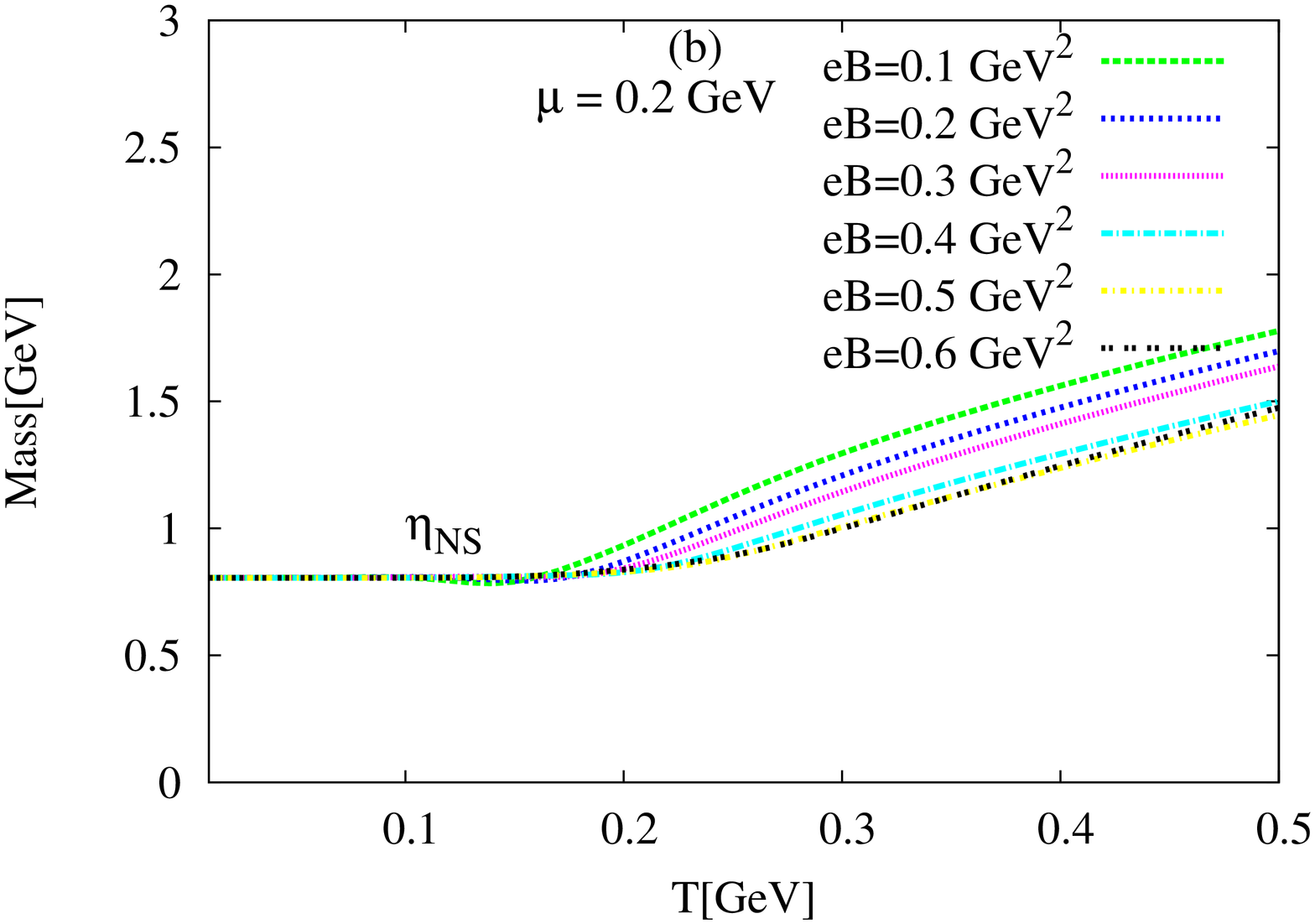}
\caption{(Color online) Left-hand panel (a): the pseudoscalar meson masses are given as function of temperature at a constant chemical potential $\mu=0.1~$GeV and different magnetic fields, $eB=0.1$, $0.2$, $0.3$, $0.4$, $0.5$ and $0.6~$GeV$^2$ from top to bottom. Right-hand panel (b) shows the same as in left-hand panel but at a constant  chemical potential $\mu=0.2~$GeV.
\label{fig:sif_f0_p}}
}
\end{figure}

In Fig. \ref{fig:sif_f0_p}, the four pseudoscalar meson masses, $m_{\eta'}$ from Eq. (\ref{mp1}), $m_{\eta}$ from Eq. (\ref{mp2}), $m_{\eta_{NS}}$ from Eq. (\ref{mp3}) and $m_{\eta_{S}}$ from Eq. (\ref{mp4}) are given as function of temperature at two constant chemical potentials, $\mu=0.1~$GeV in left-hand panel (a) and $\mu=0.2~$GeV in right-hand panel (b) and different magnetic fields, $eB=0.1$, $0.2$, $0.3$, $0.4$, $0.5$ and $0.6~$GeV$^2$ from top to bottom. Again, at low temperature, the masses remain temperature-independent.  At $T\geq T_c$, the vacuum effect is switched on. Accordingly, the masses increase rapidly with the temperature.

\begin{figure}[htb]
\centering{
\includegraphics[width=4.cm,angle=-90]{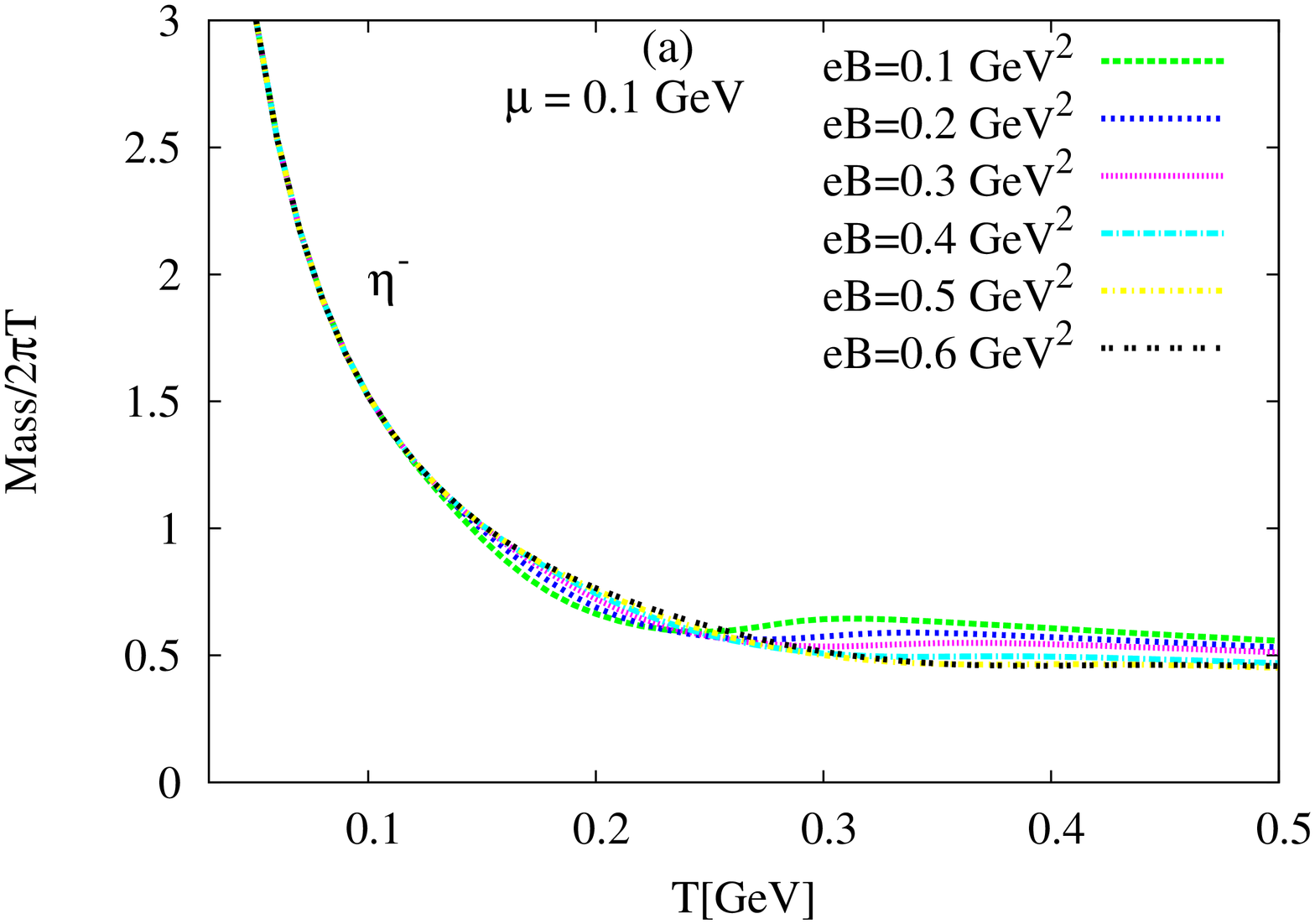}
\includegraphics[width=4.cm,angle=-90]{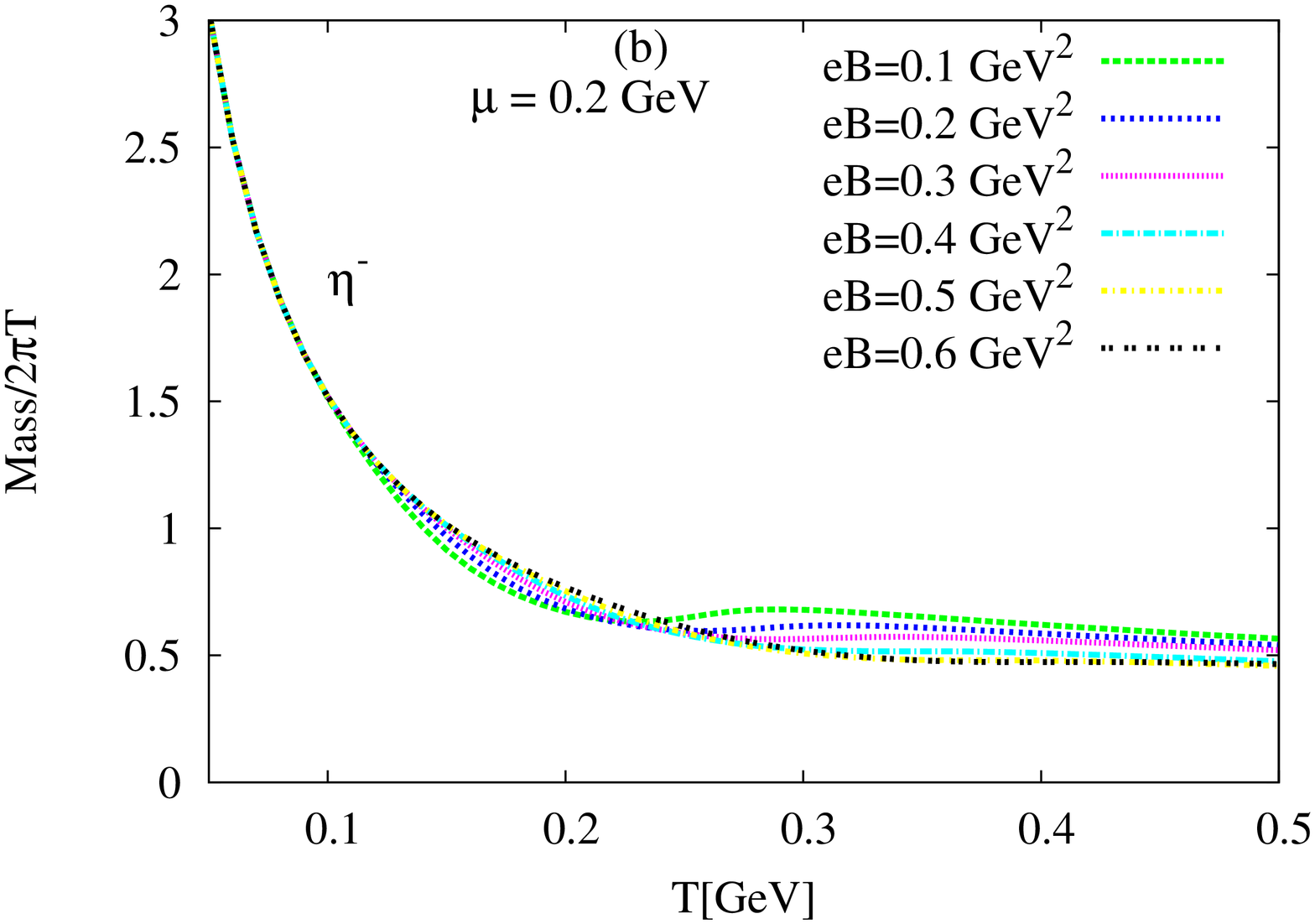}\\
\includegraphics[width=4.cm,angle=-90]{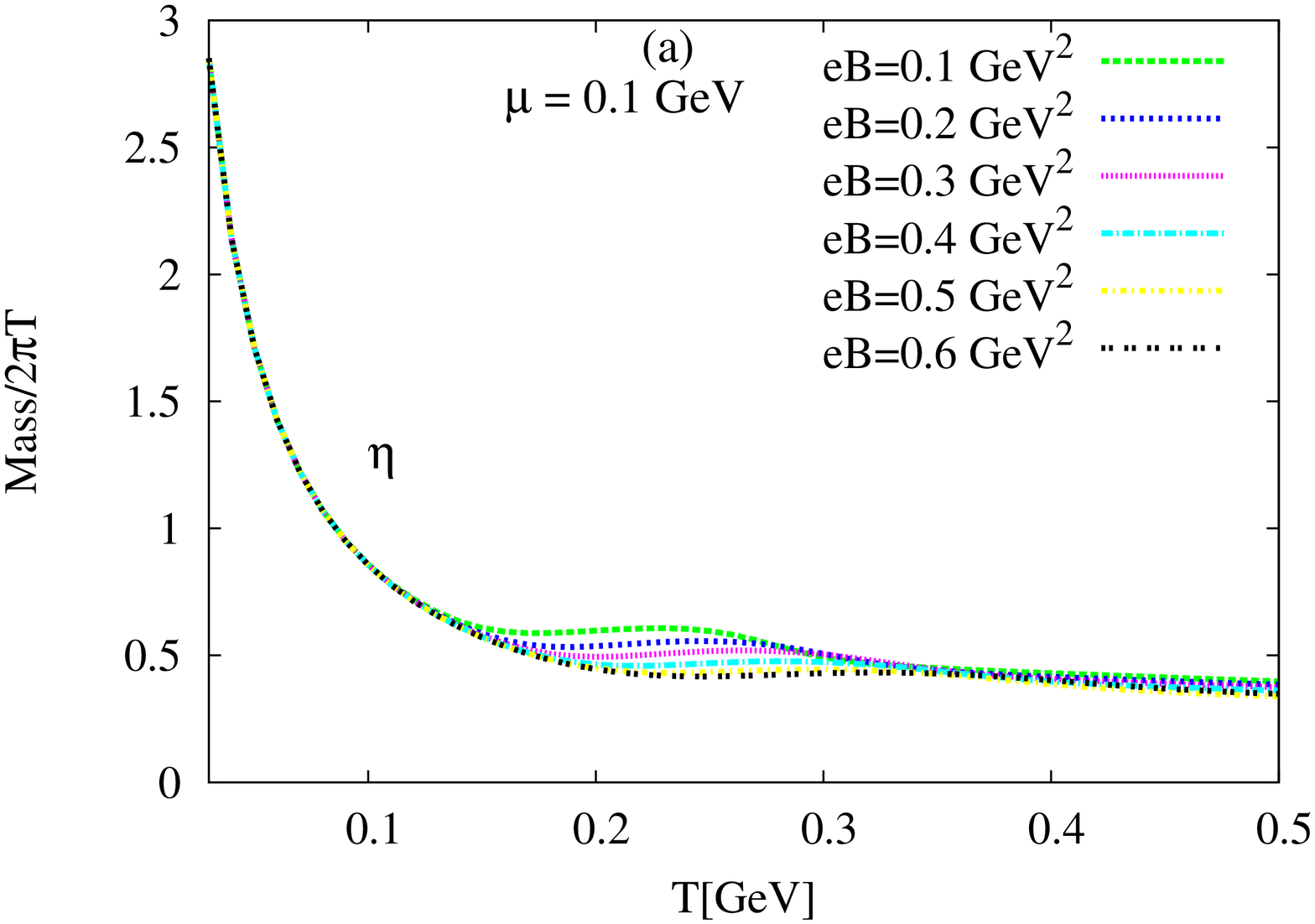}
\includegraphics[width=4.cm,angle=-90]{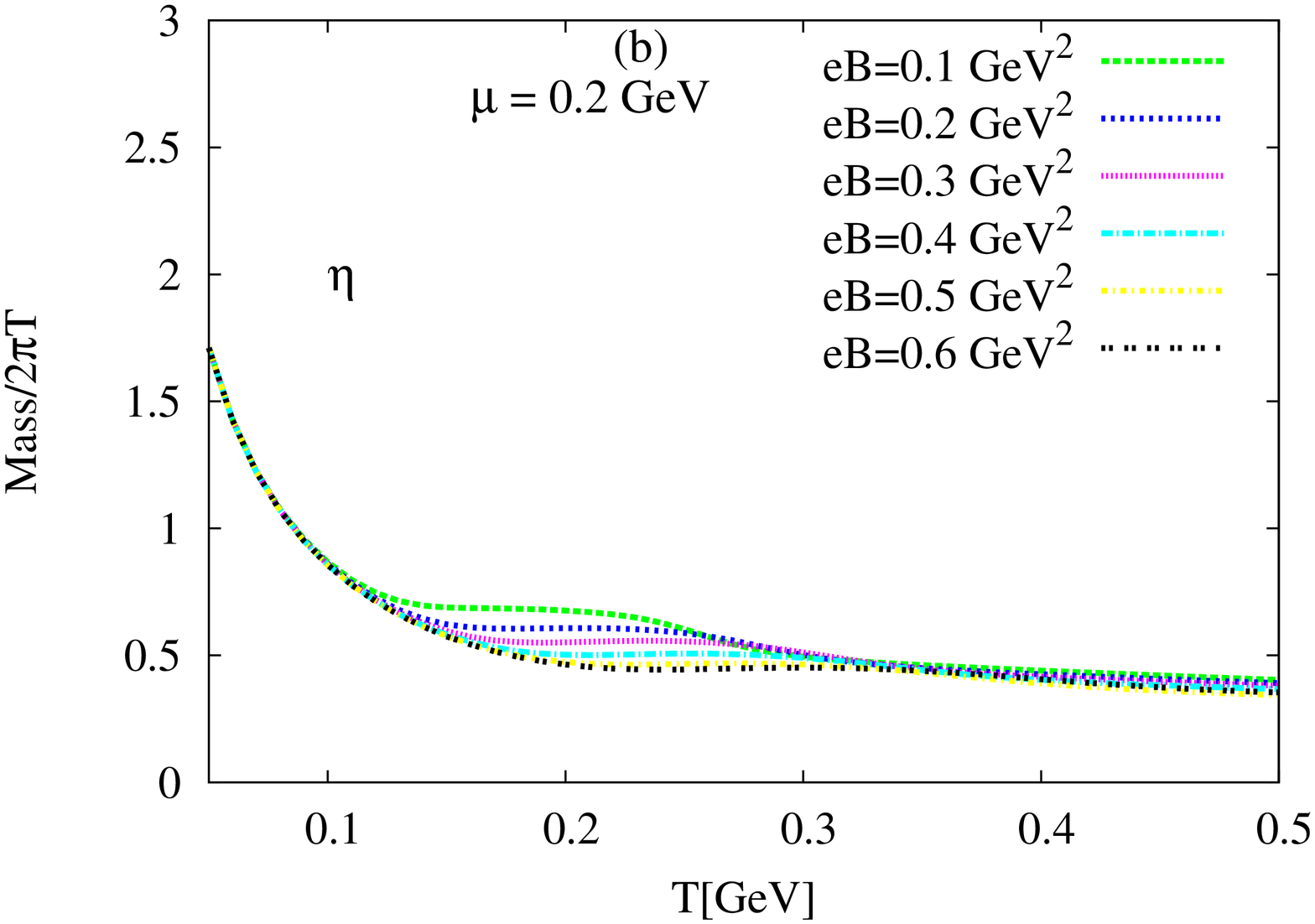}\\
\includegraphics[width=4.cm,angle=-90]{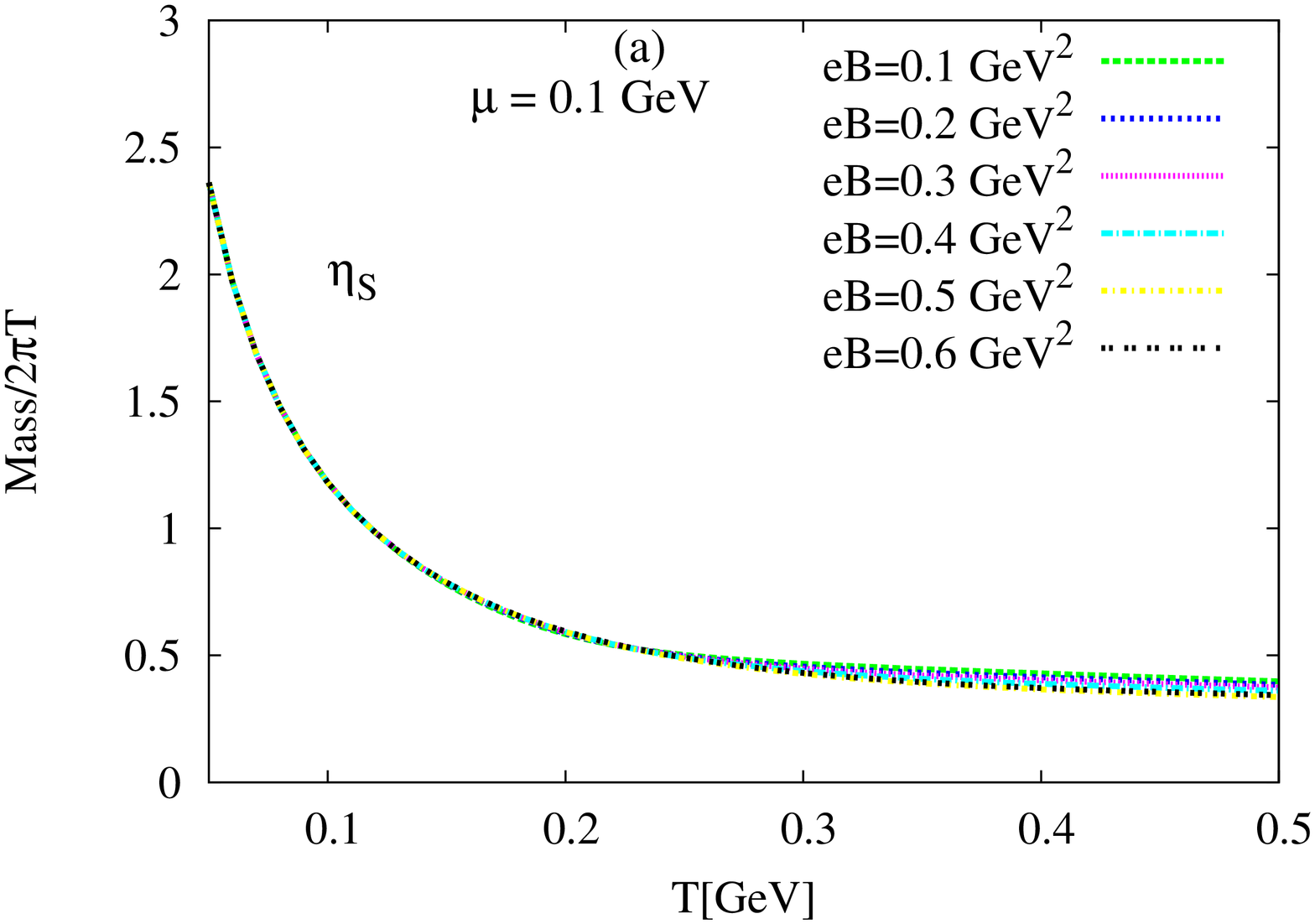}
\includegraphics[width=4.cm,angle=-90]{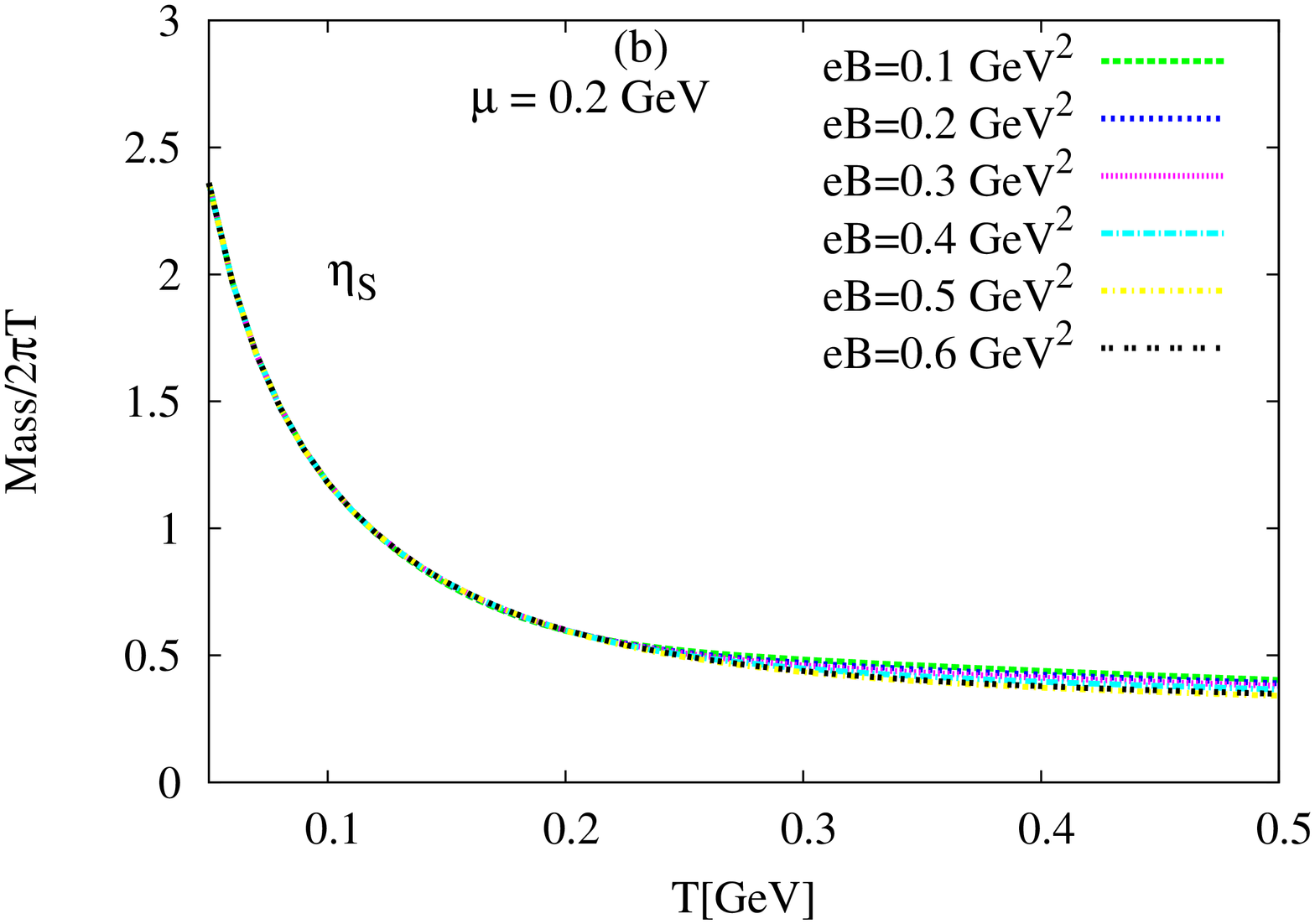}\\
\includegraphics[width=4.cm,angle=-90]{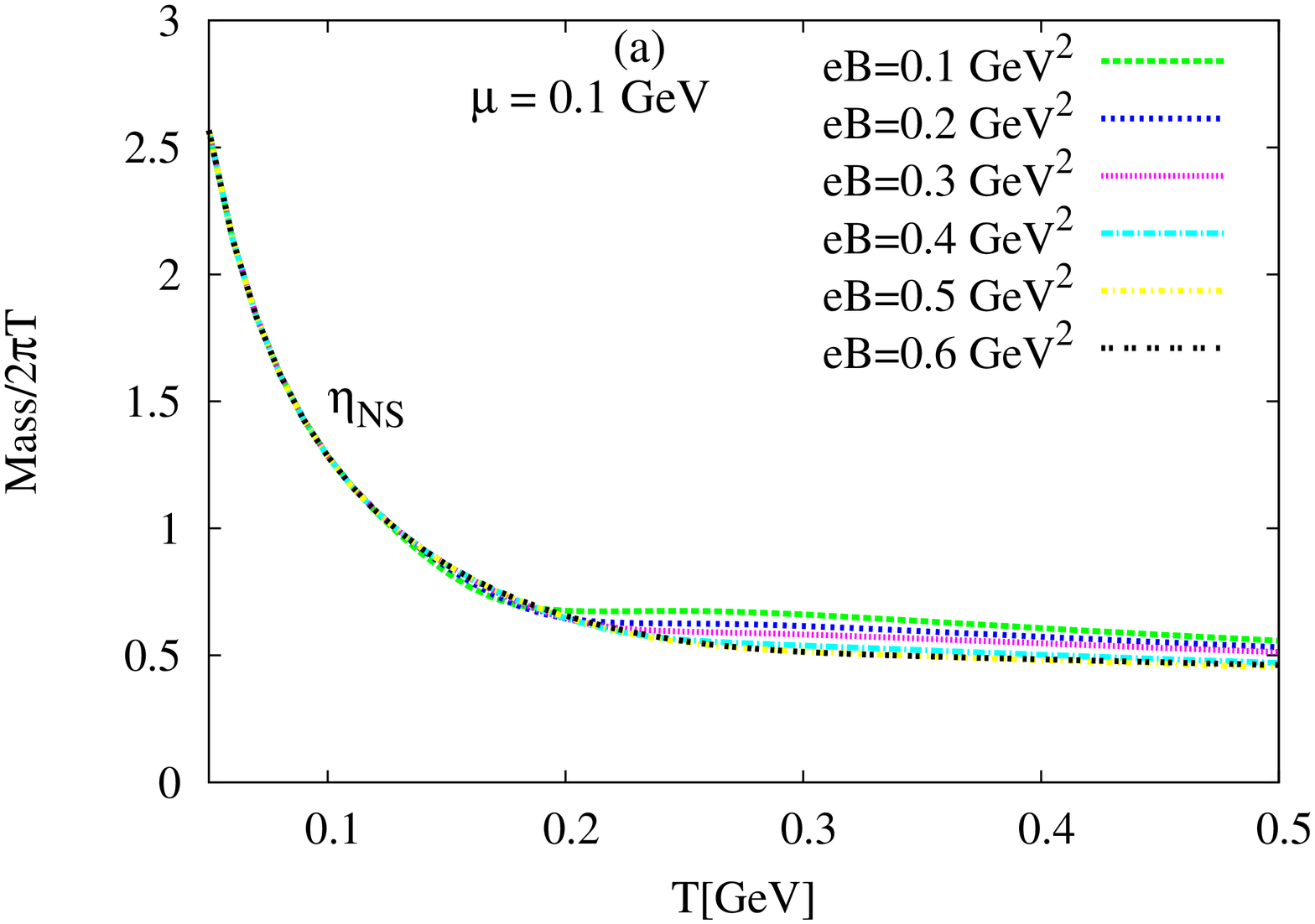}
\includegraphics[width=4.cm,angle=-90]{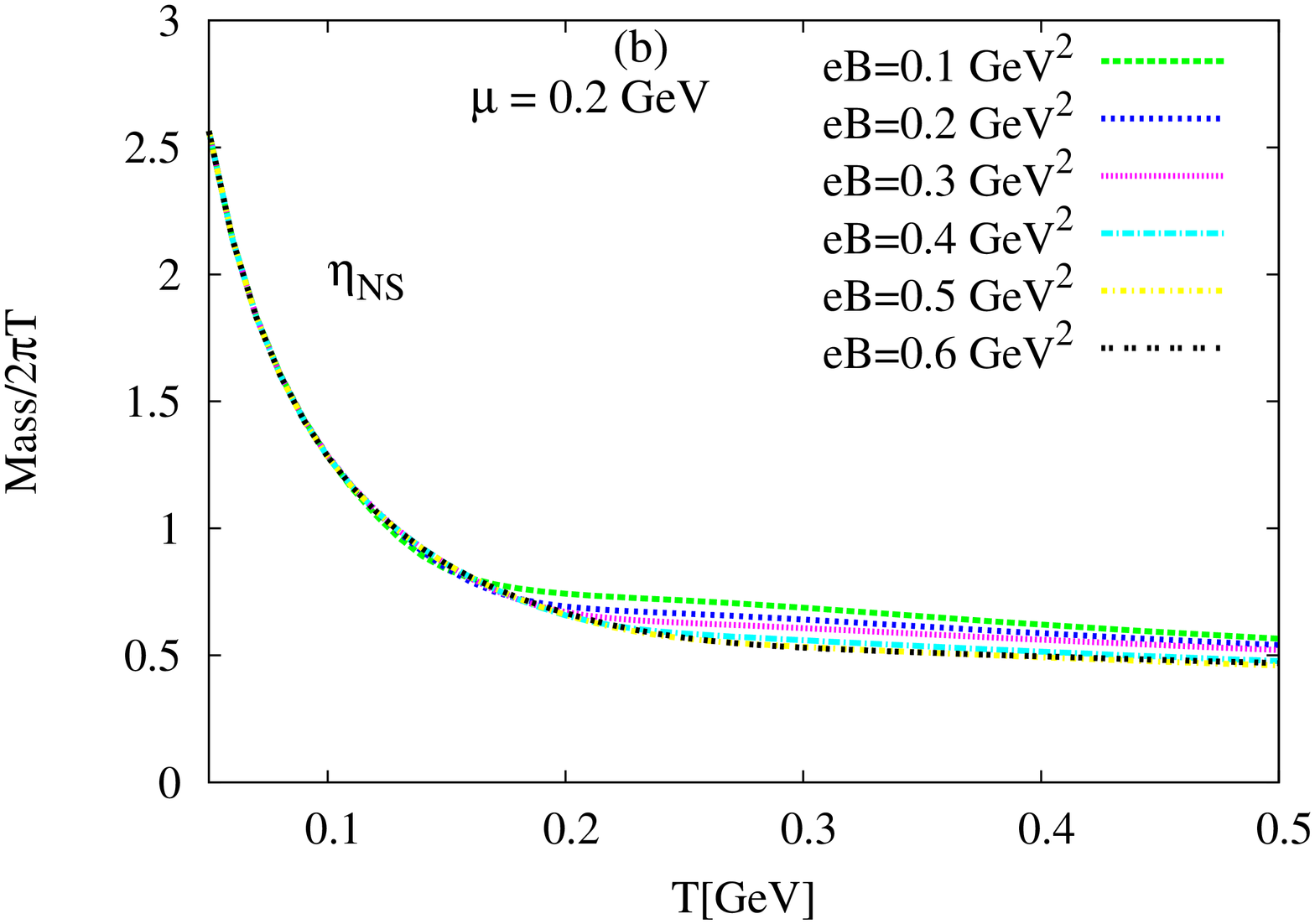}
\caption{(Color online) Left-hand panel (a): the pseudoscalar meson masses normalized to the lowest Matsubara frequency are given as function of temperature at a constant chemical potential $\mu=0.1~$GeV and different magnetic fields, $eB=0.1$, $0.2$, $0.3$, $0.4$, $0.5$ and $0.6~$GeV$^2$ from top to bottom. Right-hand panel (b): shows the same as in left-hand panel but at a constant chemical potential $\mu=0.2~$GeV.
\label{fig:sif_f0_p_norm}}
}
\end{figure}

In Fig. \ref{fig:sif_f0_p_norm}, the four pseudoscalar meson masses, $m_{\eta'}$ from Eq. (\ref{mp1}), $m_{\eta}$ from Eq. (\ref{mp2}), $m_{\eta_{NS}}$ from Eq. (\ref{mp3}) and $m_{\eta_{S}}$ from Eq. (\ref{mp4}) normalized with respect to the lowest Matsubara frequency are given as function of temperature at two constant chemical potentials, $\mu=0.1~$GeV in left-hand panel (a) and $\mu=0.2~$GeV in right-hand panel (b) and different magnetic fields, $eB=0.1$, $0.2$, $0.3$, $0.4$, $0.5$ and $0.6~$GeV$^2$ from top to bottom. The normalization should result in temperature-independent mesonic states. This would be seen as a signature for meson dissociation into quarks. It is obvious that various critical temperatures can be assigned to various pseudoscalar meson states. The normalized masses starts with high values reflecting confinement, especially at low temperatures. Then, they decrease as the temperature increases until the critical temperature, $T_c$, which differs for different meson states. At higher temperatures, the dependence of meson masses on temperature is almost entirely removed.

\begin{table}[htb]
\begin{tabular}{|c|c|c|c|c|c||}
\hline
& Symbol & PDG \cite{PDG:2012} & Present Work & \begin{tabular}{c} PNJL \cite{NJL:2013,P. Costa:PNJL} \end{tabular} & \begin{tabular}{c|c}\multicolumn{2}{ c }{Lattice QCD} \\
\hline
Hot QCD \cite{HotQCD} & PACS-CS \cite{PACS-CS} \\ \end{tabular} \\
\hline \hline
& \begin{tabular}{c}
$\pi$\\ $K $ \\ $\eta$ \\ $\eta ^{'}$ \\
\end{tabular} 
& \begin{tabular}{c}
$134.970 \pm 6.9$ \\ $497.614 \pm 24.8$ \\ $547.853 \pm 27.4$ \\ $957.78 \pm 60$ \\
\end{tabular}  
& \begin{tabular}{c}
$120$ \\ $509$ \\ $553$ \\ $965$ \\
\end{tabular}  &\begin{tabular}{c}$126$ \\ $490$ \\ $505$ \\ $949$\\ \end{tabular} &
\begin{tabular}{l}$134 \pm 6 ~~~~~~~~~~135.4 \pm 6.2 $ \\ $422.6 \pm 11.3 \, ~498 \pm 22 $\\ $579 \pm 7.3\, \;~~~~~688 \pm 32$\\ $-~~~~~~~~~~~~~~~~~~~- $~\\ \end{tabular}
 \\   
\hline
 \end{tabular}
\caption{A comparison between pseudoscalar meson masses, $J^{PC}=0^{-+}$, in MeV and the corresponding results from PNJL \cite{P. Costa:PNJL}. Both are compared with the experimental measurements, the Particle Data Group (PDG) \cite{PDG:2012} and the lattice QCD calculations \cite{HotQCD,PACS-CS}.}
 \label{masscomp}
 \end{table}

As introduced in Ref.  \cite{AD2}, Tab. \ref{masscomp} presents  a comparison between pseudoscalar meson nonets in various effective models, like PLSM ~(present work) and PNJL \cite{P. Costa:PNJL} confronted to the particle data group \cite{PDG:2012} and lattice QCD calculations \cite{HotQCD,PACS-CS}. The comparison for scalar states would be only partly possible. Some remarks are now in order.  The errors are deduced from the fitting for the parameters used in calculating the equation of states and other thermodynamics quantities.  The output results are very precise for some of the lightest hadron resonances described by the present model, PLSM.  An extended comparison is given in Ref. \cite{AD2}.

\section{Conclusions and outlook}
\label{sec:conclusion}

The QCD phase-diagram at vanishing chemical potential and finite temperature subject to an external magnetic field gained prominence among high-energy physicists,  for instance, our previous work \cite{Tawfik:2014hwa} was based on two concepts in order to explain the effects of external magnetic field on the QCD phase-diagram. Another study was done in the framework of Nambu - Jona-Lasinio (NJL) model and Polyakov NJL (PNJL) model \cite{Ferreira2014}. The main idea is that the scalar coupling parameter is taken dependent on the magnetic field intensity. Thus, the scalar coupling parameter decreases with the magnetic field increase. we also implemented the relation between the  magnetic field and the scalar coupling parameter in order to fit for the lattice QCD results \cite{Bali:2011qj}. We conclude the increase in the magnetic field increases the critical temperature. 

In the presence work, we use the Polyakov linear $\sigma$-model and assume that the external magnetic field adds some restrictions to the quarks due to the existence of free charges in the plasma phase. In doing this, we apply Landau theory (Landau quantization), which quantizes of the cyclotron orbits of charged particles in magnetic field. First, we have calculated and then analysed the thermal evolution of the chiral condensates and the deconfinement order-parameters. We notice that the Landau quantization requires additional temperature to drive the system through the chiral phase-transition. Accordingly, we find that the value of the chiral condensates increase with increasing the external magnetic field. In the contrary to various previous studies, the effects of the external magnetic field are analysed, systematically. Accordingly, the dependence of the critical temperatures of chiral and confinement phase-transitions on the magnetic field could be characterized. We deduced  $T$-$\mu$ curves using SU(3) PLSM in external magnetic field.
 
Furthermore, by using mean field approximation, we constructed the partition functions and then driven various thermodynamic quantities, like energy density and interaction rate (trace anomaly). Their dependence on temperature and chemical potential recalls to highlight that the effects of external magnetic field on the chemical potential was disregarded in all calculations at finite chemical potential. 

We have analysed the first four non-normalized higher-order moments of particle multiplicity. The thermal evolution was studied at a constant chemical potential but different magnetic fields and also at a constant magnetic field but different chemical potentials. Doing this, the chiral phase-diagram can be mapped out. We determined the irregular behavior as function of $T$ and $\mu$. We found that increasing temperature rapidly increases the four moments and the thermal dependence is obviously enhanced, when moving from lower to higher orders. The values of the moment are increasing as we increase the chemical potential. But the critical temperature $T_c$ decrease with  $\mu$. The peaks are positioned at the critical temperature.

The higher-order moments normalized to temperature are studied at a constant chemical potential  and different magnetic fields. Also they are studied at different  chemical potentials and a constant magnetic field. The statistical normalization requires scaling with respect to the standard deviation, $\sigma$, where $\sigma$ is related to the susceptibility $\chi$ or the fluctuations.  Due to the sophisticated derivations, the discussion was limited to dimensionless higher-order moments. This can be done when the normalization is done with respect to the temperature or chemical potential. We find that the higher-order moments increase with the magnetic field. We found that the moments increase with the chemical potentials. That the peaks at corresponding critical temperatures can be used to map out the chiral phase-diagrams, $T$ vs. $eB$ and $T$ vs. $\mu$. 

The effects of the magnetic field on the chiral phase-transition have been evaluated. There are different methods proposed to calculate the critical temperature and chemical potential, $\mu_c$, through implementing fluctuations in the normalized higher-order moments or by the order parameters. The latter was implemented in the present work. It is obvious that PLSM has two types of order-parameter. The first one gives the chiral phase-transition and is related to strange and non-strange chiral condensates. The second one gives hints for deconfinement phase-transition. Therefore, we can follow a procedure that at a constant magnetic field and by using strange and non-strange chiral-condensates, a dimensionless quantity would reflect the difference between the non-strange and strange condensates $\Delta_{q,s}(T)$ as a function of temperature at fixed chemical potentials, i.e. chiral phase-transition. At the same chemical potential, we can deduce the other order-parameter related to the Polyakov-loop fields as function in  temperature. Both calculations give one point, at which the two order-parameters crossing each other. It is assumed that such a point represents the transition point at the given chemical potential. We repeat this at various chemical potentials and get a set of points in a two-dimensional chart, the QCD phase-diagram. We have compared five QCD phase-diagrams, $T/T_{c0}$ vs. $\mu/\mu_{c0}$, with each others at five different values of the magnetic field. We found that both critical temperature and critical chemical potential increase with increasing the magnetic field.   

The masses can be deduced from the second derivative of the grand potential with respect to the corresponding fields, evaluated at its minimum, which is estimated at vanishing expectation values of all scalar and pseudoscalar fields. We have studied scalar and pseudoscalar meson masses as function of temperature at two different values of magnetic field and different chemical potentials. We concluded that the meson masses decrease as the temperature increases. This remains until $T$ reaches the critical value. Then, the vacuum effect becomes dominant and rapidly increases with the temperature. The decrease of the critical temperature with increasing chemical potential is represented by the bottoms (minima) in thermal behavior of meson masses before switching on the vacuum effect. At low temperatures, the scalar meson masses normalized to the lowest Matsubara frequency rapidly decrease as the temperature increases. Then, starting from the critical temperature, we find that the thermal dependence almost vanishes. Furthermore, the meson masses increase with increasing magnetic field. This characterizes $T$ vs. $eB$ phase diagram. At high temperatures, we notice that the masses of almost all meson states become temperature independent, i.e. constructing kind of a universal line. This would be seen as a signature for meson dissociation into quarks. In other words, the meson states undergo deconfimement phase-transition. It is worthwhile to highlight that the various meson states likely have different critical temperatures.

\appendix

\section{Magnetic catalysis}
\label{appnd:1}

For simplicity, we assume that the direction of the magnetic field $B$ goes along  $z$-direction. From the magnetic catalysis \cite{Shovkovy2013} and by using Landau quantization, we find that when the system is affected by a strong magnetic field, the quark dispersion relation  will be modified to be quantized by Landau quantum number, $n\geq 0$, and therefore the concept of dimensional reduction will be applied.
\begin{eqnarray}
E_u &=&\sqrt{p_{z}^{2}+m_{q}^{2}+|q_{u}|(2n+1-\sigma) B}, \label{Eu} \\
E_d &=&\sqrt{p_{z}^{2}+m_{q}^{2}+|q_{d}|(2n+1-\sigma) B}, \label{Ed} \\
E_s &=&\sqrt{p_{z}^{2}+m_{s}^{2}+|q_{s}|(2n+1-\sigma) B}, \label{Es}
\end{eqnarray} 
where $\sigma$ is related to the spin quantum number and $S$ ($\sigma=\pm S/2$). Here,  we replace $2n+1-\sigma$ by one quantum number $\nu$, where $\nu=0$ is the Lowest Landau Level (LLL) and the Maximum  Landau Level (MLL) was determined according to Eq. (\ref{MLL}) \cite{Sutapa}, $m_{f}$, where $f$ runs over $u$-, $d$- and $s$-quark mass, 
\begin{eqnarray}
m_q &=& g \frac{\sigma_x}{2}, \label{qmass} \\
m_s &=& g \frac{\sigma_y}{\sqrt{2}}.  \label{sqmass}
\end{eqnarray} 

We apply another magnetic catalysis property \cite{Shovkovy2013}, namely the dimensional reduction. As the name says, the dimensions will be reduced as $D\longrightarrow D-2$. In this situation, the three-momentum integral will transformed into a one-momentum integral
\begin{eqnarray} 
T~\int \dfrac{d^3 p}{(2 \pi)^3}~ \longrightarrow~  \dfrac{|q_{f}| B T}{2 \pi} \sum_{\nu=0}^{\infty} \int \dfrac{d p}{2 \pi} (2-1 \delta_{0\nu}). \label{DR}
\end{eqnarray} 
when $2-1 \delta_{0\nu}$ represents the degenerate in the Landau level, since for LLL we have single degenerate and doublet for the upper  Landau levels, 
\begin{eqnarray} \label{MLL}
\nu_{max} &=& \dfrac{\Lambda_{QCD}^{2}}{2 |q_{f}| B}.
\end{eqnarray} 
We use $m_q$ and $m_s$ for non-strange and strange quark mass, i.e. the masses of light quarks degenerate. This is not the case for the electric charges. In section \ref{sec:approaches}, $q_u$, $q_d$ and $q_s$ are elaborated.

\section{Minimization condition}
\label{appnd:2}

We notice that the thermodynamic potential density as given in Eq. (\ref{potential}), which has seven parameters $m^2, h_x, h_y, \lambda_1, \lambda_2, c$ and $g$, two unknown condensates $\sigma_x$ and $\sigma_y$ and the order parameters for the deconfinement, $\phi$ and $\phi^*$. The six parameters $m^2, h_x, h_y, \lambda_1, \lambda_2 $ and $ c$  are fixed in the vacuum by six experimentally known quantities \cite{Schaefer:2008hk}. In order to evaluate the unknown parameters $\sigma_x$, $ \sigma_y$, $\phi$ and $\phi^*$, we minimize the thermodynamic potential, Eq. (\ref{potential}), with respect to $\sigma_x$, $\sigma_y$, $\phi$ and $\phi^*$ or $\phi_R$ and $\phi^{*}_{R}$. Doing this, we obtain a set of four equations of motion,
\begin{eqnarray}
\left.\frac{\partial \Omega_{1}}{\partial \sigma_x} = \frac{\partial
\Omega_{1}}{\partial \sigma_y}= \frac{\partial \Omega_{1}}{\partial
\phi}= \frac{\partial \Omega_{1}}{\partial \phi^*}\right|_{min} &=& 0, \label{cond1}
\end{eqnarray}
meaning that $\sigma_x=\bar{\sigma_x}$, $\sigma_y=\bar{\sigma_y}$, $\phi=\bar{\phi}$ and $\phi^*=\bar{\phi^*}$ are the global minimum.

\section{Matsubara frequencies}
\label{sec:matsub}

In finite temperature field theory, the Matsubara frequencies are a summation over the discrete imaginary frequency, $S_{\eta}=T \sum_{i \omega_n} g(i \omega_n)$, where $g(i\, \omega_n)$ is a rational function, $\omega_n=2\, n\, \pi\, T$ for bosons and $\omega_n=(2\, n+1)\, \pi\, T$ for fermions and $n=0,\pm 1,\pm 2, \cdots$ is an integer playing the role of a quantum number. By using Matsubara weighting function $h_{\eta}(z)$, which has simple poles exactly located at $z=i\, \omega_n$, then
\bea
S_{\eta} &=& \frac{T}{2 \pi i} \oint g(z)\, h_{\eta}(z)\, dz,
\eea
where $\eta=\pm$ stands for the statistic sign for bosons and fermions, respectively. $h_{\eta}(z)$ can be chosen depending on which half plane the convergence is to be controlled, 
\begin{eqnarray}
h_{\eta}(z) &=& \left\{\begin{array}{l} \eta \frac{1+n_{\eta}(z)}{T}, \\ \\ \eta \frac{n_{\eta}(z)}{T}, \end{array} \right.
\end{eqnarray}
where $n_{\eta}(z)=\left(1+\eta\, e^{z/T} \right)^{-1}$ is the single-particle distribution function.

The mesonic masses are conjectured to have contributions from Matsubara frequencies \cite{lmf1}. Furthermore, at high temperatures, $\geq T_c$, the behavior of the thermodynamic quantities, including the quark susceptibilities, the masses is affected by the interplay between the lowest Matsubara frequency and the Polyakov  loop-correction \cite{lmf2}. We apply normalization for the different mesonic sectors  with respect to lowest Matsubara frequency \cite{Tawfik:2006B} in order to characterize the dissolving temperature of the mesonic bound states. It is found that the different mesonic states have different dissolving temperatures. This would mean that the different mesonic states have different $T_c$'s, at which the bound mesons begin to dissolve into quarks. Therefore, the normalized masses should not be different at $T>T_c$. To a large extend, their thermal and dense dependence should be removed, so that the remaining effects are defined by the free energy \cite{lmf1}, i.e. the masses of {\it free} besons are defined by $m_l$.

That the masses of almost all mesonic states become independent on $T$, i.e. constructing kind of a universal line, this would be seen as a signature for meson dissociation into quarks. It is a deconfinement phase-transition, where the quarks behave almost freely. In other words, the characteristic temperature should not be universal, as well. So far, we conclude that the universal $T_c$ characterizing the QCD phase boundary is indeed an approximative average (over various bound states).    
 
\section*{Acknowledgement}
This work is supported by the World Laboratory for Cosmology And Particle Physics \hbox{(WLCAPP)}, http://wlcapp.net/. AT would like to thank Abdel Magied Diab for the constructive discussion about the implementation of an external magnetic field in the sigma model!




\begin{thebibliography}{99}
\bibitem{Rischke:2003mt} D. R. Rischke, 
 Prog. Part. Nucl. Phys. {\bf 52}, 197 (2004).


\bibitem{Marco2010}
R. Gatto and M. Ruggieri, 
Phys. Rev. D {\bf 82}, 054027 (2010).



\bibitem{Marco2011}
M. Ruggieri, 
PoS FACES QCD  019 (2010).


\bibitem{Marco2011-2}
R. Gatto and M. Ruggieri, 
 Phys. Rev. D {\bf 83},  034016 (2011).


\bibitem{Fraga2013}
E. S. Fraga, B.W. Mintz and J. Schaffner-Bielich, 
Phys. Lett. B {\bf 731}, 154 (2014).


\bibitem{Jens13}
J. O. Andersen, W. R. Naylor, A. Tranberg, 
JHEP {\bf 04}, 187 (2014).



\bibitem{Skokov2011}
V. Skokov, 
Phys. Rev. D {\bf 85}, 034026 (2012).


\bibitem{Claudio2013}
C. Bonati, M. D'Elia, M. Mariti, F. Negro and F. Sanfilippo, 
Phys. Rev. Lett. {\bf 111}, 182001 (2013).


\bibitem{lattice-delia}
M.~D'Elia, S.~Mukherjee and F.~Sanfilippo, 
 Phys.\ Rev.\  D {\bf 82}, 051501 (2010).



\bibitem{G_Endrödi} G. Endrödi , 
JHEP {\bf 1304}, 023 (2013).


\bibitem{Klevansky1989} S.P. Klevansky and R.H. Lemmer, Phys. Rev. D {\bf 39}, 3478 (1989).

\bibitem{Gusynin1995} V.P. Gusynin, V.A. Miransky, and I.A. Shovkovy, Phys. Lett. B {\bf 349}, 477 (1995). 

\bibitem{Babansky1998} A.Y. Babansky, E.V. Gorbar, and G.V. Shchepanyuk, Phys. Lett. B 419 , 272 (1998).

\bibitem{Klimenko1999} K.G. Klimenko, in Proceedings of the 5th International Workshop On Thermal Field Theories and Their Applications, edited by U. Heinz (Regensburg Univ., Regensburg, Germany, 1999).

\bibitem{Semenoff1999} G.W. Semenoff, I.A. Shovkovy, and L.C.R. Wijewardhana, Phys. Rev. D {\bf 60}, 105024 (1999).

\bibitem{Goyal2000}  A. Goyal and M. Dahiya, Phys. Rev. D {\bf 62}, 025022 (2000).

\bibitem{Hiller2008} B. Hiller, A.A. Osipov, A.H. Blin, and J. da Providencia, SIGMAP Bulletin {\bf 4}, 024 (2008).

\bibitem{Ayala2006} A. Ayala, A. Bashir, A. Raya, and E. Rojas, Phys. Rev. D {\bf 73}, 105009 (2006). 

\bibitem{Boomsma2010} J.K. Boomsma and D. Boer, Phys. Rev. D {\bf 81, } 074005 (2010).

\bibitem{Klevansky1992} S. P. Klevansky, Rev. Mod. Phys. {\bf 64}, 649 (1992).

\bibitem{Menezes2009} D. P. Menezes, M. B. Pinto, S.S. Avancini, A. P. Martinez and C. Providencia, Phys. Rev. C {\bf 79}, 035807 (2009).

\bibitem{Menezes_2009} D. P. Menezes, M. Benghi Pinto, S. S. Avancini, and C. Providencia, Phys. Rev. C {\bf 80}, 065805 (2009).

\bibitem{Shushpanov1997} I.A. Shushpanov and A.V. Smilga, Phys. Lett. B {\bf 402}, 351(1997).

\bibitem{Agasian2000} N.O. Agasian and I.A. Shushpanov, Phys. Lett. B {\bf 472}, 143 (2000).

\bibitem{Cohen2007} T.D. Cohen, D.A. McGady, and E.S. Werbos, Phys. Rev. C {\bf 76}, 055201 (2007).

\bibitem{Kabat2002} D. Kabat, K.M. Lee, and E. Weinberg, Phys. Rev. D {\bf 66}, 014004 (2002).

\bibitem{Miransky2002} V.A. Miransky and I.A. Shovkovy, Phys. Rev. D {\bf 66}, 045006 (2002).

\bibitem{Klimenko2008} K.G. Klimenko and V.C. Zhukovsky, Phys. Lett. B {\bf 665}, 352 (2008).

\bibitem{Fukushima2010} K. Fukushima, M. Ruggieri, and R. Gatto, Phys. Rev. D {\bf 81}, 114031 (2010).


\bibitem{Braguta:2011hq}
 V.~V.~Braguta, P.~V.~Buividovich, M.~N.~Chernodub and M.~I.~Polikarpov, 
Phys. Lett. B {\bf 718}, 671 (2012).  


\bibitem{Bali:2011qj} 
G. S. Bali, F. Bruckmann, G. Endrodi, Z. Fodor, S.~D. Katz, S. Krieg, A. Schafer and K. K. Szabo, 
JHEP {\bf 1202}, 044 (2012).

   
\bibitem{Bali:2012zg} 
G. S. Bali, F. Bruckmann, G. Endrodi, Z. Fodor, S. D. Katz and A. Schafer, 
Phys. Rev. D {\bf 86}, 071502 (2012).
 
\bibitem{lattice-maxim}
P. V. Buividovich, M. N. Chernodub, E. V. Luschevskaya and M. I. Polikarpov, 
Phys.\ Lett.\  B {\bf 682}, 484 (2010).


\bibitem{Mizher:2010zb}
A. J. Mizher, M. N. Chernodub and E. S. Fraga, 
 Phys.\ Rev.\ D {\bf 82}, 105016 (2010).

 
\bibitem{Marco2012}
 M. Ruggieri, M. Tachibana and V. Greco, 
 JHEP. {\bf 1307}, 165, (2013).
 

\bibitem{Landau}
L. D. Landau and E. M. Lifshitz, {\it Quantum Mechanics}, (Butterworth Heinemann, Amesterdam, 1965).


\bibitem{Gabriel2012}
G. N. Ferrari, A. F. Garcia and M. B. Pinto, 
Phys. Rev. D {\bf 86},  096005 (2012).
 
 
 \bibitem{Tawfik:2014uka} A. Tawfik, N. Magdy and A. Diab, 
Phys. Rev. C {\bf 89}, 055210 (2014).  


\bibitem{Gell Mann:1960} M. Gell-Mann and M. Levy, 
Nuovo Cimento. {\bf 16}, 53 (1960).
 
 

\bibitem{Lenaghan:1999si} J. T. Lenaghan and D. H. Rischke, 
J. Phys. G {\bf 26}, 431 (2000).


\bibitem{Petropoulos:1998gt} N. Petropoulos,  
 J. Phys. G {\bf 25}, 2225 (1999). 


\bibitem{l} M. Levy, Nuovo Cim. {\bf 52}, 23 (1967).  


\bibitem{Hu:1974qb} B. Hu, 
Phys. Rev. D {\bf 9}, 1825 (1974). 

\bibitem{Schechter:1975ju} J. Schechter  and  M. Singer, 
Phys. Rev.  D {\bf 12}, 2781 (1975). 


\bibitem{Geddes:1979nd} H. B. Geddes, 
Phys. Rev. D {\bf 21}, 278 (1980). 


\bibitem{Polyakov:1978vu}  A.~M.~Polyakov, 
 Phys. Lett. B {\bf 72}, 477 (1978).



\bibitem{Susskind:1979up}  L.~Susskind, 
  Phys. Rev. D {\bf 20}, 2610 (1979).



\bibitem{Svetitsky:1982gs}  B. Svetitsky and L. G. Yaffe, 
  Nucl. Phys. B {\bf 210}, 423 (1982).

\bibitem{Svetitsky:1985ye}  B. Svetitsky, 
Phys. Rep. {\bf 132}, 1 (1986).
  

\bibitem{Wambach:2009ee} J. Wambach, B.-J. Schaefer and M. Wagner, 
Acta Phys. Polon. Supp. {\bf 3}, 691 (2010).

\bibitem{Kahara:2008} T. Kahara and K. Tuominen, Phys. Rev. D {\bf 78}, 034015 (2008).

\bibitem{Schaefer:2008ax} B.-J. Schaefer and M. Wagner, 
Prog. Part. Nucl. Phys. 62, 381 (2009).

\bibitem{Mao:2010} H. Mao, J. Jin and M. Huang, 
J. Phys. G{\bf 37}, 035001 (2010).

 
\bibitem{Lenaghan}
J. T. Lenaghan, D. H. Rischke and J. Schaffner-Bielich, 
Phys. Rev. D {\bf 62}, 085008 (2000).
  


\bibitem{Schaefer:2008hk} B. J. Schaefer and M. Wagner, 
Phys. Rev. D~{\bf 79}, 014018 (2009).


\bibitem{blind} 
O. Scavenius, A. Mocsy, I. N. Mishustin, and D. H. Rischke, 
Phys. Rev. C {\bf 64}, 045202 (2001). 
 
\bibitem{Fukushima:2003fw} K. Fukushima, 
Phys. Lett. B {\bf 591}, 277 (2004).
 
 
\bibitem{Roessner:2007}
 S. Rossner, C. Ratti, and W. Weise, 
 Phys. Rev. D {\bf 75}, 034007 (2007).
 
 
\bibitem{Ratti:2005jh} C. Ratti, M. A. Thaler  and W. Weise, 
Phys. Rev. D {\bf 73}, 014019 (2006).
 
\bibitem{Fukushima:2008wg} K. Fukushima, 
Phys. Rev. D {\bf 77}, 114028 (2008). 

\bibitem{Schaefer:2007d}
  B.-J. Schaefer, J. M. Pawlowski, and J. Wambach, 
 Phys. Rev. D {\bf 76}, 074023 (2007).
  
\bibitem{Tawfik:2014hwa} A. N. Tawfik and N. Magdy,  
Phys. Rev. C {\bf 90}, 015204 (2014). 

\bibitem{Tawfik_rev2014} A. Tawfik, 
Int. J. Mod. Phys. A {\bf 29}, 1430021  (2014).

\bibitem{Tawfik:2012si} A. Tawfik, 
Adv. High Energy Phys. {\bf 2013}, 574871 (2013). 

\bibitem{Vivek} Vivek Kumar Tiwari, Phys. Rev. D {\bf 88 }, 074017  (2013). 


\bibitem{Ferreira2014} M. Ferreira, P. Costa, O. Loureno, T. Frederico and C. Providencia,  Phys. Rev. D {\bf 89} 116011 (2014).


\bibitem{Shovkovy2013} I. A. Shovkovy,
Lect. Notes Phys. {\bf 871}, 13 (2013).


\bibitem{Sutapa} S. Ghosh, S. Mandal and S. Chakrabarty,
Phys. Rev. C {\bf 75}, 015805 (2007).

\bibitem{lmf1} W. Florkowski and B. L. Friman, Z. Phys. A {\bf 347}, 271 (1994).

\bibitem{lmf2} K. Dusling, C. Ratti and I. Zahed, Phys. Rev. D {\bf 79}, 034027 (2009).

\bibitem{Tawfik:2006B} A. Tawfik, Soryushiron Kenkyu {\bf 114}, B48-B50 (2006). 

\bibitem{simon} Simon Hands, Timothy J. Hollowood and Joyce C. Myers, 
JHEP {\bf 1007}, 086 (2010). 

\bibitem{PDG:2012} J. Beringer {\it et al.} (Particle Data Group), Phys. Rev. D {\bf 86}, 010001 (2012).

\bibitem{NJL:2013} T. Xia, L. He and P. Zhuang, Phys. Rev. D {\bf 88}, 056013 (2013). 

\bibitem{P. Costa:PNJL}
P. Costa, M. C. Ruivo, C. A. de Sousa and Y. L. Kalinovsky, Phys. Rev. D {\bf 70}, 116013 (2004); \\ 
P. Costa, M. C. Ruivo, C. A. de Sousa and Y. L. Kalinovsky, Phys. Rev. D {\bf 71}, 116002 (2005); \\ 
P. Costa, M. C. Ruivo, C. A. de Sousa, H. Hansen and W. M. Alberico, Phys. Rev. D {\bf 79}, 116003 (2009).  

\bibitem{HotQCD}  A. Bazavov, {\it et al.}  (HotQCD Collaboration), 
Phys. Rev. D {\bf 85}, 054503 (2012). 

\bibitem{PACS-CS} S. Aoki, {\it et al.} (PACS-CS Collaboration), 
Phys. Rev. D {\bf 81}, 074503 (2010). 

\bibitem{AD2} A. Tawfik and A. Diab, {\it Polyakov SU(3) extended linear $\sigma$-model: Sixteen mesonic states in chiral phase-structure}, to appear in PRC.


\end{thebibliography}
\end{document}